\documentclass[journal]{IEEEtran}

\pdfoutput=1

\usepackage{bobbystyle_IEEEjournal}

\usepackage{tikz}
\usetikzlibrary{spy}

\usepackage[resetlabels]{multibib}
\newcites{Supp}{References}

\newcommand{\quotes}[1]{`#1'}
\newcommand{\dquotes}[1]{``#1''}

\usepackage{chngcntr}
\counterwithout{equation}{section}

\tikzstyle{only in spy node magn 1.75}=[%
   transform canvas={%
      shift={(tikzspyinnode)},
      scale=1.75,
   }
]

\newcommand{\specialcell}[2][c]{%
  \begin{tabular}[#1]{@{}c@{}}#2\end{tabular}}

\newcolumntype{C}[1]{>{\centering\let\newline\\\arraybackslash\hspace{0pt}}m{#1}}  

\renewcommand\thetheorem{\arabic{section}.\arabic{theorem}}
\renewcommand{\theequation}{\arabic{equation}}

\newenvironment{myquote}[1]%
  {\list{}{\leftmargin=#1\rightmargin=#1}\item[]}%
  {\endlist}

\begin{document}

\title{Convolutional Analysis Operator Learning: Acceleration and Convergence}

\author{Il Yong Chun, \textit{Member}, \textit{IEEE}, and Jeffrey A. Fessler, \textit{Fellow}, \textit{IEEE}

\thanks{This work is supported in part by the Keck Foundation and NIH U01 EB018753.}

\thanks{This paper has supplementary material.
The prefix ``S'' indicates the numbers in section, equation, figure, algorithm, and footnote in the supplement.
Contact iychun@hawaii.edu for further questions about this work.}

\thanks{Il Yong Chun was with the Department of Electrical Engineering and Computer Science, The University of Michigan, Ann Arbor, MI 48019 USA,
and is now with the Department of Electrical Engineering, the University of Hawai'i at M\=anoa, Honolulu, HI 96822 USA (email: iychun@hawaii.edu).
Jeffrey A. Fessler is with the Department of Electrical Engineering and Computer Science, The University of Michigan, Ann Arbor, MI 48019 USA (email: fessler@umich.edu).}
}

\maketitle

\begin{abstract}
Convolutional operator learning is gaining attention in many signal processing and computer vision applications.
Learning kernels has mostly relied on so-called \emph{patch-domain} approaches that extract and store many overlapping patches across training signals. 
Due to memory demands, patch-domain methods have limitations when learning kernels from large datasets -- particularly with multi-layered structures, e.g., convolutional neural networks --  or when applying the learned kernels to high-dimensional signal recovery problems.
The so-called \emph{convolution} approach does not store many overlapping patches, and thus overcomes the memory problems particularly with careful algorithmic designs; it has been studied within the \dquotes{synthesis} signal model, e.g., convolutional dictionary learning.
This paper proposes a new \textit{convolutional analysis operator learning} (CAOL) framework that learns an analysis sparsifying regularizer with the convolution perspective, and develops a new convergent \textit{Block Proximal Extrapolated Gradient method using a Majorizer} (BPEG-M) to solve the corresponding block multi-nonconvex problems.
To learn diverse filters within the CAOL framework, this paper introduces an orthogonality constraint that enforces a tight-frame filter condition, and a regularizer that promotes diversity between filters.
Numerical experiments show that, with sharp majorizers, BPEG-M significantly accelerates the CAOL convergence rate compared to the state-of-the-art block proximal gradient (BPG) method.
Numerical experiments for sparse-view computational tomography show that a convolutional sparsifying regularizer learned via CAOL significantly improves reconstruction quality compared to a conventional edge-preserving regularizer. Using more and wider kernels in a learned regularizer better preserves edges in reconstructed images.
\end{abstract}

\section{Introduction} \label{sec:intro}

\IEEEPARstart{L}{earning} convolutional operators from large datasets is a growing trend in signal/image processing, computer vision, and machine learning.
The widely known \emph{patch-domain} approaches for learning kernels (e.g., filter, dictionary, frame, and transform) extract patches from training signals for simple mathematical formulation and optimization, yielding (sparse) features of training signals \cite{Aharon&Elad&Bruckstein:06TSP, Elad&Aharon:06TIP, Yaghoobi&etal:13TSP, Hawe&Kleinsteuber&Diepold:13TIP, Mairal&Bach&Ponce:14FTCGV, Cai&etal:14ACHA, Ravishankar&Bressler:15TSP, Pfister&Bresler:15SPIE, Coates&NG:bookCh}.
Due to memory demands, using many overlapping patches across the training signals hinders using large datasets
and building hierarchies on the features, 
e.g., deconvolutional neural networks \cite{Zeiler&etal:10CVPR}, 
convolutional neural network (CNN) \cite{LeCun&etal:98ProcIEEE}, 
and multi-layer convolutional sparse coding \cite{Papyan&Romano&Elad:17JMLR}.
For similar reasons, the memory requirement of patch-domain approaches discourages learned kernels from being applied to large-scale inverse problems.

To moderate these limitations of the patch-domain approach, the so-called \emph{convolution} perspective has been recently introduced by learning filters and obtaining (sparse) representations directly from the original signals without storing many overlapping patches, e.g.,  convolutional dictionary learning (CDL) \cite{Zeiler&etal:10CVPR, Bristow&etal:13CVPR, Heide&eta:15CVPR, Wohlberg:16TIP, Chun&Fessler:18TIP, Chun&Fessler:17SAMPTA}. 
For large datasets, CDL using careful algorithmic designs \cite{Chun&Fessler:18TIP} is more suitable for learning filters than patch-domain dictionary learning \cite{Aharon&Elad&Bruckstein:06TSP}; in addition, CDL can learn translation-invariant filters without obtaining highly redundant sparse representations \cite{Chun&Fessler:18TIP}.
The CDL method applies the convolution perspective for learning kernels within \dquotes{synthesis} signal models.
Within \dquotes{analysis} signal models, however, there exist no prior frameworks using the convolution perspective for learning convolutional operators, whereas patch-domain approaches for learning analysis kernels are introduced in \cite{Yaghoobi&etal:13TSP, Hawe&Kleinsteuber&Diepold:13TIP, Cai&etal:14ACHA, Ravishankar&Bressler:15TSP, Pfister&Bresler:15SPIE}.
(See brief descriptions about synthesis and analysis signal models in \cite[Sec.~\Romnum{1}]{Hawe&Kleinsteuber&Diepold:13TIP}.)

Researchers interested in dictionary learning have actively studied the structures of kernels learned by the patch-domain approach \cite{Yaghoobi&etal:13TSP, Hawe&Kleinsteuber&Diepold:13TIP, Cai&etal:14ACHA, Ravishankar&Bressler:15TSP, Pfister&Bresler:15SPIE, Barchiesi&Plumbley:13TSP, Bao&Cai&Ji:13ICCV, Ravishankar&Bressler:13ICASSP}. 
In training CNNs (see Appendix~\ref{sec:CNN}), however,
there has been less study of filter structures having non-convex constraints, e.g., orthogonality and unit-norm constraints in Section~\ref{sec:CAOL}, although it is thought that diverse (i.e., incoherent) filters can improve performance for some applications, e.g., image recognition \cite{Coates&NG:bookCh}.
On the application side, researchers have applied (deep) NNs to signal/image recovery problems.
Recent works combined model-based image reconstruction (MBIR) algorithm with image refining networks \cite{Yang&etal:16NIPS, Zhang&etal:17CVPR, Chen&Pock:17PAMI, Chen&etal:17arXiv, Wu&etal:17Fully3D, Romano&Elad&Milanfar:17SJIS, Buzzard&etal:18SJIS, Chun&Fessler:18IVMSP, Chun&etal:18arXiv:momnet, Chun&etal:18Allerton}.
In these \emph{iterative} NN methods, 
refining NNs should satisfy the non-expansiveness for fixed-point convergence 
\cite{Chun&etal:18arXiv:momnet}; 
however, their trainings lack consideration of filter diversity constraints, e.g., orthogonality constraint in Section~\ref{sec:CAOL}, 
and thus it is unclear whether the trained NNs are nonexpansive mapping \cite{Chun&etal:18Allerton}.

\begin{figure*}[!pt]
\centering

\begin{tabular}{c}
\includegraphics[trim={3cm 11.2cm 3cm 11.3cm},clip,scale=0.46]{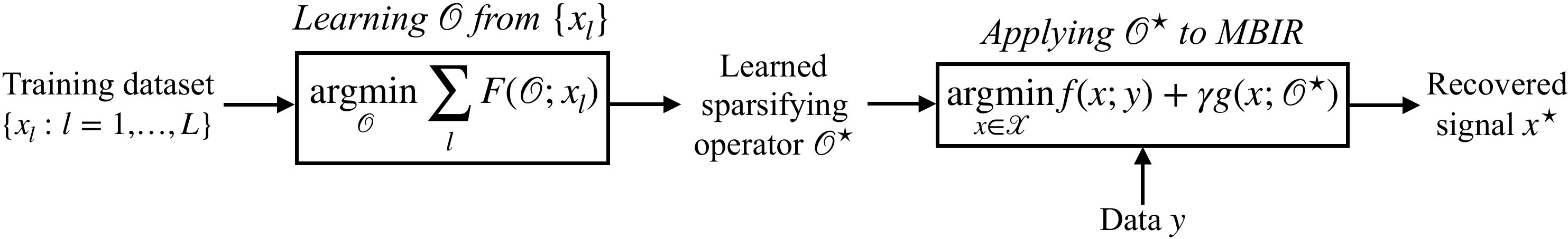}
\end{tabular}

\vspace{-0.25em}
\caption{A general flowchart from
learning sparsifying operators $\cO$ to 
solving inverse problems via MBIR using learned operators $\cO^\star$;
see Section~\ref{sec:back}.
For the $l\rth$ training sample $x_l$, 
$F(\cO; x_l)$ measures its sparse representation or sparsification errors, and
sparsity of its representation generated by $\cO$.
}
\label{diag:abstract}
\end{figure*}

This paper proposes \textit{1)} a new \textit{convolutional analysis operator learning} (CAOL) framework that learns an analysis sparsifying regularizer with the convolution perspective, and \textit{2)} a new convergent \textit{Block Proximal Extrapolated Gradient method using a Majorizer} (BPEG-M \cite{Chun&Fessler:18TIP}) for solving block multi-nonconvex problems \cite{Xu&Yin:17JSC}.
To learn diverse filters, we propose \textit{a)} CAOL with an orthogonality constraint that enforces a tight-frame (TF) filter condition in convolutional perspectives, and \textit{b)} CAOL with a regularizer that promotes filter diversity.
BPEG-M with sharper majorizers converges significantly faster than the state-of-the-art technique, Block Proximal Gradient (BPG) method \cite{Xu&Yin:17JSC} for CAOL.
This paper also introduces a new X-ray computational tomography (CT) MBIR model using a convolutional sparsifying regularizer learned via CAOL \cite{Chun&Fessler:18Asilomar}.

The remainder of this paper is organized as follows. 
Section~\ref{sec:back} reviews how learned regularizers can help solve inverse problems.
Section~\ref{sec:CAOL} proposes the two CAOL models.
Section~\ref{sec:reBPG-M} introduces BPEG-M with several generalizations, analyzes its convergence, and applies a momentum coefficient formula and restarting technique from \cite{Chun&Fessler:18TIP}.
Section~\ref{sec:CAOL+BPGM} applies the proposed BPEG-M methods to the CAOL models, designs two majorization matrices, and describes memory flexibility and applicability of parallel computing to BPEG-M-based CAOL.
Section~\ref{sec:CTrecon} introduces the CT MBIR model using a convolutional regularizer learned via CAOL \cite{Chun&Fessler:18Asilomar}, along with its properties, i.e., its mathematical relation to a convolutional autoencoder, the importance of TF filters, and its algorithmic role in signal recovery.
Section~\ref{sec:result} reports numerical experiments that show \textit{1)} the importance of sharp majorization in accelerating BPEG-M, and \textit{2)} the benefits of BPEG-M-based CAOL -- acceleration, convergence, and memory flexibility.
Additionally, Section~\ref{sec:result} reports sparse-view CT experiments that show \textit{3)} the CT MBIR using learned convolutional regularizers significantly improves the reconstruction quality compared to that using a conventional edge-preserving (EP) regularizer, and \textit{4)} more and wider filters in a learned regularizer better preserves edges in reconstructed images.
Finally, Appendix~\ref{sec:CNN} mathematically formulates unsupervised training of CNNs via CAOL, and shows that its updates attained via BPEG-M correspond to the three important CNN operators.
Appendix~\ref{sec:egs} introduces some potential applications of CAOL to image processing, imaging, and computer vision.

\section{Backgrounds: MBIR Using \emph{Learned} Regularizers} \label{sec:back}

To recover a signal $x \in \bbC^{N'}$ from a data vector $y \in \bbC^m$, 
one often considers the following MBIR optimization problem
(Appendix~\ref{sec:notation} provides mathematical notations): 
$
\argmin_{x \in \cX} f(x; y) + \gamma \, g(x),
$
where $\cX$ is a feasible set,
$f(x; y)$ is data fidelity function that models imaging physics (or image formation) and noise statistics, 
$\gamma \!>\! 0$ is a regularization parameter,
and $g(x)$ is a regularizer, such as total variation \cite[\S2--3]{Arridge&etal:19AN}. 
However, when inverse problems are extremely ill-conditioned, 
the MBIR approach using hand-crafted regularizers $g(x)$ has limitations in recovering signals.
Alternatively, there has been a growing trend in learning sparsifying regularizers
(e.g., convolutional regularizers \cite{Chun&Fessler:18TIP, Chun&Fessler:17SAMPTA, Chun&etal:19SPL, Chun&Fessler:18Asilomar, Crockett&etal:19CAMSAP}) 
from training datasets and applying the learned regularizers to the following MBIR problem \cite{Arridge&etal:19AN}:
\be{
\label{eq:mbir:learn}
\argmin_{x \in \cX} f(x; y) + \gamma g(x; \cO^\star),
\tag{B1}
}
where a learned regularizer $g(x; \cO^\star)$ quantifies consistency between 
any candidate $x$ and training data that is encapsulated in some trained sparsifying operators $\cO^\star$.
The diagram in Fig.~\ref{diag:abstract} shows the general process from
training sparsifying operators to solving inverse problems via \R{eq:mbir:learn}.
Such models \R{eq:mbir:learn} arise in a wide range of applications. 
See some examples in Appendix~\ref{sec:egs}.

This paper describes multiple aspects of learning convolutional regularizers. 
The next section first starts with proposing a new convolutional regularizer.

\section{CAOL: Models \textit{Learning} Convolutional Regularizers}  \label{sec:CAOL}

\begin{figure*}[!pt]
\centering

\begin{tabular}{c}
\includegraphics[trim={2.2cm 9.5cm 2.2cm 9.9cm},clip,scale=0.46]{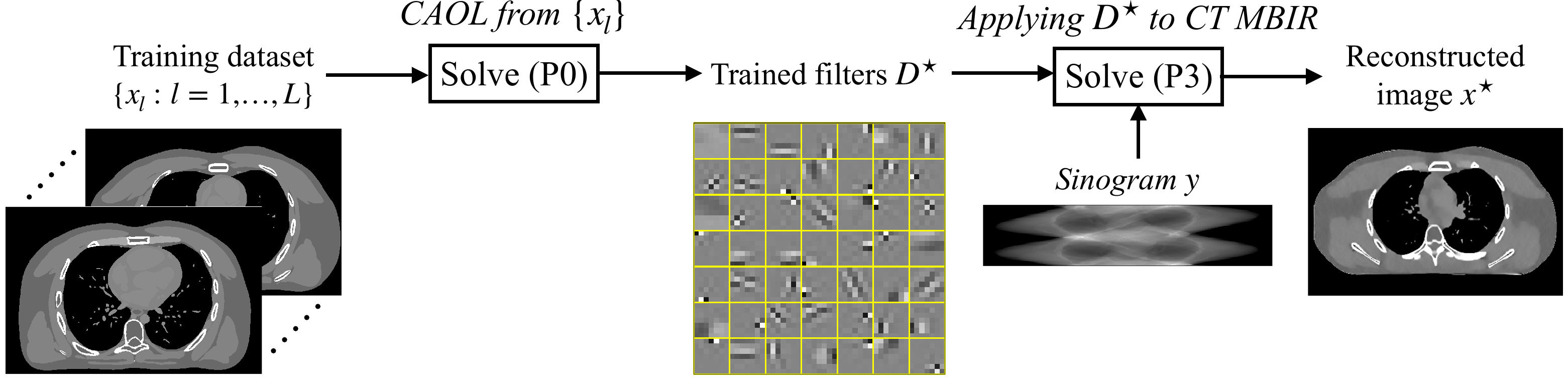}
\end{tabular}

\vspace{-0.25em}
\caption{A flowchart from CAOL \R{sys:CAOL} to 
MBIR using a convolutional sparsifying regularizer learned via CAOL \R{sys:CT&CAOL}
in sparse-view CT.
See details of the CAOL process \R{sys:CAOL} and its variants \R{sys:CAOL:orth}--\R{sys:CAOL:div}, 
and the CT MBIR process \R{sys:CT&CAOL} in 
Section~\ref{sec:CAOL} and Section~\ref{sec:CTrecon}, respectively.
}
\label{diag:caol-ctmbir}
\end{figure*}

The goal of CAOL is to find a set of filters that \dquotes{best} sparsify a set of training images.
Compared to hand-crafted regularizers,
learned convolutional regularizers 
can better extract \dquotes{true} features of estimated images and 
remove \dquotes{noisy} features with thresholding operators.
We propose the following CAOL model:
\ea{
\label{sys:CAOL}
\argmin_{D = [d_1, \ldots, d_K]} \min_{\{ z_{l,k} \}} &~ F(D, \{ z_{l,k} \}) + \beta g (D),  \tag{P0}
\\
F(D, \{ z_{l,k} \}) & :=  \sum_{l=1}^L \sum_{k=1}^K \frac{1}{2}  \left\| d_k \circledast x_l - z_{l,k} \right\|_2^2 + \alpha \| z_{l,k} \|_0, \nn
} 
where $\circledast$ denotes a convolution operator (see details about boundary conditions in the supplementary material), $\{ x_l \in \bbC^N : l =1,\ldots,L \}$ is a set of training images, $\{ d_{k} \in \bbC^{R}: k = 1,\ldots, K \}$ is a set of convolutional kernels, $\{ z_{l,k} \in \bbC^N : l = 1,\ldots,L, k=1,\ldots,K \}$ is a set of sparse codes, and $g ( D )$ is a regularizer or constraint that encourages filter diversity or incoherence, $\alpha \!>\! 0$ is a thresholding parameter controlling the sparsity of features $\{  z_{l,k} \}$, and $\beta > 0$ is a regularization parameter for $g (D)$.
We group the $K$ filters into a matrix $D \in \bbC^{R \times K}$:
\be{
\label{eq:D}
D := \left[ \arraycolsep=3pt \begin{array}{ccc} d_1 & \ldots & d_K \end{array} \right].
}
For simplicity, we fix the dimension for training signals, i.e., $\{ x_l, z_{l,k} \in \bbC^N \}$, but the proposed model \R{sys:CAOL} can use training signals of different dimension, i.e., $\{ x_l, z_{l,k} \in \bbC^{N_l} \}$.
For sparse-view CT in particular, 
the diagram in Fig.~\ref{diag:caol-ctmbir} shows the process from
CAOL \R{sys:CAOL} to solving its inverse problem via
MBIR using learned convolutional regularizers.

The following two subsections design the constraint or regularizer $g(D)$ 
to avoid redundant filters (without it, all filters could be identical).

\subsection{CAOL with Orthogonality Constraint} \label{sec:CAOL:orth}

We first propose a CAOL model with a nonconvex orthogonality constraint on the filter matrix $D$ in \R{eq:D}:
\be{
\label{sys:CAOL:orth}
\argmin_{D} \min_{\{ z_{l,k} \}} ~ F(D, \{ z_{l,k} \}) \quad \mathrm{subj.~to} ~ D D^H = \frac{1}{R} \cdot I. \tag{P1}
} 
The orthogonality condition $D D^H = \frac{1}{R} I$ in \R{sys:CAOL:orth} enforces a TF condition on the filters $\{ d_k \}$ in CAOL \R{sys:CAOL}.
Proposition~\ref{p:TFconst} below formally states this relation.

\prop{[Tight-frame filters]\label{p:TFconst}
Filters satisfying the orthogonality constraint 
$D D^H = \frac{1}{R} I$ in \R{sys:CAOL:orth} satisfy the following TF condition in a convolution perspective:
\be{
\label{eq:CAOL:TFcond}
\sum_{k=1}^K \nm{ d_k \circledast x }_2^2 = \nm{ x }_2^2, \quad \forall x \in \bbC^N,
}
for both circular and symmetric boundary conditions. 
}

\prf{\renewcommand{\qedsymbol}{}
See Section~\ref{sec:prf:p:TF} of the supplementary material.
}

Proposition~\ref{p:TFconst} corresponds to a TF result from patch-domain approaches; see~Section~\ref{sec:prf:p:TF}. (Note that the patch-domain approach in \cite[Prop.~3]{Cai&etal:14ACHA} requires $R = K$.)
However, we constrain the filter dimension to be $R \leq K$ to have an efficient solution for CAOL model \R{sys:CAOL:orth}; see Proposition~\ref{p:orth} later.
The following section proposes a more flexible CAOL model in terms of the filter dimensions $R$ and $K$.

\subsection{CAOL with Diversity Promoting Regularizer}

As an alternative to the CAOL model \R{sys:CAOL:orth}, we propose a CAOL model with a diversity promoting regularizer and a nonconvex norm constraint on the filters $\{ d_k \}$:
\ea{
\label{sys:CAOL:div}
\argmin_{D} \min_{\{ z_{l,k} \}} &~ F(D, \{ z_{l,k} \}) + \frac{\beta}{2} \overbrace{\left\| D^H D - \frac{1}{R} \cdot I \right\|_{\mathrm{F}}^2}^{\mathrm{\hbox{$=: g_{\text{div}} (D)$}}}, \nn
\\
\mathrm{subject~to} ~&~ \| d_{k} \|_2^2 = \frac{1}{R}, \quad k = 1,\ldots, K.  \tag{P2}
} 
In the CAOL model \R{sys:CAOL:div}, we consider the following:
\bulls{
\item The constraint in \R{sys:CAOL:div} forces the learned filters $\{ d_k \}$ to have uniform energy. In addition, it avoids the \dquotes{scale ambiguity} problem \cite{Rem&Karin:10TIT}. 

\item The regularizer in \R{sys:CAOL:div}, $g_{\text{div}} (D)$, promotes filter diversity, i.e., incoherence between $d_k$ and $\{ d_{k'} : k' \neq k \}$, measured by $| \ip{d_k}{d_{k'}} |^2$ for $k \neq k'$.
}

When $R = K$ and $\beta \rightarrow \infty$, the model \R{sys:CAOL:div} becomes \R{sys:CAOL:orth} since $D^H D = \frac{1}{R} I$ implies $D D^H = \frac{1}{R} I$ (for square matrices $A$ and $B$, if $AB=I$ then $BA = I$).
Thus \R{sys:CAOL:div} generalizes \R{sys:CAOL:orth} by relaxing the off-diagonal elements of the equality constraint in \R{sys:CAOL:orth}. (In other words, when $R=K$, the orthogonality constraint in \R{sys:CAOL:orth} enforces the TF condition and promotes the filter diversity.)
One price of this generalization is the extra tuning parameter $\beta$.

\R{sys:CAOL:orth}--\R{sys:CAOL:div} are challenging nonconvex optimization problems and block optimization approaches seem suitable. 
The following section proposes a new block optimization method with momentum and majorizers, 
to rapidly solve the multiple block multi-nonconvex problems proposed in this paper, while guaranteeing convergence to critical points.

\section{BPEG-M: Solving Block Multi-Nonconvex Problems with Convergence Guarantees} \label{sec:reBPG-M}

This section describes a new optimization approach, BPEG-M,
for solving block multi-nonconvex problems like 
\textit{a)} CAOL \R{sys:CAOL:orth}--\R{sys:CAOL:div},\footnote{
A block coordinate descent algorithm can be applied to CAOL \R{sys:CAOL:orth};
however, its convergence guarantee in solving CAOL \R{sys:CAOL:orth} is not yet known and might require stronger sufficient conditions than BPEG-M \cite{Tseng:2001JOTA}.
} 
\textit{b)} CT MBIR \R{sys:CT&CAOL} using learned convolutional regularizer via \R{sys:CAOL:orth} (see Section~\ref{sec:CTrecon}),
and \textit{c)} \dquotes{hierarchical} CAOL \R{sys:CNN:orth} (see Appendix~\ref{sec:CNN}).

\subsection{BPEG-M -- Setup} \label{sec:BPGM:setup}

We treat the variables of the underlying optimization problem either as a single block or multiple disjoint blocks. 
Specifically, consider the following \textit{block multi-nonconvex} optimization problem:
\ea{
\label{sys:multiConvx}
\min &~ F(x_1,\ldots, x_B) := f(x_1,\ldots,x_B) + \sum_{b=1}^B g_b (x_b),
}
where variable $x$ is decomposed into $B$ blocks $x_1 ,\ldots,x_B$ ($\{ x_b \in \bbR^{n_b} : b=1,\ldots,B \}$), $f$ is assumed to be continuously differentiable, but functions $\{ g_b : b = 1,\ldots,B \}$ are not necessarily differentiable. 
The function $g_b$ can incorporate the constraint $x_b \in \cX_b$, by allowing any $g_b$ to be extended-valued, e.g., $g_b (x_b) = \infty$ if $x_b \notin \cX_b$, for $b=1,\ldots,B$.
It is standard to assume that both $f$ and $\{ g_b \}$ are closed and proper and the sets $\{ \cX_b \}$ are closed and nonempty.
We do \textit{not} assume that $f$, $\{ g_b \}$, or $\{ \cX_b \}$ are convex. Importantly, $g_b$ can be a nonconvex $\ell^p$ quasi-norm, $p \in [0, 1)$.
The general block multi-convex problem in \cite{Chun&Fessler:18TIP, Xu&Yin:13SIAM} is a special case of \R{sys:multiConvx}.

The BPEG-M framework considers a more general concept than Lipschitz continuity of the gradient as follows:

\defn{[$M$-Lipschitz continuity] \label{d:QM}
A function $g: \bbR^n \rightarrow \bbR^{n}$ is \emph{$M$-Lipschitz continuous} on $\bbR^n$ if there exist a (symmetric) positive definite matrix $M$ such that
\bes{
\nm{g(x) - g(y)}_{M^{-1}} \leq \nm{x - y}_{M}, \quad \forall x,y,
}
where $\nm{x}_{M}^2 := x^T M x$.
}

Lipschitz continuity is a special case of $M$-Lipschitz continuity with $M$ equal to a scaled identity matrix with 
a Lipschitz constant of the gradient $\nabla f$ (e.g., for $f(x) = \frac{1}{2} \| Ax - b\|_2^2$, the (smallest) Lipschitz constant of $\nabla f$ is the maximum eigenvalue of $A^T A$).
If the gradient of a function is $M$-Lipschitz continuous, then we obtain the following quadratic majorizer (i.e., surrogate function \cite{Lange&Hunter&Yang:00JCGS, Jacobson&Fessler:07TIP}) at a given point $y$ without assuming convexity:

\lem{[Quadratic majorization (QM) via $M$-Lipschitz continuous gradients]
\label{l:QM}
Let $f : \bbR^n \rightarrow \bbR$. If $\nabla f$ is $M$-Lipschitz continuous, then
\bes{
f(x) \leq f(y) + \ip{\nabla f(y)}{x-y} + \frac{1}{2} \nm{x - y}_M^2, \quad \forall x,y \in \bbR^n.
}
}

\prf{\renewcommand{\qedsymbol}{}
See Section~\ref{sec:prf:l:QM} of the supplementary material.
}

Exploiting Definition~\ref{d:QM} and Lemma~\ref{l:QM}, the proposed method, BPEG-M, is given as follows.
To solve \R{sys:multiConvx}, we minimize a majorizer of $F$ cyclically over each block $x_1,\ldots,x_B$, while fixing the remaining blocks at their previously updated variables. Let $x_b^{(i+1)}$ be the value of $x_b$ after its $i\rth$ update, and define
\begingroup
\setlength{\thinmuskip}{1.5mu}
\setlength{\medmuskip}{2mu plus 1mu minus 2mu}
\setlength{\thickmuskip}{2.5mu plus 2.5mu}
\bes{
f_b^{(i+1)}(x_b) := f \Big( x_1^{(i+1)}, \ldots, x_{b-1}^{(i+1)}, x_b, x_{b+1}^{(i)}, \ldots, x_{B}^{(i)} \Big), \quad \forall b,i.
}
\endgroup
At the $b\rth$ block of the $i\rth$ iteration, we apply Lemma~\ref{l:QM} to functional $f_b^{(i+1)}(x_b)$ with a $M_b^{(i+1)}$-Lipschitz continuous gradient, and minimize the majorized function.\footnote{The quadratically majorized function allows a unique minimizer if $g_b^{(i+1)} (x_b)$ is convex and $\cX_b^{(i+1)}$ is a convex set (note that $M_b^{(i+1)} \!\succ\! 0$).} Specifically, BPEG-M uses the updates
\begingroup
\setlength{\thinmuskip}{1.5mu}
\setlength{\medmuskip}{2mu plus 1mu minus 2mu}
\setlength{\thickmuskip}{2.5mu plus 2.5mu}
\fontsize{9.5pt}{11.4pt}\selectfont
\allowdisplaybreaks
\ea{
x_b^{(i+1)} 
& = \argmin_{ x_b } \hspace{0.1em} \ip{ \nabla_{x_b} f_b^{(i+1)} (\acute{x}_b^{(i+1)}) }{ x_b - \acute{x}_b^{(i+1)} } \nn
\\
& \hspace{4.5em} + \frac{1}{2} \nm{ x_b - \acute{x}_b^{(i+1)} }_{\widetilde{M}_b^{(i+1)}}^2 + g_b (x_b) \label{update:x} 
\\
&= \argmin_{ x_b } \hspace{0.1em} \frac{1}{2} \bigg\| x_b - \bigg( \acute{x}_b^{(i+1)} - \Big( \widetilde{M}_b^{(i+1)} \Big)^{\!\!\!-1} \nn
\\
& \hspace{8.0em} \cdot \nabla_{x_b} f_b^{(i+1)} (\acute{x}_b^{(i+1)}) \bigg) \bigg\|_{\widetilde{M}_b^{(i+1)}}^2 \!\! + g_b (x_b) \nn
\\
&= \mathrm{Prox}_{g_b}^{\widetilde{M}_b^{(i+1)}} \!\! \bigg( \! \underbrace{ \acute{x}_b^{(i+1)} - \Big( \! \widetilde{M}_b^{(i+1)} \! \Big)^{\!\!\!-1} \! \nabla_{x_b} f_b^{(i+1)} (\acute{x}_b^{(i+1)}) }_{\text{\emph{extrapolated gradient step using a majorizer} of $f_b^{(i+1)}$}} \! \bigg), \nn
}
\endgroup
where
\be{
\label{update:xacute}
\acute{x}_b^{(i+1)} = x_{b}^{(i)} + E_b^{(i+1)} \left( x_b^{(i)} - x_b^{(i-1)} \right),
}
the proximal operator is defined by
\bes{
\mathrm{Prox}_g^M (y) := \argmin_{x} \, \frac{1}{2} \nm{x - y}_M^2 + g(x),
}
$\nabla f_b^{(i+1)} (\acute{x}_b^{(i+1)})$ is the block-partial gradient of $f$ at $\acute{x}_b^{(i+1)}$, an \textit{upper-bounded majorization matrix} is updated by
\be{
\label{update:Mtilde}
\widetilde{M}_b^{(i+1)} = \lambda_b \cdot M_b^{(i+1)} \succ 0, \qquad \lambda_b > 1,
}
and $M_b^{(i+1)} \!\in\! \bbR^{n_b \times n_b}$ is a symmetric positive definite \textit{majorization matrix} of $\nabla f_b^{(i+1)}$.
In \R{update:xacute}, the $\bbR^{n_b \times n_b}$ matrix $E_b^{(i+1)} \succeq 0$ is an \textit{extrapolation matrix} that accelerates convergence in solving block multi-convex problems \cite{Chun&Fessler:18TIP}.
We design it in the following form:
\be{
\label{update:Eb}
E_b^{(i+1)} = e_b^{(i)} \cdot \frac{\delta (\lambda_b - 1)}{2 (\lambda_b + 1)} \cdot \left( M_b^{(i+1)} \right)^{-1/2} \left( M_b^{(i)} \right)^{1/2}, 
}
for some $\{ 0 \leq e_b^{(i)} \leq 1 : \forall b, i \}$ and $\delta < 1$, to satisfy condition \R{cond:Wb} below.
In general, choosing $\lambda_b$ values in \R{update:Mtilde}--\R{update:Eb} to accelerate convergence is application-specific.
Algorithm~\ref{alg:BPGM} summarizes these updates.

\begin{algorithm}[t!]
\caption{BPEG-M}
\label{alg:BPGM}

\begin{algorithmic}
\REQUIRE $\{ x_b^{(0)} = x_b^{(-1)} : \forall b \}$, $\{ E_b^{(i)} \in [0, 1], \forall b,i \}$, $i=0$

\WHILE{a stopping criterion is not satisfied}

\FOR{$b = 1,\ldots,B$}

\begingroup
\setlength{\thinmuskip}{1.5mu}
\setlength{\medmuskip}{2mu plus 1mu minus 2mu}
\setlength{\thickmuskip}{2.5mu plus 2.5mu}
\fontsize{9.5pt}{11.4pt}\selectfont
\STATE Calculate  $M_b^{(i+1)}$, $\displaystyle \widetilde{M}_b^{(i+1)}$ by \R{update:Mtilde}, and $E_b^{(i+1)}$ by \R{update:Eb}
\STATE $\displaystyle \acute{x}_b^{(i+1)} = \, x_{b}^{(i)} + E_b^{(i+1)} \! \left( x_b^{(i)} - x_b^{(i-1)} \right)$
\STATE $\displaystyle x_b^{(i+1)} = \, \ldots$ \\ $\displaystyle \mathrm{Prox}_{g_b}^{\widetilde{M}_b^{(i+1)}} \!\! \bigg( \!\! \acute{x}_b^{(i+1)} \!-\! \Big( \! \widetilde{M}_b^{(i+1)} \! \Big)^{\!\!\!\!-1} \!\! \nabla f_b^{(i+1)} (\acute{x}_b^{(i+1)}) \!\! \bigg)$
\endgroup

\ENDFOR

\STATE $i = i+1$

\ENDWHILE

\end{algorithmic}
\end{algorithm}

The majorization matrices $M_b^{(i)}$ and $\widetilde{M}_b^{(i+1)}$ in \R{update:Mtilde} influence the convergence rate of BPEG-M.
A tighter majorization matrix (i.e., a matrix giving tighter bounds in the sense of Lemma~\ref{l:QM}) provided faster convergence rate \cite[Lem.~1]{Fessler&etal:93TNS}, \cite[Fig.~2--3]{Chun&Fessler:18TIP}.
An interesting observation in Algorithm~\ref{alg:BPGM} is that there exists a tradeoff between majorization sharpness via \R{update:Mtilde} and extrapolation effect via \R{update:xacute} and \R{update:Eb}. For example, increasing $\lambda_b$ (e.g., $\lambda_b = 2$) allows more extrapolation but results in looser majorization; setting $\lambda_b \rightarrow 1$ results in sharper majorization but provides less extrapolation.

\rem{\label{r:BPGM}
The proposed BPEG-M framework -- with key updates \R{update:x}--\R{update:xacute} -- generalizes the BPG method \cite{Xu&Yin:17JSC}, and has several benefits over BPG \cite{Xu&Yin:17JSC} and BPEG-M introduced earlier in \cite{Chun&Fessler:18TIP}:
\begin{itemize}
\item The BPG setup in \cite{Xu&Yin:17JSC} is a particular case of BPEG-M using a scaled identity majorization matrix $M_b$ with a Lipschitz constant of $\nabla f_b^{(i+1)} (\acute{x}_b^{(i+1)})$. The BPEG-M framework can significantly accelerate convergence by allowing sharp majorization; see \cite[Fig.~2--3]{Chun&Fessler:18TIP} and Fig.~\ref{fig:Comp:diffBPGM}. 
This generalization was first introduced for block multi-convex problems in \cite{Chun&Fessler:18TIP}, 
but the proposed BPEG-M in this paper addresses the more general problem, block multi-(non)convex optimization.

\item BPEG-M is useful for controlling the tradeoff between majorization sharpness and extrapolation effect in different blocks, by allowing each block to use different $\lambda_b$ values. If tight majorization matrices can be designed for a certain block $b$, then it could be reasonable to maintain the majorization sharpness by setting $\lambda_b$ very close to 1. 
When setting $\lambda_b = 1 + \epsilon$ (e.g., $\epsilon$ is a machine epsilon) and using $E_b^{(i+1)} = 0$ (no extrapolation), solutions of the original and its upper-bounded problem become (almost) identical.  In such cases, it is unnecessary to solve the upper bounded problem \R{update:x}, and the proposed BPEG-M framework allows using the solution of $f_b^{(i+1)}(x_b)$ without QM; see Section~\ref{sec:CAOL:spCd}. This generalization was not considered in \cite{Xu&Yin:17JSC}.

\item The condition for designing the extrapolation matrix \R{update:Eb}, i.e., \R{cond:Wb} in Assumption 3, is more general than that in \cite[(9)]{Chun&Fessler:18TIP} (e.g., \R{eg:Wb:eigen}).
Specifically, the matrices $E_b^{(i+1)}$ and $M_b^{(i+1)}$ in \R{update:Eb} need not be diagonalized by the same basis.
\end{itemize}
}

The first two generalizations lead to the question, \dquotes{Under the sharp QM regime (i.e., having tight bounds in Lemma~\ref{l:QM}), what is the best way in controlling $\{ \lambda_b \}$ in \R{update:Mtilde}--\R{update:Eb} in Algorithm~\ref{alg:BPGM}?}
Our experiments show that, if sufficiently sharp majorizers are obtained for partial or all blocks, then giving more weight to sharp majorization provides faster convergence compared to emphasizing extrapolation; for example, $\lambda_{b} = 1+\epsilon$ gives faster convergence than $\lambda_b = 2$.

\subsection{BPEG-M -- Convergence Analysis} \label{sec:convg-analysis}

This section analyzes the convergence of Algorithm~\ref{alg:BPGM} under the following assumptions.

\begin{itemize}
\item[] {\em Assumption 1)} $F$ is proper and lower bounded in $\dom(F)$, $f$ is continuously differentiable, $g_b$ is proper lower semicontinuous, $\forall b$.\footnote{
$F : \bbR^n \rightarrow (- \infty, + \infty]$ is proper if $\dom F \neq \emptyset$. 
$F$ is lower bounded in $\dom(F) := \{ x : F(x) < \infty \}$ if $\inf_{x \in \dom(F)} F(x) > -\infty$.
$F$ is lower semicontinuous at point $x_0$ if $\liminf_{x \rightarrow x_0} F(x) \geq F(x_0)$.
}
\R{sys:multiConvx} has a critical point $\bar{x}$, i.e., $0 \in \partial F(\bar{x})$,
where $\partial F(x)$ denotes the limiting subdifferential of $F$ at $x$ (see \cite[\S 1.9]{Kruger:03JMS}, \cite[\S 8]{Rockafellar&Wets:book}).
\item[] {\em Assumption 2)} The block-partial gradients of $f$, $\nabla f_b^{(i+1)}$, are $M_b^{(i+1)}$-Lipschitz continuous, i.e.,
\ea{
\label{eq:QMbound}
&~ \nm{ \nabla_{x_b} f_b^{(i+1)} (u) - \nabla_{x_b} f_b^{(i+1)} (v)  }_{\big( \! M_b^{(i+1)} \! \big)^{\!-1}} 
\nn
\\
& \leq \nm{u - v}_{M_b^{(i+1)}},
}
for $u,v \in \bbR^{n_b}$, and (unscaled) majorization matrices satisfy $m_{b} I_{n_b} \preceq M_b^{(i+1)}$ with $0 < m_{b} < \infty$, $\forall b, i$.
\item[] {\em Assumption 3)} The extrapolation matrices $E_b^{(i+1)} \succeq 0$ satisfy
\be{
\label{cond:Wb}
\Big( E_b^{(i+1)} \Big)^T M_b^{(i+1)} E_b^{(i+1)} \preceq \frac{\delta^2 (\lambda_b - 1)^2}{4 (\lambda_b + 1)^2} \cdot M_b^{(i)},
}
for any $\delta < 1$, $\forall b,i$.
\end{itemize}

Condition \R{cond:Wb} in Assumption 3 generalizes that in \cite[Assumption~3]{Chun&Fessler:18TIP}.
If eigenspaces of $E_b^{(i+1)}$ and $M_b^{(i+1)}$ coincide (e.g., diagonal and circulant matrices), $\forall i$ \cite[Assumption~3]{Chun&Fessler:18TIP}, \R{cond:Wb} becomes
\begingroup
\setlength{\thinmuskip}{1.5mu}
\setlength{\medmuskip}{2mu plus 1mu minus 2mu}
\setlength{\thickmuskip}{2.5mu plus 2.5mu}
\be{
\label{eg:Wb:eigen}
E_b^{(i+1)} \preceq \frac{\delta (\lambda_b - 1)}{2 (\lambda_b + 1)} \cdot  \left( M_b^{(i)} \right)^{1/2} \left( M_b^{(i+1)} \right)^{-1/2},
} 
\endgroup
as similarly given in \cite[(9)]{Chun&Fessler:18TIP}. 
This generalization allows one to consider arbitrary structures of $M_b^{(i)}$ across iterations.

\lem{[Sequence bounds]
\label{l:seqBound}
Let $\{ \widetilde{M}_b : b = 1,\ldots,B \}$ and $\{ E_b : b = 1,\ldots,B \}$ be as in \R{update:Mtilde}--\R{update:Eb}, respectively. 
The cost function decrease for the $i\rth$ update satisfies:
\ea{
\label{eq:l:seqBound}
F_b (x_b^{(i)}) - F_b (x_b^{(i+1)}) 
& \geq \frac{\lambda_b - 1}{4} \nm{ x_b^{(i)} - x_b^{(i+1)} }_{M_b^{(i+1)}}^2 \nn
\\
& \hspace{1.2em} - \frac{ ( \lambda_b - 1 ) \delta^2}{4} \nm{ x_b^{(i-1)} - x_b^{(i)} }_{ M_b^{(i)}}^2
}
}
\prf{\renewcommand{\qedsymbol}{}
See Section~\ref{sec:prf:l:seqBound} of the supplementary material.
}

Lemma~\ref{l:seqBound} generalizes \cite[Lem.~1]{Xu&Yin:17JSC} using $\{ \lambda_b = 2 \}$. 
Taking the majorization matrices in \R{eq:l:seqBound} to be scaled identities with Lipschitz constants, i.e., $M_b^{(i+1)} \!=\! L_b^{(i+1)} \cdot I$ and $M_b^{(i)} \!=\! L_b^{(i)} \cdot I$, where $L_b^{(i+1)}$ and $L_b^{(i)}$ are Lipschitz constants, the bound \R{eq:l:seqBound} becomes equivalent to that in \cite[(13)]{Xu&Yin:17JSC}.
Note that BPEG-M for block multi-convex problems in \cite{Chun&Fessler:18TIP} can be viewed within BPEG-M in Algorithm~\ref{alg:BPGM}, by similar reasons in \cite[Rem.~2]{Xu&Yin:17JSC} -- bound \R{eq:l:seqBound} holds for the block multi-convex problems by taking $E_b^{(i+1)}$ in \R{eg:Wb:eigen} as $E_b^{(i+1)} \preceq \delta \cdot ( M_b^{(i)} )^{1/2} ( M_b^{(i+1)} )^{-1/2}$ in \cite[Prop.~3.2]{Chun&Fessler:18TIP}.

\prop{[Square summability]
\label{p:sqSum}
Let $\{ x^{(i+1)} : i \geq 0 \}$ be generated by Algorithm \ref{alg:BPGM}.
We have
\be{
\label{eq:p:sqSum}
\sum_{i=0}^{\infty} \nm{ x^{(i)} - x^{(i+1)} }_2^2 < \infty.
}
}

\prf{\renewcommand{\qedsymbol}{}
See Section~\ref{sec:prf:p:sqSum} of the supplementary material.
}

Proposition~\ref{p:sqSum} implies that
\be{
\label{eq:convg_to0}
\nm{ x^{(i)} - x^{(i+1)} }_2^2 \rightarrow 0,
}
and \R{eq:convg_to0} is used to prove the following theorem:

\thm{[A limit point is a critical point]
\label{t:subseqConv}
Under Assumptions 1--3, let $\{ x^{(i+1)} : i \geq 0 \}$ be generated by Algorithm \ref{alg:BPGM}.
Then any limit point $\bar{x}$ of $\{ x^{(i+1)} : i \geq 0 \}$ is a critical point of \R{sys:multiConvx}. 
If the subsequence $\{ x^{(i_j+1)} \}$ converges to $\bar{x}$, then
\bes{
\lim_{j \rightarrow \infty} F(x^{(i_j + 1 )}) = F(\bar{x}).
} 
}

\prf{\renewcommand{\qedsymbol}{}
See Section~\ref{sec:prf:t:subseqConv} of the supplementary material.
}

Finite limit points exist if the generated sequence $\{ x^{(i+1)} : i \geq 0 \}$ is bounded; see, for example, \cite[Lem.~3.2--3.3]{Bao&Ji&Shen:14ACHA}. For some applications, the boundedness of $\{ x^{(i+1)} : i \geq 0 \}$ can be satisfied by choosing appropriate regularization parameters, e.g., \cite{Chun&Fessler:18TIP}.

\subsection{Restarting BPEG-M} \label{sec:reBPGM}

BPG-type methods \cite{Chun&Fessler:18TIP, Xu&Yin:17JSC, Xu&Yin:13SIAM} can be further accelerated by applying \textit{1)} a momentum coefficient formula similar to those used in fast proximal gradient (FPG) methods \cite{Beck&Teboulle:09SIAM, Nesterov:07CORE, Tseng:08techRep}, and/or \textit{2)} an adaptive momentum restarting scheme \cite{ODonoghue&Candes:15FCM, Giselsson&Boyd:14CDC}; see \cite{Chun&Fessler:18TIP}. 
This section applies these two techniques to further accelerate BPEG-M in Algorithm~\ref{alg:BPGM}.

First, we apply the following increasing momentum-coefficient formula to \R{update:Eb} \cite{Beck&Teboulle:09SIAM}: 
\be{
\label{eq:mom_coeff}
e_b^{(i+1)} = \frac{\theta^{(i)}  - 1}{\theta^{(i+1)}}, \quad \theta^{(i+1)} = \frac{1 + \sqrt{1 + 4 (\theta^{(i)})^2}}{2}.
}
This choice guarantees fast convergence of FPG method \cite{Beck&Teboulle:09SIAM}.
Second, we apply a momentum restarting scheme \cite{ODonoghue&Candes:15FCM, Giselsson&Boyd:14CDC}, when the following \textit{gradient-mapping} criterion is met \cite{Chun&Fessler:18TIP}:
\be{
\label{eq:restart:grad}
\cos \! \left( \Theta \! \left( M_b^{(i+1)} \!\! \left( \acute{x}_b^{(i+1)} - x_b^{(i+1)} \right),  x_b^{(i+1)} - x_b^{(i)} \right) \right) > \omega,
}
where the angle between two nonzero real vectors $\vartheta$ and $\vartheta'$ is $\Theta (\vartheta, \vartheta') := \ip{\vartheta}{\vartheta'} / ( \nm{\vartheta}_2 \nm{\vartheta'}_2 )$ and $\omega \in [-1, 0]$.
This scheme restarts the algorithm whenever the momentum, i.e., $x_b^{(i+1)} - x_b^{(i)}$, is likely to lead the algorithm in an unhelpful direction, as measured by the gradient mapping at the $x_b^{(i+1)}$-update.
We refer to BPEG-M combined with the methods \R{eq:mom_coeff}--\R{eq:restart:grad} as restarting BPEG-M (reBPEG-M). Section~\ref{sec:reBPGM-supp} in the supplementary material summarizes the updates of reBPEG-M.

To solve the block multi-nonconvex problems proposed in this paper (e.g., \R{sys:CAOL:orth}--\R{sys:CT&CAOL}), we apply reBPEG-M (a variant of Algorithm~\ref{alg:BPGM}; see Algorithm~\ref{alg:reBPGM}), promoting fast convergence to a critical point.

\section{Fast and Convergent CAOL via BPEG-M} \label{sec:CAOL+BPGM}

This section applies the general BPEG-M approach to CAOL.
The CAOL models \R{sys:CAOL:orth} and \R{sys:CAOL:div} satisfy the assumptions of BPEG-M; see Assumption 1--3 in Section~\ref{sec:convg-analysis}.
CAOL models \R{sys:CAOL:orth} and \R{sys:CAOL:div} readily satisfy Assumption~1 of BPEG-M.
To show the continuously differentiability of $f$ and the lower boundedness of $F$, consider that \textit{1)} $\sum_{l} \sum_{k} \frac{1}{2}  \left\| d_k \circledast x_l - z_{l,k} \right\|_2^2$ in \R{sys:CAOL} is continuously differentiable with respect to $D$ and $\{ z_{l,k} \}$;
\textit{2)} the sequences $\{ D^{(i+1)} \}$ are bounded, because they are in the compact set $\cD_{\text{\R{sys:CAOL:orth}}} = \{ D: D D^H = \frac{1}{R} I \}$ and $\cD_{\text{\R{sys:CAOL:div}}} = \{ d_k : \| d_k \|_2^2 = \frac{1}{R}, \forall k \}$ in \R{sys:CAOL:orth} and \R{sys:CAOL:div}, respectively; 
and \textit{3)} the positive thresholding parameter $\alpha$ ensures that the sequence $\{ z_{l,k}^{(i+1)} \}$ is bounded (otherwise the cost would diverge).
In addition, for both \R{sys:CAOL:orth} and \R{sys:CAOL:div}, the lower semicontinuity of regularizer $g_b$ holds, $\forall b$. 
For $D$-optimization, the indicator function of the sets $\cD_{\text{\R{sys:CAOL:orth}}}$ and $\cD_{\text{\R{sys:CAOL:div}}}$ is lower semicontinuous, because the sets are compact.
For $\{z_{l,k}\}$-optimization, the $\ell^0$-quasi-norm is a lower semicontinuous function.
Assumptions~2 and 3 are satisfied with the majorization matrix designs in this section -- see Sections~\ref{sec:CAOL:filt}--\ref{sec:CAOL:spCd} later -- and the extrapolation matrix design in \R{update:Eb}, respectively.

Since CAOL models \R{sys:CAOL:orth} and \R{sys:CAOL:div} satisfy the BPEG-M conditions, we solve \R{sys:CAOL:orth} and \R{sys:CAOL:div} by the reBPEG-M method with a two-block scheme, i.e., we alternatively update all filters $D$ and all sparse codes $\{z_{l,k} : l = 1,\ldots,L, k=1,\ldots,K\}$. 
Sections~\ref{sec:CAOL:filt} and \ref{sec:CAOL:spCd} describe details of $D$-block and $\{ z_{l,k} \}$-block optimization within the BPEG-M framework, respectively.
The BPEG-M-based CAOL algorithm is particularly useful for learning convolutional regularizers from large datasets because of its memory flexibility and parallel computing applicability, as described in Section~\ref{sec:CAOL:memory} and Sections~\ref{sec:CAOL:filt}--\ref{sec:CAOL:spCd}, respectively.

\subsection{Filter Update: $D$-Block Optimization} \label{sec:CAOL:filt}

We first investigate the structure of the system matrix in the filter update for \R{sys:CAOL}.
This is useful for \textit{1)} accelerating majorization matrix computation in filter updates (e.g., Lemmas~\ref{l:MDdiag}--\ref{l:MDscaleI}) and \textit{2)} applying $R \!\times\! N$-sized adjoint operators (e.g., $\Psi_l^H$ in \R{eq:Psi_l} below) to an $N$-sized vector without needing the Fourier approach \cite[Sec.~\Romnum{5}-A]{Chun&Fessler:18TIP} that uses commutativity of convolution and Parseval's relation.
Given the current estimates of $\{ z_{l,k} : l=1,\ldots,L, k=1,\ldots,K \}$, the filter update problem of \R{sys:CAOL} is equivalent to
\be{
\label{sys:filter}
\argmin_{\{d_k\}} \frac{1}{2} \sum_{k=1}^K \sum_{l=1}^L \left\| \Psi_l d_k - z_{l,k} \right\|_2^2 + \beta g (D),
}
where $D$ is defined in \R{eq:D}, $\Psi_l \in \bbC^{N \times R}$ is defined by
\be{
\label{eq:Psi_l}
\Psi_l := \left[ \begin{array}{ccc} P_{B_1} \hat{x}_l & \ldots & P_{B_R} \hat{x}_l \end{array} \right],
}
$P_{B_r} \in \bbC^{N \times \hat{N}}$ is the $r\rth$ (rectangular) selection matrix that selects $N$ rows corresponding to the indices $B_r = \{ r, \ldots, r+N-1 \}$ from $I_{\hat{N}}$, $\{ \hat{x}_l \in \bbC^{\hat{N}} : l=1,\ldots,L \}$ is a set of padded training data, $\hat{N} = N+R-1$.
Note that applying $\Psi_l^H$ in \R{eq:Psi_l} to a vector of size $N$ is analogous to calculating cross-correlation between $\hat{x}_l$ and the vector, i.e., $( \Psi_l^H \hat{z}_{l,k} )_{r} = \sum_{n=1}^N \hat{x}_{n+r-1}^{*} (\hat{z}_{l,k})_n$, $r=1,\ldots,R$.
In general, $\hat{(\cdot)}$ denotes a padded signal vector.

\subsubsection{Majorizer Design} \label{sec:filt:maj}

This subsection designs multiple majorizers for the $D$-block optimization and compares their required computational complexity and tightness. 
The next proposition considers the structure of $\Psi_l$ in \R{eq:Psi_l} to obtain the Hessian $\sum_{l=1}^L \Psi^H_l \Psi_l \in \bbC^{R \times R}$ in \R{sys:filter} for an arbitrary boundary condition.
\prop{[Exact Hessian matrix $M_D$] \label{p:MDexhess}
The following matrix $M_D \in \bbC^{R \times R}$ is identical to $\sum_{l=1}^L \Psi^H_l \Psi_l$:
\be{
\label{eq:soln:filterHess}
\left[ M_D \right]_{r,r'} = \sum_{l=1}^L \ip{P_{B_r} \hat{x}_l}{P_{B_{r'}} \hat{x}_l}, \quad r,r'  = 1,\ldots,R.
}
}

A sufficiently large number of training signals (with $N \geq R$), $L$, can guarantee $M_D = \sum_{l=1}^L \Psi^H_l \Psi_l \succ 0$ in Proposition~\ref{p:MDexhess}.
The drawback of using Proposition~\ref{p:MDexhess} is its polynomial computational complexity, i.e., $O (L R^2 N)$ -- see Table \ref{tab:MD:compt}. 
When $L$ (the number of training signals) or $N$ (the size of training signals) are large, the quadratic complexity with the size of filters -- $R^2$ -- can quickly increase the total computational costs when multiplied by $L$ and $N$. (The BPG setup in \cite{Xu&Yin:17JSC} additionally requires $O (R^3)$ because it uses the eigendecomposition of \R{eq:soln:filterHess} to calculate the Lipschitz constant.)

\begin{figure}[!t]
\centering
\begin{tabular}{c}
\vspace{-0.25em}\includegraphics[scale=0.5, trim=0 0.2em 2.2em 1.2em, clip]{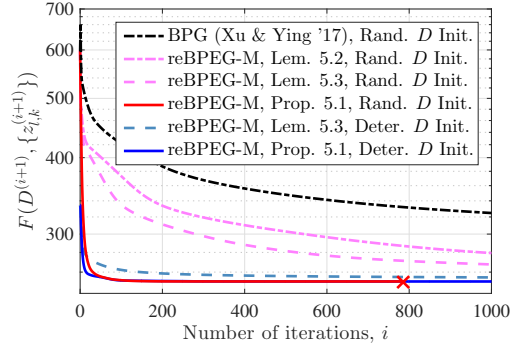} \\
{\small (a) The fruit dataset ($L = 10$, $N = 100 \!\times\! 100$)} \\
\vspace{-0.25em}\includegraphics[scale=0.5, trim=0 0.2em 2.2em 1em, clip]{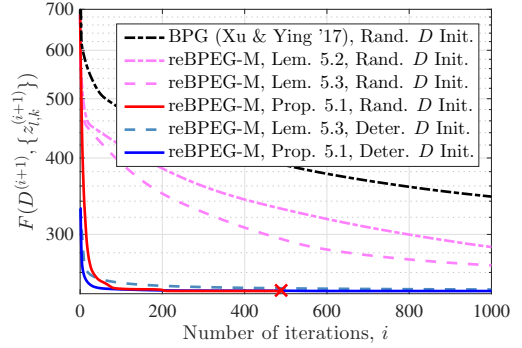}  \\
{\small (b) The city dataset ($L = 10$, $N = 100 \!\times\! 100$)} \\
\end{tabular}

\vspace{-0.25em}
\caption{Cost minimization comparisons in CAOL \R{sys:CAOL:orth} with different BPG-type algorithms and datasets ($R \!=\! K \!=\! 49$ and $\alpha \!=\! 2.5 \!\times\! 10^{-4}$; solution \R{eq:soln:spCode:exact} was used for sparse code updates; BPG (Xu \& Ying '17) \cite{Xu&Yin:17JSC} used the maximum eigenvalue of Hessians for Lipschitz constants; the cross mark \text{\sffamily x} denotes a termination point).
A sharper majorization leads to faster convergence of BPEG-M;
for all the training datasets considered in this paper, the majorization matrix in Proposition~\ref{p:MDexhess} is sharper than those in Lemmas~\ref{l:MDdiag}--\ref{l:MDscaleI}. 
}
\label{fig:Comp:diffBPGM}
\end{figure}

Considering CAOL problems \R{sys:CAOL} themselves, different from CDL  \cite{Chun&Fessler:18TIP, Chun&Fessler:17SAMPTA, Wohlberg:16TIP, Heide&eta:15CVPR, Bristow&etal:13CVPR}, the complexity $O (L R^2 N)$ in applying Proposition~\ref{p:MDexhess} is reasonable. In BPEG-M-based CDL \cite{Chun&Fessler:18TIP, Chun&Fessler:17SAMPTA}, a majorization matrix for kernel update is calculated every iteration because it depends on updated sparse codes; however, in CAOL, one can precompute $M_D$ via Proposition~\ref{p:MDexhess} (or Lemmas~\ref{l:MDdiag}--\ref{l:MDscaleI} below) without needing to change it every kernel update.
The polynomial computational cost in applying Proposition~\ref{p:MDexhess} becomes problematic only when the training signals change.
Examples include  \textit{1)} hierarchical CAOL, e.g., CNN in Appendix~\ref{sec:CNN}, \textit{2)} \dquotes{adaptive-filter MBIR} particularly with high-dimensional signals \cite{Cai&etal:14ACHA, Xu&etal:12TMI, Elad&Aharon:06TIP}, and \textit{3)} online learning \cite{Liu&etal:17arXiv, Mairal&etal:09ICML}. Therefore, we also describe a more efficiently computable majorization matrix at the cost of looser bounds (i.e., slower convergence; see Fig~\ref{fig:Comp:diffBPGM}). 
Applying Lemma~\ref{l:diag(|At|W|A|1)}, we first introduce a diagonal majorization matrix $M_D$ for the Hessian $\sum_{l} \Psi_l^H \Psi_l$ in \R{sys:filter}:

\begin{table}[!pt]	

\centering
\renewcommand{\arraystretch}{1.1}
	
\caption{Computational complexity of different majorization matrix designs for the filter update problem \R{sys:filter}}	
\label{tab:MD:compt}
	
\begin{tabular}{C{2.2cm}C{2cm}}
\hline \hline
Lemmas~\ref{l:MDdiag}--\ref{l:MDscaleI} & Proposition~\ref{p:MDexhess} \\ 
\hline
$O( L R N )$ & $O( L R^2 N )$ \\
\hline \hline
\end{tabular}
\end{table}

\lem{[Diagonal majorization matrix $M_D$] \label{l:MDdiag}
The following matrix $M_D \in \bbC^{R \times R}$ satisfies $M_D \succeq \sum_{l=1}^L \Psi^H_l \Psi_l$:
\be{
\label{eq:soln:filterMaj:loose}
M_D = \diag \! \left( \sum_{l=1}^L | \Psi_l^H | | \Psi_l | 1_{R} \right),
}
where $|\cdot|$ takes the absolute values of the elements of a matrix. 
}
The majorization matrix design in Lemma~\ref{l:MDdiag} is more efficient to compute than that in Proposition~\ref{p:MDexhess}, because no $R^2$-factor is needed for calculating $M_D$ in Lemma~\ref{l:MDdiag}, i.e., $O( L R N )$; see Table~\ref{tab:MD:compt}. 
Designing $M_D$ in Lemma~\ref{l:MDdiag} takes fewer calculations than \cite[Lem.~5.1]{Chun&Fessler:18TIP} using Fourier approaches, when $R \!<\! \log ( \hat{N} )$.
Using Lemma~\ref{l:diag(|A|1)}, we next design a potentially sharper majorization matrix than \R{eq:soln:filterMaj:loose}, while maintaining the cost $O( L R N )$:
\lem{[Scaled identity majorization matrix $M_D$] \label{l:MDscaleI}
The following matrix $M_D \in \bbC^{R \times R}$ satisfies $M_D \succsim \sum_{l=1}^L \Psi^H_l \Psi_l$:
\be{
\label{eq:soln:filterMaj}
M_D = \sum_{r=1}^R \left| \sum_{l=1}^L \ip{P_{B_1} \hat{x}_l}{P_{B_{r}} \hat{x}_l} \right| \cdot I_R,
}
for a circular boundary condition.
}
\prf{\renewcommand{\qedsymbol}{}
See Section~\ref{sec:prf:l:MD} of the supplementary material.
}

\begin{figure}[!tp]
\centering
\begin{tabular}{c}
\vspace{-0.25em}\includegraphics[scale=0.58, trim=0 0.2em 2em 1em, clip]{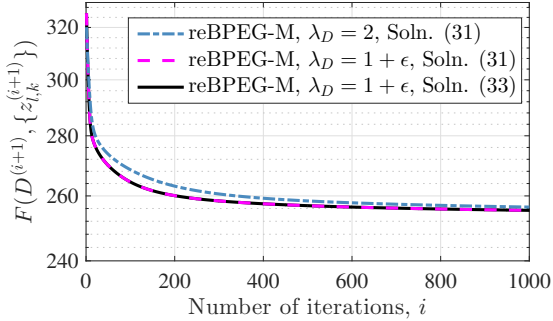} \\
{\small (a) The fruit dataset ($L = 10$, $N = 100 \!\times\! 100$)} \\
\vspace{-0.25em}\includegraphics[scale=0.58, trim=0 0.2em 2em 1em, clip]{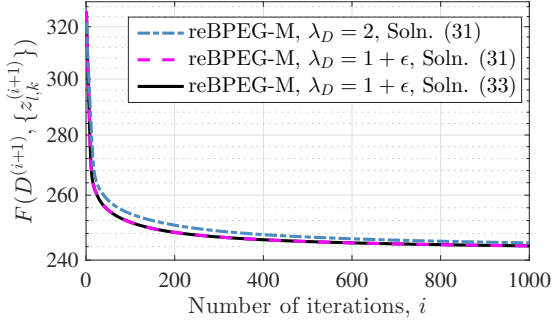}  \\
{\small (b) The city dataset ($L = 10$, $N = 100 \!\times\! 100$)} \\
\end{tabular}

\vspace{-0.25em}
\caption{Cost minimization comparisons in CAOL \R{sys:CAOL:orth} with different BPEG-M algorithms and datasets (Lemma~\ref{l:MDdiag} was used for $M_D$; $R \!=\! K \!=\! 49$; deterministic filter initialization and random sparse code initialization). 
Under the sharp majorization regime, maintaining sharp majorization (i.e., $\lambda_D \!=\! 1+\epsilon$) provides faster convergence than giving more weight on extrapolation (i.e., $\lambda_D \!=\! 2$).
(The same behavior was found in sparse-view CT application \cite[Fig.~3]{Chun&Fessler:18Asilomar}.)
There exist no differences in convergence between solution \R{eq:soln:spCode:exact} and solution \R{eq:soln:spCode} using $\{ \lambda_Z \!=\! 1+\epsilon \}$. 
}
\label{fig:Comp:exact_vs_approx}
\end{figure}

For all the training datasets used in this paper, we observed that the tightness of majorization matrices in Proposition~\ref{p:MDexhess} and Lemmas~\ref{l:MDdiag}--\ref{l:MDscaleI} for the Hessian $\sum_{l} \Psi^H_l \Psi_l$ is given by
\be{
\label{eq:MD:tight}
\sum_{l=1}^L \Psi^H_l \Psi_l = \text{\R{eq:soln:filterHess}} \preceq \text{\R{eq:soln:filterMaj}} \preceq \text{\R{eq:soln:filterMaj:loose}}.
}
(Note that \R{eq:soln:filterHess}\,$\preceq$\,\R{eq:soln:filterMaj:loose} always holds regardless of training data.)
Fig.~\ref{fig:Comp:diffBPGM} illustrates the effects of the majorizer sharpness in \R{eq:MD:tight} on CAOL convergence rates.
As described in Section~\ref{sec:BPGM:setup}, selecting $\lambda_D$ (see \R{sys:prox:filter:orth} and \R{sys:prox:filter:div} below) controls the tradeoff between majorization sharpness and extrapolation effect. 
We found that using fixed $\lambda_D = 1+\epsilon$ 
gives faster convergence than $\lambda_D = 2 $; see Fig.~\ref{fig:Comp:exact_vs_approx} (this behavior is more obvious in solving the CT MBIR model in \R{sys:CT&CAOL} via BPEG-M -- see \cite[Fig.~3]{Chun&Fessler:18Asilomar}).
The results in Fig.~\ref{fig:Comp:exact_vs_approx} and \cite[Fig.~3]{Chun&Fessler:18Asilomar} show that, under the sharp majorization regime, maintaining sharper majorization is more critical in accelerating the convergence of BPEG-M than giving more weight to extrapolation.

Sections \ref{sec:prox:filter:orth} and \ref{sec:prox:filter:div} below apply the majorization matrices designed in this section to proximal mappings of $D$-optimization in \R{sys:CAOL:orth} and \R{sys:CAOL:div}, respectively.

\subsubsection{Proximal Mapping with Orthogonality Constraint} \label{sec:prox:filter:orth}

The corresponding proximal mapping problem of \R{sys:filter} using the orthogonality constraint in \R{sys:CAOL:orth} is given by
\ea{
\label{sys:prox:filter:orth}
\{ d_k^{(i+1)} \} = \argmin_{\{ d_k \}} &~ 
\sum_{k=1}^K \frac{1}{2}  \left\| d_k - \nu_k^{(i+1)} \right\|_{\widetilde{M}_D}^2, \nn
\\
\mathrm{subject~to} &~ D D^H = \frac{1}{R} \cdot I,
}
where
\ea{
\nu_{k}^{(i+1)} &= \textstyle \acute{d}_k^{(i+1)} - \widetilde{M}_D^{-1} \sum_{l=1}^L \Psi_l^H \Big( \Psi_l \acute{d}_k^{(i+1)} - z_{l,k} \Big),
\label{eq:nuk} 
\\
\acute{d}_k^{(i+1)} &= d_k^{(i)} + E_D^{(i+1)} \Big( d_k^{(i)} - d_k^{(i-1)} \Big), \label{eq:dk_ac}
}
for $k=1,\ldots,K$, and $\widetilde{M}_D = \lambda_D M_D$ by \R{update:Mtilde}. 
One can parallelize over $k = 1,\ldots,K$ in computing $\{ \nu_k^{(i+1)} \}$ in \R{eq:nuk}.
The proposition below provides an optimal solution to \R{sys:prox:filter:orth}:

\prop{
\label{p:orth}
Consider the following constrained minimization problem:
\begingroup
\setlength{\thinmuskip}{1.5mu}
\setlength{\medmuskip}{2mu plus 1mu minus 2mu}
\setlength{\thickmuskip}{2.5mu plus 2.5mu}
\ea{
\label{p:eq:TF}
\min_{D} ~ \nm{ \widetilde{M}_D^{1/2} D  - \widetilde{M}_D^{1/2} \mathcal{V} }_{\mathrm{F}}^2, \quad 
\mathrm{subj.~to} ~ D D^H = \frac{1}{R} \cdot I,
}
\endgroup
where $D$ is given as \R{eq:D}, $\mathcal{V} = [\nu_{1}^{(i+1)} \cdots \nu_{K}^{(i+1)}] \in \bbC^{R \times K}$, $\widetilde{M}_D = \lambda_D M_D$, and $M_D \in \bbR^{R \times R}$ is given by \R{eq:soln:filterHess}, \R{eq:soln:filterMaj:loose}, or \R{eq:soln:filterMaj}.
The optimal solution to \R{p:eq:TF} is given by
\bes{
 D^\star = \frac{1}{\sqrt{R}} \cdot U \left[ \arraycolsep=1pt \begin{array}{cc} I_R, & 0_{R \times (K-R)} \end{array} \right] V^H, \quad \mbox{for}~ R \leq K,
}
where $\widetilde{M}_D \mathcal{V}$ has (full) singular value decomposition, $\widetilde{M}_D \mathcal{V} = U \Lambda V^H$.
}
\prf{\renewcommand{\qedsymbol}{}
See Section~\ref{sec:prf:p:orth} of the supplementary material.
}

When using Proposition~\ref{p:MDexhess}, $\widetilde{M}_D \nu_{k}^{(i+1)}$ of $\widetilde{M}_D \mathcal{V}$ in Proposition~\ref{p:orth} simplifies to the following update:
\bes{
\widetilde{M}_D \nu_{k}^{(i+1)} = (\lambda_D - 1) M_D \acute{d}_k^{(i+1)} + \sum_{l=1}^L \Psi_l^H z_{l,k}.
}
Similar to obtaining $\{ \nu_k^{(i+1)} \}$ in \R{eq:nuk}, computing $\{ \widetilde{M}_D \nu_k^{(i+1)} : k=1,\ldots,K \}$ is parallelizable over $k$.

\subsubsection{Proximal Mapping with Diversity Promoting Regularizer}  \label{sec:prox:filter:div}

The corresponding proximal mapping problem of \R{sys:filter} using the norm constraint and diversity promoting regularizer in \R{sys:CAOL:div} is given by
\ea{
\label{sys:prox:filter:div}
\{ d_k^{(i+1)} \} = \argmin_{\{ d_k \}} &~ 
\sum_{k=1}^K \frac{1}{2}  \left\| d_k - \nu_k^{(i+1)} \right\|_{\widetilde{M}_D}^2 + \frac{\beta}{2} g_{\text{div}}( D ), \nn
\\
\mathrm{subject~to} &~ \left\| d_k \right\|_2^2 = \frac{1}{R}, \quad k = 1,\ldots, K,
}
where $g_{\text{div}}( D )$, $\nu_{k}^{(i+1)}$, and $\acute{d}_k^{(i+1)}$ are given as in \R{sys:CAOL:div}, \R{eq:nuk}, and \R{eq:dk_ac}, respectively.
We first decompose the regularization term $g_{\text{div}}(D)$ as follows:
\begingroup
\allowdisplaybreaks
\ea{
\label{eq:DtD}
g_{\text{div}} (D)
& = \sum_{k=1}^K \sum_{k'=1}^K \big( d_k^H d_{k'} d_{k'}^H d_k - R^{-1} \big) \nn
\\
& = \sum_{k=1}^K d_k^H \bigg( \sum_{k' \neq k} d_{k'} d_{k'}^H \bigg) d_k + \big( d_k^H d_k - R^{-1} \big)^2 \nn
\\
& = \sum_{k=1}^K d_k^H \Gamma_k  d_k, 
}
\endgroup
where the equality in \R{eq:DtD} holds by using the constraint in \R{sys:prox:filter:div}, and the Hermitian matrix $\Gamma_k \in \bbC^{R \times R}$ is defined by
\be{
\label{eq:Ek}
\Gamma_k := \sum_{k' \neq k} d_{k'} d_{k'}^H.
}
Using \R{eq:DtD} and \R{eq:Ek}, we rewrite \R{sys:prox:filter:div} as
\ea{
\label{sys:prox:filter2}
d_k^{(i+1)} = \argmin_{d_k} &~ 
\frac{1}{2} \left\| d_k - \nu_k^{(i)} \right\|_{\widetilde{M}_D}^2 + \frac{\beta}{2} d_k^H \Gamma_k d_k, \nn
\\
\mathrm{subject~to} &~ \left\| d_k \right\|_2^2 = \frac{1}{R}, \quad k = 1,\ldots,K.
}
This is a quadratically constrained quadratic program with $\{ \widetilde{M}_D + \beta \Gamma_k \succ 0 : k = 1,\ldots,K\}$. 
We apply an accelerated Newton's method to solve \R{sys:prox:filter2}; see Section~\ref{sec:Newton}.
Similar to solving \R{sys:prox:filter:orth} in Section~\ref{sec:prox:filter:orth}, solving \R{sys:prox:filter:div} is a small-dimensional problem ($K$ separate problems of size $R$).

\subsection{Sparse Code Update: $\{z_{l,k} \}$-Block Optimization} \label{sec:CAOL:spCd}

Given the current estimate of $D$, the sparse code update problem for \R{sys:CAOL} is given by
\be{
\label{sys:spCode}
\argmin_{\{ z_{l,k} \}} \sum_{l=1}^L \sum_{k=1}^K \frac{1}{2} \left\| d_k \circledast x_{l} - z_{l,k} \right\|_2^2  +  \alpha \left\| z_{l,k} \right\|_0.
}
This problem separates readily, 
allowing parallel computation with $LK$ threads.
An optimal solution to \R{sys:spCode} is efficiently obtained by the well-known hard thresholding:
\be{
\label{eq:soln:spCode:exact}
z_{l,k}^{(i+1)}  = \cH_{\!\sqrt{2\alpha}} \left( d_k \circledast x_{l} \right), 
}
for $k=1,\ldots,K$ and $l=1,\ldots,L$, where 
\be{
\label{eq:def:hardthr}
\cH_a (x)_{n} := \left\{ \begin{array}{cc} 0, & | x_{n} | < a_{n}, \\ x_{n}, & | x_{n} | \geq a_{n}. \end{array} \right.
}
for all $n$. 
Considering $\lambda_Z$ (in $\widetilde{M}_Z = \lambda_Z M_Z$) as $\lambda_Z \!\rightarrow\! 1$, the solution obtained by the BPEG-M approach becomes equivalent to \R{eq:soln:spCode:exact}. 
To show this, observe first that the BPEG-M-based solution (using $M_Z = I_{N}$) to \R{sys:spCode} is obtained by
\ea{
\label{eq:soln:spCode}
z_{l,k}^{(i+1)}  &= \cH_{\!\sqrt{ \frac{2\alpha}{\lambda_Z}}} \Big( \zeta_{l,k}^{(i+1)} \Big),
\\
\zeta_{l,k}^{(i+1)} &= \left( 1 - \lambda_Z^{-1} \right) \cdot \acute{z}_{l,k}^{(i+1)} + \lambda_Z^{-1} \cdot d_k \circledast x_{l},
\nn \\
\acute{z}_{l,k}^{(i+1)} &= z_{l,k}^{(i)} + E_Z^{(i+1)} \Big( z_{l,k}^{(i)} - z_{l,k}^{(i-1)} \Big). \nn
}
The downside of applying solution \R{eq:soln:spCode} is that it would require additional memory to store the corresponding extrapolated points -- $\{ \acute{z}_{l,k}^{(i+1)} \}$ -- and the memory grows with $N$, $L$, and $K$.
Considering the sharpness of the majorizer in \R{sys:spCode}, i.e.,  $M_Z = I_{N}$, and the memory issue, it is reasonable to consider the solution \R{eq:soln:spCode} with no extrapolation, i.e., $\{ E_Z^{(i+1)} = 0 \}$:
\bes{
z_{l,k}^{(i+1)} 
= \cH_{\! \sqrt{\frac{2\alpha}{\lambda_Z}}} \Big( (\lambda_Z - 1)^{-1}\lambda_Z \cdot z_{l,k}^{(i)} + \lambda_Z^{-1} \cdot d_k \circledast x_{l} \Big)
}
becoming equivalent to \R{eq:soln:spCode:exact} as $\lambda_Z \!\rightarrow\! 1$.

Solution \R{eq:soln:spCode:exact} has two benefits over \R{eq:soln:spCode}: compared to \R{eq:soln:spCode}, \R{eq:soln:spCode:exact} requires only half the memory to update all $z_{l,k}^{(i+1)}$ vectors and no additional computations related to $\acute{z}_{l,k}^{(i+1)}$.
While having these benefits, empirically \R{eq:soln:spCode:exact} has equivalent convergence rates as \R{eq:soln:spCode} using $\{ \lambda_Z \!=\! 1+\epsilon \}$; see Fig.~\ref{fig:Comp:exact_vs_approx}. Throughout the paper, we solve the sparse coding problems (e.g., \R{sys:spCode} and $\{ z_k \}$-block optimization in \R{sys:CT&CAOL}) via optimal solutions in the form of \R{eq:soln:spCode:exact}.

\subsection{Lower Memory Use than Patch-Domain Approaches} \label{sec:CAOL:memory}

The convolution perspective in CAOL \R{sys:CAOL} requires much less memory than conventional patch-domain approaches; 
thus, it is more suitable for learning filters from large datasets or applying the learned filters to high-dimensional MBIR problems.
First, consider the training stage (e.g., \R{sys:CAOL}).
The patch-domain approaches, e.g., \cite{Cai&etal:14ACHA, Ravishankar&Bressler:15TSP, Aharon&Elad&Bruckstein:06TSP}, require about $R$ times more memory to store training signals. For example, 2D patches extracted by $\sqrt{R} \!\times\! \sqrt{R}$-sized windows (with \dquotes{stride} one and periodic boundaries \cite{Cai&etal:14ACHA, Papyan&Romano&Elad:17JMLR}, as used in convolution) require about $R$ (e.g., $R = 64$ \cite{Aharon&Elad&Bruckstein:06TSP, Ravishankar&Bressler:15TSP}) times more memory than storing the original image of size $\sqrt{N} \!\times\! \sqrt{N}$. For $L$ training images, their memory usage dramatically increases with a factor $L R N$.
This becomes even more problematic in forming hierarchical representations, e.g., CNNs -- see Appendix~\ref{sec:CNN}.
Unlike the patch-domain approaches, the memory use of CAOL \R{sys:CAOL} only depends on the $L N$-factor to store training signals.
As a result, the BPEG-M algorithm for CAOL \R{sys:CAOL:orth} requires about two times less memory than the patch-domain approach \cite{Cai&etal:14ACHA} (using BPEG-M).
See Table~\ref{tab:AOL}-B.
(Both the corresponding BPEG-M algorithms use identical computations per iteration that scale with $L R^2 N$; see Table~\ref{tab:AOL}-A.)

Second, consider solving MBIR problems. Different from the training stage, the memory burden depends on how one applies the learned filters.
In \cite{Pfister&Bresler:17ICASSP}, the learned filters are applied with the conventional convolutional operators -- e.g., $\circledast$ in \R{sys:CAOL} -- and, thus, there exists no additional memory burden. 
However, in \cite{Elad&Aharon:06TIP, Chun&etal:17Fully3D, Zheng&etal:19TCI}, the $\sqrt{R} \!\times\! \sqrt{R}$-sized learned kernels are applied with a matrix constructed by many overlapping patches extracted from the updated image at each iteration. 
In adaptive-filter MBIR problems \cite{Elad&Aharon:06TIP, Cai&etal:14ACHA, Pfister&Bresler:15SPIE}, the memory issue pervades the patch-domain approaches.

\begin{table}[!t]	

\centering
\renewcommand{\arraystretch}{1.1}
	
\caption{
Comparisons of computational complexity and memory usages between CAOL and patch-domain approach
}	
\label{tab:AOL}
	
\begin{tabular}{C{2.15cm}|C{2.85cm}|C{2.3cm}}
\hline \hline
\multicolumn{3}{c}{A.~Computational complexity per BPEG-M iteration}
\\
\hline
& Filter update & Sparse code update
\\
\hline
CAOL \R{sys:CAOL:orth} & $O(L K R N) + O(R^2 K)$ & $O(L K R N)$ 
\\ 
\hline
Patch-domain \cite{Cai&etal:14ACHA}$^\dagger$ & $O (L R^2 N) + O (R^3)$ & $O (L R^2 N)$
\\
\hline \hline
\end{tabular}

\vspace{0.75pc}

\begin{tabular}{C{2.15cm}|C{2.3cm}|C{2.3cm}}
\hline \hline
\multicolumn{3}{c}{B.~Memory usage for BPEG-M algorithm}
\\
\hline
& Filter update & Sparse code update
\\
\hline
CAOL \R{sys:CAOL:orth} & $O(LN) + O(RK)$ &  $O(L K N)$
\\ 
\hline
Patch-domain \cite{Cai&etal:14ACHA}$^\dagger$ & $O(L R N) + O(R^2)$ & $O(L R N)$
\\
\hline \hline
\end{tabular}

\medskip
\begin{myquote}{0.1in}
$^\dagger$
The patch-domain approach \cite{Cai&etal:14ACHA} considers the orthogonality constraint in \R{sys:CAOL:orth} with $R \!=\! K$; see Section~\ref{sec:CAOL:orth}.
The estimates consider all the extracted overlapping patches of size $R$ with the stride parameter $1$ and periodic boundaries, as used in convolution.
\end{myquote}
\end{table}

\section{Sparse-View CT MBIR using Convolutional Regularizer Learned via CAOL, and BPEG-M} \label{sec:CTrecon}

This section introduces a specific example of applying the learned convolutional regularizer, i.e., $F(D^\star, \{ z_{l,k} \})$ in \R{sys:CAOL}, from a representative dataset to recover images in \textit{extreme} imaging that collects highly undersampled or noisy measurements.
We choose a sparse-view CT application since it has interesting challenges in reconstructing images that include Poisson noise in measurements, nonuniform noise or resolution properties in reconstructed images, and complicated (or no) structures in the system matrices. 
For CT, undersampling schemes can significantly reduce the radiation dose and cancer risk from CT scanning. 
The proposed approach can be applied to other applications (by replacing the data fidelity and spatial strength regularization terms in \R{sys:CT&CAOL} below).

We pre-learn TF filters $\{ d_k^\star \in \bbR^K : k = 1,\ldots,K \}$ via CAOL \R{sys:CAOL:orth} with a set of high-quality (e.g., normal-dose) CT images $\{ x_l : l = 1,\ldots,L \}$.
To reconstruct a linear attenuation coefficient image $x \in \bbR^{N'}$ from post-log measurement $y \in \bbR^m$ \cite{Chun&Talavage:13Fully3D, Chun&etal:17Fully3D}, we apply the learned convolutional regularizer to CT MBIR and solve the following block multi-nonconvex problem \cite{Chun&Fessler:18Asilomar, Crockett&etal:19CAMSAP}:
\begingroup
\allowdisplaybreaks
\setlength{\thinmuskip}{1.5mu}
\setlength{\medmuskip}{2mu plus 1mu minus 2mu}
\setlength{\thickmuskip}{2.5mu plus 2.5mu}
\ea{
\label{sys:CT&CAOL}
\argmin_{x \geq 0} & \underbrace{ \frac{1}{2} \left\| y - A x \right\|_W^2 }_{\text{data fidelity $f(x;y)$}} + 
\nn \\
\gamma \cdot & \underbrace{ \min_{\{ z_k \}} \sum_{k=1}^K \frac{1}{2} \left\| d_{k}^\star \circledast x - z_k \right\|_2^2 
+ \alpha' \sum_{n=1}^{N'} \psi_{j} \phi( (z_{k})_n ) }_{\text{learned convolutional regularizer $g(x,\{z_k\}; \{ d_k \})$}}. \tag{P3}
}
\endgroup
\!\!Here, $A \in \bbR^{m \times N'}$ is a CT system matrix, $W \in \bbR^{m \times m}$ is a (diagonal) weighting matrix with elements $\{ W_{l,l} = \rho_l^2 / ( \rho_l + \sigma^2 ) : l  = 1,\ldots,m \}$ based on a Poisson-Gaussian model for the pre-log measurements $\rho \in \bbR^m$ with electronic readout noise variance $\sigma^2$ \cite{Chun&Talavage:13Fully3D, Chun&etal:17Fully3D, Zheng&etal:19TCI}, $\psi \in \bbR^{N'}$ is a pre-tuned spatial strength regularization vector \cite{Fessler&Rogers:96TIP} with non-negative elements
$\{ \psi_{n} =  ( \sum_{l=1}^m A_{l,n}^2 W_{l,l} )^{1/2} / ( \sum_{l=1}^m A_{l,n}^2 )^{1/2} : n = 1,\ldots,N' \}$\footnote{
See details of computing $\{ A_{l,j}^2 : \forall l,j \}$ in \cite{Chun&Fessler:18Asilomar}.
} 
that promotes uniform resolution or noise properties in the reconstructed image \cite[Appx.]{Chun&etal:17Fully3D}, an indicator function $\phi(a)$ is equal to $0$ if $a = 0$, and is $1$ otherwise, $z_k \in \bbR^{N'}$ is unknown sparse code for the $k\text{th}$ filter, 
and $\alpha' \!>\! 0$ is a thresholding parameter.

We solved \R{sys:CT&CAOL} via reBPEG-M in Section~\ref{sec:reBPG-M} with a two-block scheme \cite{Chun&Fessler:18Asilomar}, and summarize the corresponding BPEG-M updates as
\begingroup
\allowdisplaybreaks
\ea{
\label{eq:soln:CT:bpgm}
x^{(i+1)} &= \bigg[ \big( \widetilde{M}_A + \gamma I_R \big)^{-1} \cdot \bigg( \widetilde{M}_A \eta^{(i+1)} + 
\nn \\
& \hspace{1.95em} \gamma \sum_{k=1}^K ( P_f d_k^\star ) \circledast  \cH_{\!\sqrt{2 \alpha' \psi}} \big( d_k^\star \circledast x^{(i)} \big) \bigg) \bigg]_{\geq 0},
}
where 
\ea{
\eta^{(i+1)} &= \acute{x}^{(i+1)} - \widetilde{M}_A^{-1} A^T W \Big( A \acute{x}^{(i+1)} -y \Big),
\label{eq:soln:CT:bpgm:eta} \\
\acute{x}^{(i+1)} &= x^{(i)} + E_A^{(i+1)} \Big( x^{(i)} - x^{(i-1)} \Big),
\nn
}
\endgroup
$\widetilde{M}_A = \lambda_A M_A$ by \R{update:Mtilde}, a diagonal majorization matrix $M_A \succeq A^T W A$ is designed by Lemma~\ref{l:diag(|At|W|A|1)}, 
and $P_f \in \bbC^{R \times R}$ flips a column vector in the vertical direction (e.g., it rotates 2D filters by $180^\circ$).
Interpreting the update \R{eq:soln:CT:bpgm} leads to the following two remarks:

\rem{
\label{r:autoenc}
When the convolutional regularizer learned via CAOL \R{sys:CAOL:orth} is applied to MBIR, it works as an autoencoding CNN: 
\be{
\label{eq:autoenc}
\cM ( x ) = \sum_{k=1}^K (P_f d_k^\star) \circledast \cH_{\!\sqrt{2 \alpha_k'}} \left( d_k^\star \circledast x \right)
}
(setting $\psi = 1_{N'}$ and generalizing $\alpha'$ to $\{ \alpha'_k : k=1,\ldots,K \}$ in \R{sys:CT&CAOL}).
This is an explicit mathematical motivation
for constructing architectures of iterative regression CNNs for MBIR, 
e.g., BCD-Net \cite{Chun&Fessler:18IVMSP, Chun&etal:19MICCAI, Lim&etal:19TMI, Lim&etal:18NSSMIC}
and Momentum-Net \cite{Chun&etal:18Allerton, Chun&etal:18arXiv:momnet}. 
Particularly when the learned filters $\{ d_k^\star \}$ in \R{eq:autoenc} satisfy the TF condition, they are useful for compacting energy of an input signal $x$ and removing unwanted features via the non-linear thresholding in \R{eq:autoenc}.
}

\rem{
\label{r:autoenc-bpgm}
Update \R{eq:soln:CT:bpgm} improves the solution $x^{(i+1)}$ by weighting between \textit{a)} the extrapolated point considering the data fidelity, i.e., $\eta^{(i+1)}$ in \R{eq:soln:CT:bpgm:eta}, and \textit{b)} the \dquotes{refined} update via the ($\psi$-weighting) convolutional autoencoder, i.e., $\sum_{k} ( P_f d_k^\star ) \circledast  \cH_{\!\sqrt{2 \alpha' \psi}} ( d_k^\star \circledast x^{(i)} )$.
}

\section{Results and Discussion} \label{sec:result}

\subsection{Experimental Setup} \label{sec:exp}

This section examines the performance (e.g., scalability, convergence, and acceleration) and behaviors (e.g., effects of model parameters on filters structures and effects of dimensions of learned filter on MBIR performance) of the proposed CAOL algorithms and models, respectively.

\subsubsection{CAOL} \label{sec:exp:CAOL}

We tested the introduced CAOL models/algorithms for four datasets: 
\textit{1)} the fruit dataset with $L = 10$ and $N = 100 \!\times\! 100$ \cite{Zeiler&etal:10CVPR}; 
\textit{2)} the city dataset with $L = 10$ and $N = 100 \!\times\! 100$ \cite{Heide&eta:15CVPR}; 
\textit{3)} the CT dataset of $L = 80$ and $N = 128 \!\times\! 128$, created by dividing down-sampled $512 \!\times\! 512$ XCAT phantom slices \cite{Segars&etal:08MP} into $16$ sub-images \cite{Olshausen&Field:96Nature, Bristow&etal:13CVPR} -- referred to the CT-(\romnum{1}) dataset; 
\textit{4)} the CT dataset of with $L = 10$ and $N = 512 \!\times\! 512$ from down-sampled $512 \!\times\! 512$ XCAT phantom slices \cite{Segars&etal:08MP} -- referred to the CT-(\romnum{2}) dataset.
The preprocessing includes intensity rescaling to $[0,1]$ \cite{Zeiler&etal:10CVPR, Bristow&etal:13CVPR, Heide&eta:15CVPR} and/or (global) mean substraction \cite[\S 2]{Jarrett&etal:09ICCV}, \cite{Aharon&Elad&Bruckstein:06TSP}, as conventionally used in many sparse coding studies, e.g., \cite{Aharon&Elad&Bruckstein:06TSP, Jarrett&etal:09ICCV, Zeiler&etal:10CVPR, Bristow&etal:13CVPR, Heide&eta:15CVPR}. 
For the fruit and city datasets, we trained $K = 49$ filters of size $R = 7 \!\times\! 7$.
For the CT dataset~(\romnum{1}), we trained filters of size $R = 5 \!\times\! 5$, with $K = 25$ or $K = 20$.
For CT reconstruction experiments, we learned the filters from the CT-(\romnum{2}) dataset; however, we did not apply mean subtraction because it is not modeled in \R{sys:CT&CAOL}. 

The parameters for the BPEG-M algorithms were defined as follows.\footnote{The remaining BPEG-M parameters not described here are identical to those in \cite[\Romnum{7}-A2]{Chun&Fessler:18TIP}.}
We set the regularization parameters $\alpha, \beta$ as follows:
\begin{itemize}
\item {CAOL \R{sys:CAOL:orth}}: To investigate the effects of $\alpha$, we tested \R{sys:CAOL:orth} with different $\alpha$'s in the case $R = K$. For the fruit and city datasets, we used $\alpha = 2.5 \!\times\! \{ 10^{-5}, 10^{-4} \}$; for the CT-(\romnum{1}) dataset, we used $\alpha = \{ 10^{-4}, 2 \!\times\! 10^{-3} \}$. 
For the CT-(\romnum{2}) dataset (for CT reconstruction experiments), see details in \cite[Sec.~\Romnum{5}1]{Chun&Fessler:18Asilomar}.

\item {CAOL \R{sys:CAOL:div}}: Once $\alpha$ is fixed from the CAOL \R{sys:CAOL:orth} experiments above, we tested \R{sys:CAOL:div} with different $\beta$'s to see its effects in the case $R > K$. For the CT-(\romnum{1}) dataset, we fixed $\alpha = 10^{-4}$, and used $\beta = \{ 5 \!\times\! 10^6, 5 \!\times\! 10^4 \}$.
\end{itemize}
We set $\lambda_D = 1+\epsilon$ as the default.
We initialized filters in either deterministic or random ways.
The deterministic filter initialization follows that in \cite[Sec.~3.4]{Cai&etal:14ACHA}.
When filters were randomly initialized, we used a scaled one-vector for the first filter.
We initialize sparse codes mainly with a deterministic way that applies \R{eq:soln:spCode:exact} based on $\{ d_k^{(0)} \}$.
If not specified, we used the random filter and deterministic sparse code initializations.
For BPG \cite{Xu&Yin:17JSC}, we used the maximum eigenvalue of Hessians for Lipschitz constants in \R{sys:filter}, and applied the gradient-based restarting scheme in Section~\ref{sec:reBPGM}.
We terminated the iterations if the relative error stopping criterion (e.g., \cite[(44)]{Chun&Fessler:18TIP}) is met before reaching the maximum number of iterations.
We set the tolerance value as $10^{-13}$ for the CAOL algorithms using Proposition~\ref{p:MDexhess}, and $10^{-5}$ for those using Lemmas~\ref{l:MDdiag}--\ref{l:MDscaleI}, and the maximum number of iterations to $2 \!\times\! 10^4$.

The CAOL experiments used the convolutional operator learning toolbox~\cite{chun:19:convolt}.

\subsubsection{Sparse-View CT MBIR with Learned Convolutional Regularizer via CAOL} \label{sec:exp:CT}

We simulated sparse-view sinograms of size $888 \times 123$  (\quotes{detectors or rays} $\times$ \quotes{regularly spaced projection views or angles}, where $984$ is the number of full views) with GE LightSpeed fan-beam geometry corresponding to a monoenergetic source with $10^5$ incident photons per ray and no background events, and electronic noise variance $\sigma^2 \!=\! 5^2$. 
We avoided an inverse crime in our imaging simulation and reconstructed images with a coarser grid with $\Delta_x \!=\! \Delta_y \!=\! 0.9766$ mm; see details in \cite[Sec.~\Romnum{5}-A2]{Chun&Fessler:18Asilomar}.

For EP MBIR, we finely tuned its regularization parameter to achieve both good root mean square error (RMSE) and structural similarity index measurement \cite{Wang&etal:04TIP} values. 
For the CT MBIR model \R{sys:CT&CAOL}, we chose the model parameters $\{ \gamma, \alpha' \}$ that showed a good tradeoff between the data fidelity term and the learned convolutional regularizer, and set $\lambda_A \!=\! 1+\epsilon$.
We evaluated the reconstruction quality by the RMSE (in a modified Hounsfield unit, HU, where air is $0$ HU and water is $1000$ HU) in a region of interest.
See further details in \cite[Sec.~\Romnum{5}-A2]{Chun&Fessler:18Asilomar} and Fig.~\ref{fig:CTrecon}.

The imaging simulation and reconstruction experiments used the Michigan image reconstruction toolbox~\cite{fessler:16:irt}.

\subsection{CAOL with BPEG-M} \label{sec:result:CAOL}

Under the sharp majorization regime (i.e., partial or all blocks have sufficiently tight bounds in Lemma~\ref{l:QM}), the proposed convergence-guaranteed BPEG-M can achieve significantly faster CAOL convergence rates compared with the state-of-the-art BPG algorithm \cite{Xu&Yin:17JSC} for solving block multi-nonconvex problems, by several generalizations of  BPG (see Remark~\ref{r:BPGM}) and two majorization designs (see Proposition~\ref{p:MDexhess} and Lemma~\ref{l:MDscaleI}). See Fig.~\ref{fig:Comp:diffBPGM}.
In controlling the tradeoff between majorization sharpness and extrapolation effect of BPEG-M (i.e., choosing $\{ \lambda_b \}$ in \R{update:Mtilde}--\R{update:Eb}), maintaining majorization sharpness is more critical than gaining stronger extrapolation effects to accelerate convergence under the sharp majorization regime.
See Fig.~\ref{fig:Comp:exact_vs_approx}.

While using about two times less memory (see Table~\ref{tab:AOL}), CAOL \R{sys:CAOL} learns TF filters corresponding to those given by the patch-domain TF learning in \cite[Fig.~2]{Cai&etal:14ACHA}.
See Section~\ref{sec:CAOL:memory} and Fig.~\ref{fig:filters_BPGM} with deterministic $\{ d_k^{(0)} \}$.
Note that BPEG-M-based CAOL \R{sys:CAOL} requires even less memory than BPEG-M-based CDL in \cite{Chun&Fessler:18TIP}, by using exact sparse coding solutions (e.g., \R{eq:soln:spCode:exact} and \R{eq:soln:CT:bpgm}) without saving their extrapolated points.
In particular,  
when tested with the large CT dataset of $\{ L \!=\! 40, N \!=\! 512 \!\times\! 512 \}$,
the BPEG-M-based CAOL algorithm ran fine,
while BPEG-M-based CDL \cite{Chun&Fessler:18TIP} and patch-domain AOL \cite{Cai&etal:14ACHA} were terminated due to exceeding available memory.\footnote{
Their double-precision MATLAB implementations were tested on 3.3 GHz Intel Core i5 CPU with 32 GB RAM.
}
In addition, the CAOL models \R{sys:CAOL:orth} and \R{sys:CAOL:div} are easily parallelizable with $K$ threads.
Combining these results, the BPEG-M-based CAOL is a reasonable choice for learning filters from large training datasets. 
Finally, \cite{Chun&etal:19SPL} shows theoretically how using many samples can improve CAOL, 
accentuating the benefits of the low memory usage of CAOL.

\begin{figure}[!t]
\small\addtolength{\tabcolsep}{-4pt}
\centering

\begin{tabular}{cc}

\vspace{-0.1em}\hspace{-0.05em}\includegraphics[scale=0.5, trim=2em 5.2em 0em 4em, clip]{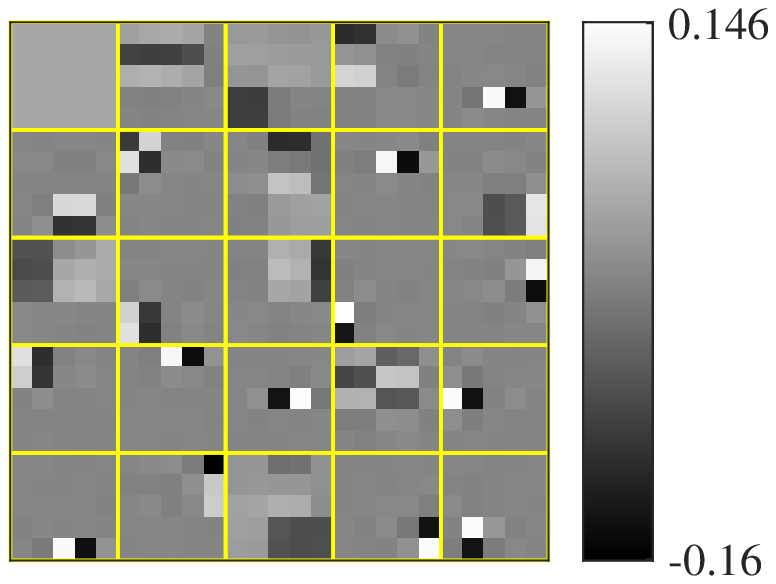} &
\vspace{-0.1em}\hspace{-0.4em}\includegraphics[scale=0.5, trim=2em 5.2em 0.1em 4em, clip]{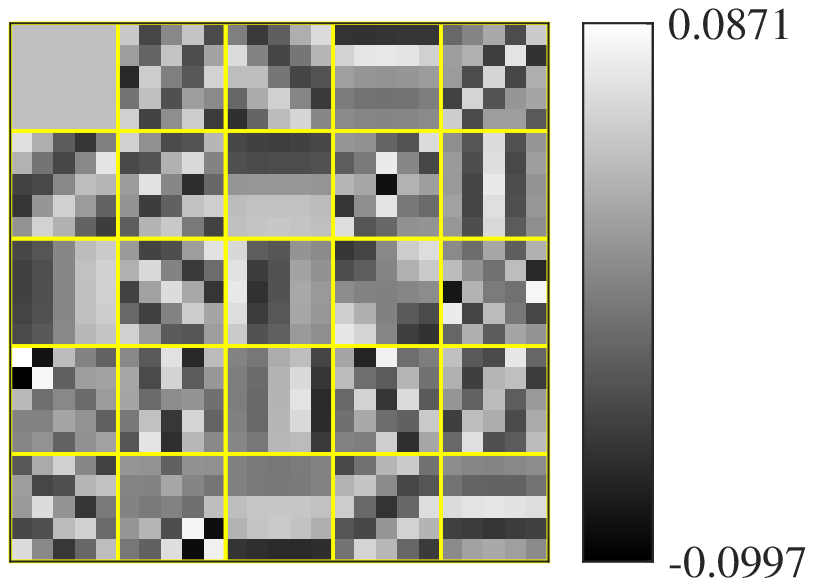} \\

{\small (a1) $\alpha = 10^{-4}$} & {\small (a2) $\alpha = 2 \!\times\! 10^{-3}$} \\

\multicolumn{2}{c}{\small (a) Learned filters via CAOL \R{sys:CAOL:orth} ($R = K = 25$)} \\

\includegraphics[scale=0.506, trim=2em 1.0em 0em 0.4em, clip]{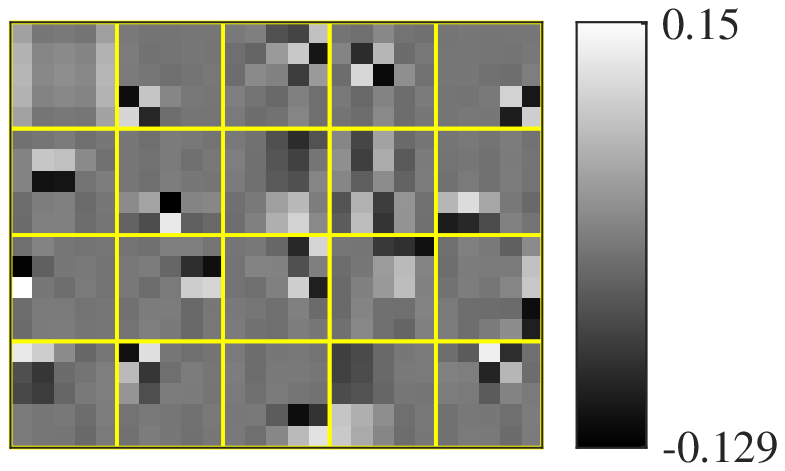} &
\includegraphics[scale=0.506, trim=2em 1.0em 0em 0.4em, clip]{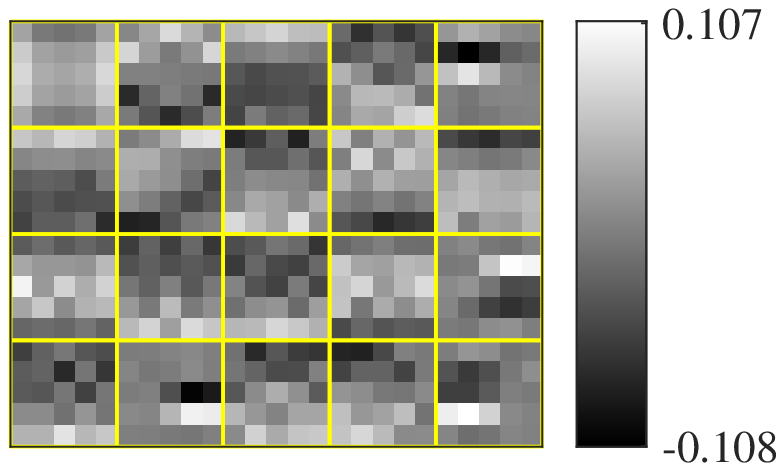} \vspace{-0.3em} \\

\small{$g_{\text{div}}(D) = \mathbf{8.96 \!\times\! 10^{-6}}$} & \small{$g_{\text{div}}(D) = \mathbf{0.12}$} \\

\small{(b1) $\alpha = 10^{-4}$, $\beta = 5 \!\times\! 10^{6}$} & {\small (b2) $\alpha = 10^{-4}$, $\beta = 5 \!\times\! 10^{4}$}  \\

\multicolumn{2}{c}{\small (b) Learned filters via CAOL \R{sys:CAOL:div} ($R = 25, K = 20$) } 

\end{tabular}

\vspace{-0.25em}
\caption{Examples of learned filters with different CAOL models and parameters (Proposition~\ref{p:MDexhess} was used for $M_D$; the CT-(\romnum{1}) dataset with a symmetric boundary condition).}
\label{fig:CAOL:filters:CT}
\end{figure}

\begin{figure*}[!t]
\centering
\small\addtolength{\tabcolsep}{-6.5pt}
\renewcommand{\arraystretch}{0.1}

    \begin{tabular}{ccccc}

      \small{(a) Ground truth} 
      & 
      \small{(b) Filtered back-projection} & \small{(c) EP} 
      & 
      \specialcell[c]{\small (d) Proposed MBIR \mbox{\R{sys:CT&CAOL}}, \\ \small with \R{eq:autoenc} of $R \!=\! K \!=\! 25$} 
      & 
      \specialcell[c]{\small (e) Proposed MBIR \mbox{\R{sys:CT&CAOL}}, \\ \small with \R{eq:autoenc} of $R \!=\! K \!=\! 49$} \\
       
       \begin{tikzpicture}
       		\begin{scope}[spy using outlines={rectangle,yellow,magnification=1.25,size=15mm,connect spies}]
       			\node {\includegraphics[viewport=35 105 275 345, clip, width=3.4cm,height=3.4cm]{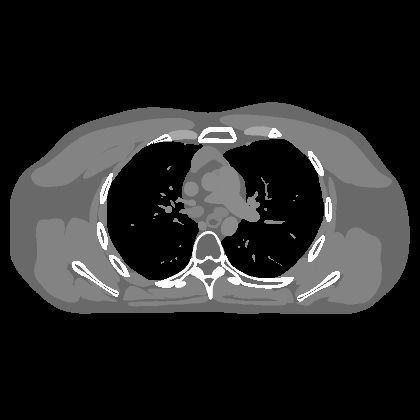}};
			\spy on (0.8,-0.95) in node [left] at (-0.3,-2.2);
                		\draw [->, thick, red] (0.1,0.85+0.3) -- (0.1,0.85);
                		\draw [->, thick, red] (-0.65+0.3,0.1) -- (-0.65,0.1);
                		\draw [->, thick, red] (-1.1,-1.3) -- (-1.1+0.3,-1.3);  
			\draw [red] (-1.35,0.15) circle [radius=0.28];
			\draw [red] (-1.35,-0.45) circle [radius=0.28];
		\end{scope}
        \end{tikzpicture}   
        &
        \begin{tikzpicture}
        		\begin{scope}[spy using outlines={rectangle,yellow,magnification=1.25,size=15mm,connect spies}]
       			\node {\includegraphics[viewport=35 105 275 345, clip, width=3.4cm,height=3.4cm]{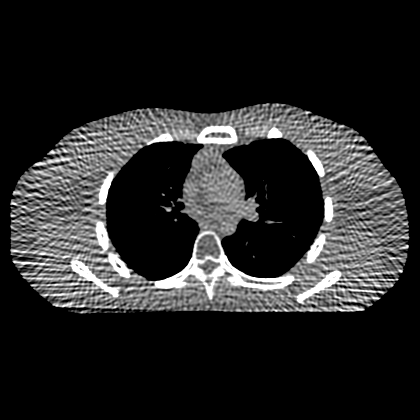}};
			\spy on (0.8,-0.95) in node [left] at (-0.3,-2.2);
			\node [black] at (0.7,-1.95) {\small $\mathrm{RMSE} = 82.8$};
		\end{scope}
        \end{tikzpicture}   
        &
        \begin{tikzpicture}
        		\begin{scope}[spy using outlines={rectangle,yellow,magnification=1.25,size=15mm,connect spies}]
       			\node {\includegraphics[viewport=35 105 275 345, clip, width=3.4cm,height=3.4cm]{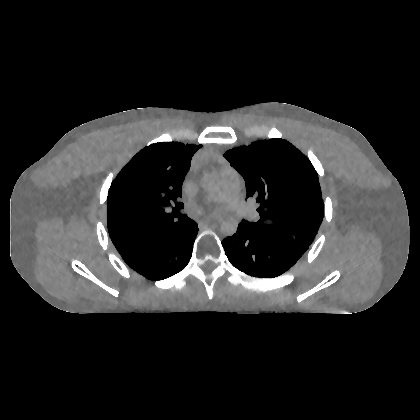}};
			\spy on (0.8,-0.95) in node [left] at (-0.3,-2.2);
                		\draw [->, thick, red] (0.1,0.85+0.3) -- (0.1,0.85);
                		\draw [->, thick, red] (-0.65+0.3,0.1) -- (-0.65,0.1);
                		\draw [->, thick, red] (-1.1,-1.3) -- (-1.1+0.3,-1.3);  
			\node [black] at (0.7,-1.95) {\small $\mathrm{RMSE} = 40.8$};
		\end{scope}
        \end{tikzpicture}   
        &
        \begin{tikzpicture}
        		\begin{scope}[spy using outlines={rectangle,yellow,magnification=1.25,size=15mm,connect spies}]
       			\node {\includegraphics[viewport=35 105 275 345, clip, width=3.4cm,height=3.4cm]{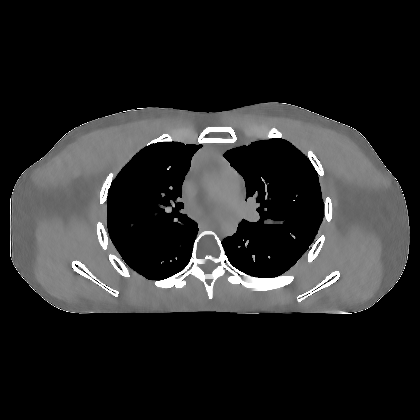}};
			\spy on (0.8,-0.95) in node [left] at (-0.3,-2.2);
                		\draw [->, thick, red] (0.1,0.85+0.3) -- (0.1,0.85);
                		\draw [->, thick, red] (-0.65+0.3,0.1) -- (-0.65,0.1);
                		\draw [->, thick, red] (-1.1,-1.3) -- (-1.1+0.3,-1.3);  
			\draw [red] (-1.35,0.15) circle [radius=0.28];
			\draw [red] (-1.35,-0.45) circle [radius=0.28];
			\node [blue] at (0.7,-1.95) {\small $\mathrm{RMSE} = 35.2$};
		\end{scope}
        \end{tikzpicture}   
        &
        \begin{tikzpicture}
        		\begin{scope}[spy using outlines={rectangle,yellow,magnification=1.25,size=15mm,connect spies}]
       			\node {\includegraphics[viewport=35 105 275 345, clip, width=3.4cm,height=3.4cm]{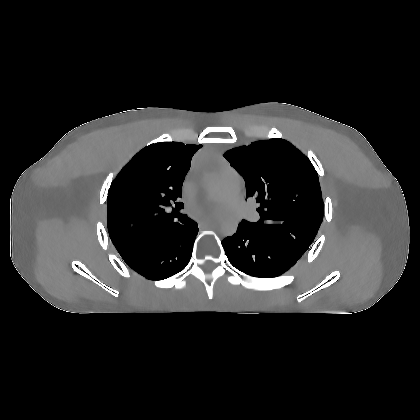}};
			\spy on (0.8,-0.95) in node [left] at (-0.3,-2.2);
                		\draw [->, thick, red] (0.1,0.85+0.3) -- (0.1,0.85);
                		\draw [->, thick, red] (-0.65+0.3,0.1) -- (-0.65,0.1);
                		\draw [->, thick, red] (-1.1,-1.3) -- (-1.1+0.3,-1.3);  
			\draw [red] (-1.35,0.15) circle [radius=0.28];
			\draw [red] (-1.35,-0.45) circle [radius=0.28];
			\node [red] at (0.7,-1.95) {\small $\mathrm{RMSE} = 34.7$};
		\end{scope}
        \end{tikzpicture}          

    \end{tabular}

 \vspace{-0.25em}
\caption{
Comparisons of reconstructed images from different reconstruction methods for sparse-view CT ($123$ views ($12.5$\% sampling); for the MBIR model \R{sys:CT&CAOL}, convolutional regularizers were trained by CAOL \R{sys:CAOL:orth} -- see \cite[Fig.~2]{Chun&Fessler:18Asilomar}; display window is within $[800, 1200]$ HU) \cite{Chun&Fessler:18Asilomar}.  
The MBIR model \R{sys:CT&CAOL} using convolutional sparsifying regularizers trained via CAOL \R{sys:CAOL:orth} shows higher image reconstruction accuracy compared to the EP reconstruction; see red arrows and magnified areas. 
For the MBIR model \R{sys:CT&CAOL}, the autoencoder (see Remark~\ref{r:autoenc}) using the filter dimension $R \!=\! K \!=\! 49$ improves reconstruction accuracy of that using $R \!=\! K \!=\! 25$; compare the results in (d) and (e).
In particular, the larger dimensional filters improve the edge sharpness of reconstructed images; see circled areas. The corresponding error maps are shown in Fig.~\ref{fig:CTrecon:err} of the supplementary material.
}
\label{fig:CTrecon}
\end{figure*}

The effects of parameters for the CAOL models are shown as follows.
In CAOL~\R{sys:CAOL:orth}, as the thresholding parameter $\alpha$ increases, the learned filters have more elongated structures; see Figs.~\ref{fig:CAOL:filters:CT}(a) and \ref{fig:CAOL:filters:fruit&city&ct-ii}.
In CAOL~\R{sys:CAOL:div}, when $\alpha$ is fixed, increasing the filter diversity promoting regularizer $\beta$ successfully lowers coherences between filters (e.g., $g_{\text{div}}(D)$ in \R{sys:CAOL:div}); see Fig.~\ref{fig:CAOL:filters:CT}(b).

In adaptive MBIR (e.g., \cite{Elad&Aharon:06TIP, Cai&etal:14ACHA, Pfister&Bresler:15SPIE}), one may apply adaptive image denoising \cite{Donoho:95TIT, Donoho&Johnstone:95JASA, Chang&Yu&Vetterli:00TIP, Blu&Luisier:07TIP, Liu&etal:15CVPR, Pfister&Bresler:17ICASSP} to optimize thresholding parameters. 
However, if CAOL \R{sys:CAOL} and testing the learned convolutional regularizer to MBIR (e.g., \R{sys:CT&CAOL}) are separated, selecting \dquotes{optimal} thresholding parameters in (unsupervised) CAOL is challenging -- similar to existing dictionary or analysis operator learning methods.
Our strategy to select the thresholding parameter $\alpha$ in CAOL~\R{sys:CAOL:orth} (with $R=K$) is given as follows.
We first apply the first-order finite difference filters $\{ d_k : \| d_k \|_2^2 = 1/R, \forall k \}$ (e.g., $\frac{1}{\sqrt{2R}} [1, -1]^T$ in 1D) to all training signals and find their sparse representations, and then find $\alpha_{\mathrm{est}}$ that corresponds to the largest $95 (\pm 1)\%$ of non-zero elements of the sparsified training signals. This procedure defines the range $[ \frac{1}{10} \alpha_{\mathrm{est}}, \alpha_{\mathrm{est}}]$ to select desirable $\alpha^\star$ and its corresponding filter $D^\star$. We next ran CAOL~\R{sys:CAOL:orth} with multiple $\alpha$ values within this range. Selecting $\{ \alpha^\star, D^\star \}$ depends on application.  
For CT MBIR, $D^\star$ that both has (short) first-order finite difference filters and captures diverse (particularly diagonal) features of training signals, gave good RMSE values and well preserved edges; see Fig.~\ref{fig:CAOL:filters:fruit&city&ct-ii}(c) and \cite[Fig.~2]{Chun&Fessler:18Asilomar}.

\subsection{Sparse-View CT MBIR with Learned Convolutional Sparsifying Regularizer (via CAOL) and BPEG-M} \label{sec:result:CT}

In sparse-view CT using only $12.5$\% of the full projections views, the CT MBIR \R{sys:CT&CAOL} using the learned convolutional regularizer via CAOL \R{sys:CAOL:orth} outperforms EP MBIR; it reduces RMSE by approximately $5.6$--$6.1$HU.
See the results in Figs.~\ref{fig:CTrecon}(c)--(e).
The model \R{sys:CT&CAOL} can better recover high-contrast regions (e.g., bones) -- see red arrows and magnified areas in Fig.~\ref{fig:CTrecon}(c)--(e). 
Nonetheless, the filters with $R\!=\!K\!=\!5^2$ in the ($\psi$-weighting) autoencoding CNN, i.e., $\sum_{k} ( P_f d_k^\star ) \circledast  \cH_{\!\sqrt{2 \alpha' \psi}} ( d_k^\star \circledast (\cdot) )$ in \R{eq:autoenc}, can blur edges in low-contrast regions (e.g., soft tissues) while removing noise. See Fig.~\ref{fig:CTrecon}(d) -- the blurry issues were similarly observed in \cite{Chun&etal:17Fully3D, Zheng&etal:19TCI}. 
The larger dimensional kernels (i.e., $R\!=\!K\!=\!7^2$) in the convolutional autoencoder can moderate this issue, while further reducing RMSE values; compare the results in Fig.~\ref{fig:CTrecon}(d)--(e).
In particular, the larger dimensional convolutional kernels capture more diverse features -- see \cite[Fig.~2]{Chun&Fessler:18Asilomar}) -- and the diverse features captured in kernels are useful to further improve the performance of the proposed MBIR model \R{sys:CT&CAOL}. (The importance of diverse features in kernels was similarly observed in CT experiments with the learned autoencoders having a fixed kernel dimension; see Fig.~\ref{fig:CAOL:filters:fruit&city&ct-ii}(c).)
The RMSE reduction over EP MBIR is comparable to that of CT MBIR \R{sys:CT&CAOL} 
using the $\{ R, K \!=\! 8^2 \}$-dimensional filters trained via the patch-domain AOL \cite{Ravishankar&Bressler:15TSP}; 
however, at each BPEG-M iteration, this MBIR model using the trained (non-TF) filters via patch-domain AOL \cite{Ravishankar&Bressler:15TSP} requires more computations than the proposed CT MBIR model \R{sys:CT&CAOL} using the learned convolutional regularizer via CAOL \R{sys:CAOL:orth}.
See related results and discussion in 
Fig.~\ref{fig:CTrecon:TL} and Section~\ref{sec:result:supp}, respectively.

On the algorithmic side, the BPEG-M framework can guarantee the convergence of CT MBIR \R{sys:CT&CAOL}.
Under the sharp majorization regime in BPEG-M, maintaining the majorization sharpness is more critical than having stronger extrapolation effects -- see \cite[Fig.~3]{Chun&Fessler:18Asilomar}, as similarly shown in CAOL experiments (see Section~\ref{sec:result:CAOL}).

\section{Conclusion} \label{sec:conclusion}

Developing rapidly converging and memory-efficient CAOL engines is important, since it is a basic element in training CNNs in an unsupervised learning manner (see Appendix~\ref{sec:CNN}). 
Studying structures of convolutional kernels is another fundamental issue, since it can avoid learning redundant filters or provide energy compaction properties to filters.
The proposed BPEG-M-based CAOL framework has several benefits.
First, the orthogonality constraint and diversity promoting regularizer in CAOL are useful in learning filters with diverse structures.
Second, the proposed BPEG-M algorithm significantly accelerates CAOL over the state-of-the-art method, BPG \cite{Xu&Yin:17JSC}, with our sufficiently sharp majorizer designs.
Third, BPEG-M-based CAOL uses much less memory compared to patch-domain AOL methods \cite{Yaghoobi&etal:13TSP, Hawe&Kleinsteuber&Diepold:13TIP, Ravishankar&Bressler:15TSP}, and easily allows parallel computing.
Finally, the learned convolutional regularizer provides the autoencoding CNN architecture in MBIR, and outperforms EP reconstruction in sparse-view CT.

Similar to existing unsupervised synthesis or analysis operator learning methods, the biggest remaining challenge of CAOL is optimizing its model parameters. 
This would become more challenging when one applies CAOL to train CNNs (see Appendix~\ref{sec:CNN}).
Our first future work is developing \dquotes{task-driven} CAOL that is particularly useful to train thresholding values.
Other future works include further acceleration of BPEG-M in Algorithm~\ref{alg:BPGM}, designing sharper majorizers requiring only $O( L R N )$ for the filter update problem of CAOL~\R{sys:CAOL}, and applying the CNN model learned via \R{sys:CNN:orth} to MBIR.

\section*{Appendix}
\renewcommand{\thesubsection}{\Alph{subsection}}

\subsection{Training CNN in a unsupervised manner via CAOL} \label{sec:CNN}

This section mathematically formulates an unsupervised training cost function for classical CNN (e.g., LeNet-5 \cite{LeCun&etal:98ProcIEEE} and AlexNet \cite{Krizhevsky&etal:12NIPS}) and solves the corresponding optimization problem, via the CAOL and BPEG-M frameworks studied in Sections~\ref{sec:CAOL}--\ref{sec:CAOL+BPGM}.
We model the three core modules of CNN: \textit{1)} convolution, \textit{2)} pooling, e.g., average \cite{LeCun&etal:98ProcIEEE} or max \cite{Jarrett&etal:09ICCV}, and \textit{3)} thresholding, e.g., RELU \cite{Nair&Hinton:10ICML}, while considering the TF filter condition in Proposition~\ref{p:TFconst}. 
Particularly, the orthogonality constraint in CAOL \R{sys:CAOL:orth} leads to a sharp majorizer, and BPEG-M is useful to train CNNs with convergence guarantees.
Note that it is unclear how to train such diverse (or incoherent) filters described in Section~\ref{sec:CAOL} by the most common CNN optimization method, the stochastic gradient method in which gradients are computed by back-propagation. The major challenges include \textit{a)} the non-differentiable hard thresholding operator related to $\ell^0$-norm in \R{sys:CAOL}, \textit{b)} the nonconvex filter constraints in \R{sys:CAOL:orth} and \R{sys:CAOL:div}, \textit{c)} using the identical filters in both encoder and decoder (e.g., $W$ and $W^H$ in Section~\ref{sec:prf:p:TF}), and \textit{d)} vanishing gradients.

For simplicity, we consider a two-layer CNN with a single training image, but one can extend the CNN model \R{sys:CNN:orth} (see below) to \dquotes{deep} layers with multiple images. The first layer consists of \textit{1c)} convolutional, \textit{1t)} thresholding, and \textit{1p)} pooling layers; the second layer consists of \textit{2c)} convolutional and \textit{2t)} thresholding layers.
Extending CAOL \R{sys:CAOL:orth}, we model two-layer CNN training as the following optimization problem:
\begingroup
\setlength{\thinmuskip}{1.5mu}
\setlength{\medmuskip}{2mu plus 1mu minus 2mu}
\setlength{\thickmuskip}{2.5mu plus 2.5mu}
\allowdisplaybreaks
\ea{
\label{sys:CNN:orth}
\argmin_{\{ d_{k}^{[1]}, d_{k,k'}^{[2]} \}} \min_{\{ z_{k}^{[1]}, z_{k'}^{[2]} \}}
&~ \sum_{k=1}^{K_1} \frac{1}{2}  \left\| d_{k}^{[1]} \circledast x - z_{k}^{[1]} \right\|_2^2 + \alpha_1 \left\| z_{k}^{[1]} \right\|_0 \nn
\\
+ &\, \frac{1}{2} \left\| \left( \sum_{k=1}^{K_1} \left[ \arraycolsep=2pt \begin{array}{c} d_{k,1}^{[2]} \circledast P z_{k}^{[1]} \\ \vdots \\ d_{k,K_2}^{^{[2]}} \circledast P z_{k}^{[1]} \end{array} \right] \right) - \left[ \arraycolsep=2pt \begin{array}{c} z_{1}^{[2]} \\ \vdots \\ z_{K_2}^{[2]} \end{array} \right] \right\|_2^2 \nn
\\
+ &\, \alpha_2 \sum_{k'=1}^{K_2} \left\| z_{k'}^{[2]} \right\|_0 \nn
\\
\mathrm{subject~to} ~~~~&~ D^{[1]} \big( D^{[1]} \big)^H = \frac{1}{R_1} \cdot I, \nn
\\
&~ D_k^{[2]} \big( D_k^{[2]} \big)^H = \frac{1}{R_2} \cdot I, \quad k = 1,\ldots,K_1,  
\tag{A1}
} 
\endgroup
where $x \in \bbR^{N}$ is the training data, $\{ d_{k}^{[1]} \in \bbR^{R_1} : k = 1,\ldots,K_1 \}$ is a set of filters in the first convolutional layer, $\{ z_{k}^{[1]} \in \bbR^{N} : k = 1,\ldots,K_1\}$ is a set of features after the first thresholding layer, $\{ d_{k,k'}^{[2]} \in \bbR^{R_2} : k' = 1,\ldots,K_2 \}$ is a set of filters for each of $\{ z_{k}^{[1]} \}$ in the second convolutional layer, $\{ z_{k'}^{[2]} \in \bbR^{N/\omega} : k = 1,\ldots,K_2\}$ is a set of features after the second thresholding layer, $D^{[1]}$ and $\{ D_k^{[2]} \}$ are similarly given as in \R{eq:D}, $P \in \bbR^{N/\omega \times \omega}$ denotes an average pooling \cite{LeCun&etal:98ProcIEEE} operator (see its definition below), and $\omega$ is the size of pooling window. 
The superscripted number in the bracket of vectors and matrices denotes the $(\cdot)\rth$ layer.
Here, we model a simple average pooling operator $P \in \bbR^{(N/\omega) \times \omega}$ by a block diagonal matrix with row vector $\frac{1}{\omega} 1_\omega^T \in \bbR^{\omega}$: $P := \frac{1}{\omega} \bigoplus_{j=1}^{N/\omega} 1_\omega^T$.
We obtain a majorization matrix of $P^T P$ by $P^T P \preceq \diag( P^T P 1_N ) = \frac{1}{\omega} I_N$ (using Lemma~\ref{l:diag(|At|W|A|1)}).
For 2D case, the structure of $P$ changes, but $P^T P \preceq \frac{1}{\omega} I_N$ holds.

We solve the CNN training model in \R{sys:CNN:orth} via the BPEG-M techniques in Section~\ref{sec:CAOL+BPGM}, and relate the solutions of \R{sys:CNN:orth} and modules in the two-layer CNN training. The symbols in the following items denote the CNN modules.
\begin{itemize}
[\setlength{\IEEElabelindent}{\IEEEilabelindentA}]
\item[\textit{1c})] Filters in the first layer, $\{ d_{k}^{[1]} \}$: Updating the filters is straightforward via the techniques in Section~\ref{sec:prox:filter:orth}. 

\item[\textit{1t})] Features at the first layers, $\{ z_{k}^{[1]} \}$: Using BPEG-M with the $k\rth$ set of TF filters $\{ d_{k,k'}^{[2]} : k' \}$ and $P^T P \preceq \frac{1}{\omega} I_N$ (see above), the proximal mapping for $z_{k}^{[1]}$ is
\begingroup
\setlength{\thinmuskip}{1.5mu}
\setlength{\medmuskip}{2mu plus 1mu minus 2mu}
\setlength{\thickmuskip}{2.5mu plus 2.5mu}
\be{
\label{eq:CNN:layer1:spCd}
\min_{z_{k}^{[1]}} \frac{1}{2} \left\| d_{k}^{[1]} \circledast x - z_{k}^{[1]} \right\|_2^2 + \frac{1}{2\omega'} \left\| z_{k}^{[1]} - \zeta_{k}^{[k]} \right\|_2^2 + \alpha_1  \left\| z_{k}^{[1]} \right\|_0,
}
\endgroup
where $\omega' = \omega / \lambda_Z$ and $\zeta_{k}^{[k]}$ is given by \R{update:x}. Combining the first two quadratic terms in \R{eq:CNN:layer1:spCd} into a single quadratic term leads to an optimal update for \R{eq:CNN:layer1:spCd}:
\bes{
z_{k}^{[1]} = \cH_{\! \sqrt{ 2 \frac{\omega' \alpha_1}{\omega' + 1} }} \left( d_{k}^{[1]} \circledast x  + \frac{1}{\omega'} \zeta_{k}^{[k]}  \right), \quad k \in [K],
} 
where the hard thresholding operator $\cH_a (\cdot)$ with a thresholding parameter $a$ is defined in \R{eq:def:hardthr}.

\item[\textit{1p})] Pooling, $P$: Applying the pooling operator $P$ to $\{ z_{k}^{[1]} \}$ gives input data -- $\{ P z_{k}^{[1]} \}$ -- to the second layer.

\item[\textit{2c})] Filters in the second layer, $\{ d_{k,k'}^{[2]} \}$: We update the $k\rth$ set filters $\{ d_{k,k'}^{[2]} : \forall k' \}$ in a sequential way. Updating the $k\rth$ set filters is straightforward via the techniques in Section~\ref{sec:prox:filter:orth}. 

\item[\textit{2t})] Features at the second layers, $\{  z_{k'}^{[2]} \}$: The corresponding update is given by
\bes{
z_{k'}^{[2]} = \cH_{\!\sqrt{2 \alpha_2}} \left( \sum_{k=1}^{K_1} d_{k,k'}^{[1]} \circledast P z_{k}^{[1]} \right), \quad k' \in [K_2].
}
\end{itemize}

Considering the introduced mathematical formulation of training CNNs \cite{LeCun&etal:98ProcIEEE} via CAOL, BPEG-M-based CAOL has potential to be a basic engine to rapidly train CNNs with big data (i.e., training data consisting of many (high-dimensional) signals).

\subsection{Examples of $\{ f(x;y), \cX \}$ in MBIR model \R{eq:mbir:learn} using learned regularizers} \label{sec:egs}

This section introduces some potential applications 
of using MBIR model \R{eq:mbir:learn} using learned regularizers
in imaging processing, imaging, and computer vision.
We first consider quadratic data fidelity function in the form of $f(x;y) = \frac{1}{2} \| y - A x \|_W^2$.
Examples include
\begin{itemize}
\item Image debluring (with $W \!=\! I$ for simplicity), 
where $y$ is a blurred image, $A$ is a blurring operator, and $\cX$ is a box constraint; 
\item Image denoising (with $A \!=\! I$),
where $y$ is a noisy image corrupted by additive white Gaussian noise (AWGN),
$W$ is the inverse covariance matrix corresponding to AWGN statistics,
and $\cX$ is a box constraint;
\item Compressed sensing (with $\{ W \!=\! I, \cX \!\in\! \bbC^{N'} \}$ for simplicity) \cite{Chun&Adcock:17TIT, Chun&Adcock:18ACHA},
where $y$ is a measurement vector, 
and $A$ is a compressed sensing operator,
e.g., subgaussian random matrix, bounded orthonormal system, subsampled isometries, certain types of random convolutions;
\item Image inpainting (with $W \!=\! I$ for simplicity),
where $y$ is an image with missing entries, $A$ is a masking operator, and $\cX$ is a box constraint;
\item Light-field photography from focal stack data 
with $f(x;y) = \sum_{c} \| y_c - \sum_{s}  A_{c,s} x_{s} \|_2^2$,
where $y_c$ denotes measurements collected at the $c\rth$ sensor, 
$A_{c,s}$ models camera imaging geometry at the $s\rth$ angular position for the $c\rth$ detector, 
$x_s$ denotes the $s\rth$ sub-aperture image, $\forall c,s$,
and $\cX$ is a box constraint  \cite{Block&Chun&Fessler:18IVMSP, Chun&etal:18arXiv:momnet}.
\end{itemize}
Examples that use nonlinear data fidelity function include 
image classification using the logistic function \cite{Mairal&etal:09NIPS}, 
magnetic resonance imaging considering unknown magnetic field variation \cite{Fessler:10SPM}, 
and positron emission tomography \cite{Lim&etal:19TMI}.

\subsection{Notation} \label{sec:notation}

We use $\nm{\cdot}_{p}$ to denote the $\ell^p$-norm and write $\ip{\cdot}{\cdot}$ for the standard inner product on $\bbC^N$.  
The weighted $\ell^2$-norm with a Hermitian positive definite matrix $A$ is denoted by $\nm{\cdot}_{A} = \nm{ A^{1/2} (\cdot) }_2$.
$\nm{\cdot}_{0}$ denotes the $\ell^0$-quasi-norm, i.e., the number of nonzeros of a vector.  
The Frobenius norm of a matrix is denoted by $\| \cdot \|_{\mathrm{F}}$. 
$( \cdot )^T$, $( \cdot )^H$, and $( {\cdot} )^*$ indicate the transpose, complex conjugate transpose (Hermitian transpose), and complex conjugate, respectively. 
$\diag(\cdot)$ denotes the conversion of a vector into a diagonal matrix or diagonal elements of a matrix into a vector.
$\bigoplus$ denotes the matrix direct sum of matrices.
$[C]$ denotes the set $\{1,2,\ldots,C\}$.  
Distinct from the index $i$, we denote the imaginary unit $\sqrt{-1}$ by $\imath$. 
For (self-adjoint) matrices $A,B \in \bbC^{N \times N}$, the notation $B \preceq A$ denotes that $A-B$ is a positive semi-definite matrix.

\section*{Acknowledgment}

We thank Xuehang Zheng for providing CT imaging simulation setup, and Dr. Jonghoon Jin for constructive feedback on CNNs.

\bibliographystyle{IEEEtran}
\bibliography{referenceBibs_Bobby}


\renewcommand{\thefigure}{S.\arabic{figure}}
\renewcommand{\thetable}{S.\Roman{table}}
\renewcommand{\thesection}{S.\Roman{section}}
\renewcommand{\thefootnote}{S.\arabic{footnote}}
\renewcommand{\thetheorem}{S.\arabic{theorem}}
\renewcommand{\theequation}{S.\arabic{equation}}
\renewcommand{\thealgorithm}{S.\arabic{algorithm}} 

\setcounter{section}{0}
\setcounter{equation}{0}
\setcounter{theorem}{0}
\setcounter{figure}{0}
\setcounter{table}{0}
\setcounter{algorithm}{0}
\setcounter{footnote}{0}

{
\twocolumn[
\begin{center}
\fontsize{23.4}{23.4}\selectfont Convolutional Analysis Operator Learning: Acceleration and Convergence 
\\
(Supplementary Material)
\\
\vspace{0.2in}
\fontsize{11}{11}\selectfont Il Yong Chun, \textit{Member}, \textit{IEEE}, and Jeffrey A. Fessler, \textit{Fellow}, \textit{IEEE}
\vspace{0.45in}
\end{center}]
}


This supplementary material for \cite{Chun&Fessler:18arXiv:supp} provides mathematical proofs, detailed descriptions, and additional experimental results that support several arguments in the main manuscript.  We use the prefix ``S'' for the numbers in section, equation, figure, algorithm, and footnote in the supplementary material.\footnote{Supplementary material dated August $22$, 2019.}

\subsubsection*{Comments on Convolutional Operator $\circledast$}

Throughout the paper, we fix the dimension of $d_k \circledast x_l$ by $N$ (e.g., \dquotes{$\mathrm{same}$} option in convolution functions in MATLAB) for simplicity. However, one can generalize it to $P_B (d_k \circledast x_l)$ for considering arbitrary boundary truncations (e.g., \dquotes{$\mathrm{full}$} or \dquotes{$\mathrm{valid}$} options) and conditions (e.g., zero boundary). Here, $d_k \circledast x_l \in \bbC^{N\!+\!R\!-\!1}$, $P_{B} \in \bbC^{N' \times (N\!+\!R\!-\!1)}$ is a selection matrix with $| B | = N'$ and $N' \leq N\!+\!R\!-\!1$, and $B$ is a list of distinct indices from the set $\{1, \ldots, N\!+\!R\!-\!1\}$ that correspond to truncating the boundaries of the padded convolution.

\section{Proofs of Proposition~\ref{p:TFconst} and Its Relation to Results Derived by Local Approaches} \label{sec:prf:p:TF}

We consider the following 1D setup for simplicity.
A non-padded signal $x \in \bbC^{N}$ has support in the set $\{ 0,1, \ldots, N-1 \}$.
The odd-sized filters $\{ d_k \in \bbC^R : k \in [K] \}$ have finite support in the set $\{  -\Delta, -\Delta+1, \ldots, \Delta \}$ and padded signal $\hat{x} \in \bbC^{N+2\Delta}$ has finite support in the set $\{ 0,1, \ldots, N-1+2\Delta \}$, where $\Delta$ is a half width of odd-sized filters $d_k$'s, e.g., $\Delta = \lfloor R/2 \rfloor$.
We aim to find conditions of $\{ d_k : k \in [K] \}$ to show
\begingroup
\allowdisplaybreaks
\ea{
\label{eq:prf:p:TF:design}
\sum_{k=1}^K \nm{ d_k \circledast x }_2^2 &= \nm{ x }_2^2 \nn
\\
\leftrightarrow  \sum_{k=1}^K \sum_{n=\Delta}^{N-1+\Delta} \left| \sum_{r=-\Delta}^{\Delta}  \hat{x} (n-r) d_k (r) \right|^2 
& = \sum_{n'=0}^{N-1} | x( n' ) |^2,
}
for any $x \in \bbC^N$.
We first rewrite the term $\sum_{k} \nm{ d_k \circledast x }_2^2$ by
\eas{
&\, \sum_{k=1}^K \nm{ d_k \circledast x }_2^2
\\
&= \sum_{k=1}^K \sum_{n=\Delta}^{N-1+\Delta} \left( \sum_{r=-\Delta}^{\Delta} \hat{x} (n-r) d_k (r) \right)^{\!\!*} \sum_{r'=-\Delta}^{\Delta}  \hat{x} (n-r') d_k (r')
\\
&= \sum_{k=1}^K \sum_{n=\Delta}^{N-1+\Delta} \sum_{r=-\Delta}^{\Delta}  \sum_{r'=-\Delta}^{\Delta}  \hat{x}^{*} (n-r) d_k^{*} (r) \hat{x} (n-r') d_k (r')
\\
&= \sum_{k=1}^K \sum_{n=\Delta}^{N-1+\Delta} \sum_{r=-\Delta}^{\Delta} \left| \hat{x} (n-r) \right|^2 \left| d_k (r) \right|^2 
\\
& \hspace{8.4em} + \sum_{r' \neq r}  \hat{x}^{*} (n-r) d_k^{*} (r) \hat{x} (n-r') d_k (r').
}
The second summation term further simplifies to
\eas{
&\, \sum_{n=\Delta}^{N-1+\Delta} \sum_{r=-\Delta}^{\Delta}  \sum_{r' \neq r}  \sum_{k=1}^K  \hat{x}^{*} (n-r) d_k^{*} (r) \hat{x} (n-r') d_k (r')
\\
&= \sum_{n=\Delta}^{N-1+\Delta} \sum_{r=-\Delta}^{\Delta}  \sum_{r' \neq r}  \hat{x}^{*} (n-r)  \hat{x} (n-r')  \sum_{k=1}^K d_k^{*} (r) d_k (r').
}
If $d_k$'s satisfy the orthogonality condition in Proposition~\ref{p:TFconst}, i.e.,
\be{
\label{eq:CAOL:TFfilt}
\sum_{k=1}^K d_k (r) d_k^{*} (r') = \frac{1}{R} \delta_{r-r'}, \quad \forall r,r' \in \bbZ^1~\mathrm{or}~\bbZ^2,
}
where $\delta_n$ denotes the Kronecker impulse,
then the equality in \R{eq:prf:p:TF:design} holds:
\eas{
\sum_{k=1}^K \nm{ d_k \circledast x }_2^2 
&=  \sum_{n=\Delta}^{N-1+\Delta} \sum_{r=-\Delta}^{\Delta} \left| \hat{x} (n-r) \right|^2 \sum_{k=1}^K \left| d_k (r) \right|^2 
\\
&= \frac{1}{R} \sum_{n=\Delta}^{N-1+\Delta} \sum_{r=-\Delta}^{\Delta} \left| \hat{x} (j-r) \right|^2
\\
&= \sum_{n'=0}^{N-1} | x( n' ) |^2
}
\endgroup
where the last equality holds by periodic or mirror-reflective signal padding. 
It is straightforward to extend the proofs to even-sized filters and 2D case. 

We next explain the relation between the TF condition in Proposition~\ref{p:TFconst} and that given by the local approach.
Reformulate $[ (d_1 \!\circledast x)^H, \ldots, (d_K \!\circledast x)^H ]^H$ as $[ (W P_1)^H, \ldots, (W P_N)^H ]^H x$, where the $k\rth$ row of $W \in \bbC^{K \times R}$ corresponds to the $k\rth$ filter's coefficients, $\{ P_n \in \bbC^{R \times N} \}$ is a set of patch extraction operators (with a circular boundary condition and the sliding parameter $1$), and $x \in \bbC^N$. 
To enforce a TF condition with this local perspective, the matrix $W$ (in \cite{Cai&etal:14ACHA:supp, Ravishankar&Bressler:15TSP:supp}) should satisfy $\sum_{n=1}^N P_n^H W^H W P_n = I$. 
This is satisfied when $W^H W \!\!=\!\! \frac{1}{R} \!\cdot\! I$, considering that $\sum_{n=1}^N P_n^H P_n \!=\! R \!\cdot\! I$ with the patch extraction assumptions above. 
Thus, the orthogonality constraint $D D^H \!=\! \frac{1}{R} I$ in Proposition~\ref{p:TFconst}, i.e., \R{eq:CAOL:TFfilt}, corresponds to the TF condition derived by the local approach.

\section{Proofs of Lemma~\ref{l:QM}} \label{sec:prf:l:QM}

By the $1\mathrm{st}$-order Taylor integral, observe that
\bes{
f(x) - f(y) = \int_0^1 \ip{\nabla f (y + t(x-y))}{x - y}  \D t.
}
In addition, we attain
\ea{
\label{eq:newIneq}
\ip{x}{y} &= x^T M^{-1/2} M^{1/2} y = \ip{M^{-1/2} x}{M^{1/2} y}
\nn \\
&\leq \nm{x}_{M^{-1}} \nm{y}_{M},
}
for any $x,y \in \bbR^n$ and $M =  M^T \succ 0$, where the second equality hold by $M^{-1/2} = ( M^{-1/2} )^T$ due to the assumption of $M$ and the inequality holds by Cauchy-Schwarz inequality and the definition of $\nm{x}_{M}^2$ in Definition~\ref{d:QM}.
For $x,y \in \bbR^n$, we now obtain that
\begingroup
\setlength{\thinmuskip}{1.5mu}
\setlength{\medmuskip}{2mu plus 1mu minus 2mu}
\setlength{\thickmuskip}{2.5mu plus 2.5mu}
\allowdisplaybreaks
\eas{
&\,\, f(x) 
\\
&= f(y) + \int_0^1 \ip{\nabla f (y + t(x-y))}{x - y} \D t
\\
&= f(y) + \ip{\nabla f(y)}{x - y} \hspace{0.2em}+
\\
& \quad \int_0^1 \ip{\nabla f (y + t(x-y)) - \nabla f(y)}{x - y} \D t 
\\
&\leq f(y) + \ip{\nabla f(y)}{x - y} \hspace{0.2em}+
\\
& \quad \int_0^1 \nm{ \nabla f (y + t(x-y)) - \nabla f(y) }_{M^{-1}} \nm{x - y}_{M} \D t 
\\
&\leq f(y) + \ip{\nabla f(y)}{x - y}  + \int_0^1 t \nm{  x - y }_{M}^2 \D t 
\\
&= f(y) + \ip{\nabla f(y)}{x-y} + \frac{1}{2} \| x-y \|_{M}^2,
}
\endgroup
where the first inequality holds by \R{eq:newIneq}, and the second inequality holds by $M$-Lipschitz continuity of $\nabla f$ (see Definition~\ref{d:QM}).
This completes the proof.

\section{Proofs of Lemma~\ref{l:seqBound}} \label{sec:prf:l:seqBound}

The following proof extends that given in \cite[Lem.~1]{Xu&Yin:17JSC}.
By the $M$-Lipschitz continuity of $\nabla_{x_b} f_b^{(i+1)} (x_b)$ about $x_b$ and Proposition \ref{l:QM}, it holds that (e.g., see \cite[Lem. S.1]{Chun&Fessler:18TIP})
\ea{
\label{eq:seqBound:1}
&~ f_b^{(i+1)} (x_b^{(i+1)}) \nn
\\
& \leq f_b^{(i+1)} (x_b^{(i)}) + \ip{\nabla_{x_b} f_b^{(i+1)} (x_b^{(i)})}{x_b^{(i+1)}-x_b^{(i)}} \nn
\\
& \quad + \frac{1}{2} \nm{x_b^{(i+1)} - x_b^{(i)}}_{M_b^{(i+1)}}^2.
}
Considering that $x_b^{(i+1)}$ is a minimizer of \R{update:x}, we have
\ea{
\label{eq:seqBound:2}
&\,\, \ip{ \nabla_{x_b} f_b^{(i+1)} (\acute{x}_b^{(i+1)}) }{ x_b^{(i+1)} - \acute{x}_b^{(i)} } \nn  
\\
&\, + \frac{1}{2} \nm{ x_b^{(i+1)} - \acute{x}_b^{(i+1)} }_{\widetilde{M}_b^{(i+1)}}^2 + g_b (x_b^{(i+1)}) \nn
\\
&\leq \ip{ \nabla_{x_b} f_b^{(i+1)} (\acute{x}_b^{(i+1)}) }{ x_b^{(i)} - \acute{x}_b^{(i)} } \nn
\\
& \quad\, + \frac{1}{2} \nm{ x_b^{(i)} - \acute{x}_b^{(i+1)} }_{\widetilde{M}_b^{(i+1)}}^2 + g_b (x_b^{(i)}) 
}

Summing \R{eq:seqBound:1} and \R{eq:seqBound:2}, we obtain
\begingroup
\setlength{\thinmuskip}{1.5mu}
\setlength{\medmuskip}{2mu plus 1mu minus 2mu}
\setlength{\thickmuskip}{2.5mu plus 2.5mu}
\allowdisplaybreaks
\ea{
&\,\, F_b (x_b^{(i)}) - F_b (x_b^{(i+1)}) \nn
\\ 
&= f_b^{(i+1)} (x_b^{(i)}) + g_b (x_b^{(i)}) -  f_b^{(i+1)} (x_b^{(i+1)}) - g_b (x_b^{(i+1)}) \nn
\\
& \geq \ip{ \nabla_{x_b} f_b^{(i+1)} (\acute{x}_b^{(i+1)}) }{x_b^{(i+1)} - x_b^{(i)}} \nn
\\
& \quad - \ip{ \nabla_{x_b} f_b^{(i+1)} (x_b^{(i)}) }{x_b^{(i+1)} - x_b^{(i)}} \nn
\\
& \quad - \frac{1}{2} \nm{x_b^{(i+1)} - x_b^{(i)}}_{M_b^{(i+1)}}^2 + \frac{1}{2} \nm{ x_b^{(i+1)} - \acute{x}_b^{(i+1)} }_{\widetilde{M}_b^{(i+1)}}^2 \nn 
\\
& \quad + \frac{1}{2} \nm{ x_b^{(i)} - \acute{x}_b^{(i+1)} }_{\widetilde{M}_b^{(i+1)}}^2 \nn
\\
& \geq \ip{  \nabla_{x_b} f_b^{(i+1)} (\acute{x}_b^{(i+1)}) - \nabla_{x_b} f_b^{(i+1)} (x_b^{(i)}) }{x_b^{(i+1)} - x_b^{(i)}} \nn
\\
& \quad + \ip{ \widetilde{M}_b^{(i+1)} ( x_b^{(i)} - \acute{x}_b^{(i+1)} ) }{ x_b^{(i+1)} - x_b^{(i)} } \nn
\\
& \quad +  \frac{1}{2} \nm{ x_b^{(i+1)} - x_b^{(i)} }_{\widetilde{M}_b^{(i+1)} - M_b^{(i+1)}}^2 \nn
\\
& \geq - \nm{ x_b^{(i+1)} - x_b^{(i)} }_2 \cdot \nm{ \nabla_{x_b} f_b^{(i+1)} (\acute{x}_b^{(i+1)}) - \nabla_{x_b} f_b^{(i+1)} (x_b^{(i)}) }_2 \label{eq:seqBound:Cauchy}
\\
& \quad - \nm{  x_b^{(i+1)} - x_b^{(i)} }_{\widetilde{M}_b^{(i+1)}} \cdot \nm{ x_b^{(i)} - \acute{x}_b^{(i+1)} }_{\widetilde{M}_b^{(i+1)}} \nn
\\
& \quad + \frac{1}{2} \nm{ x_b^{(i+1)} - x_b^{(i)} }_{\widetilde{M}_b^{(i+1)} - M_b^{(i+1)}}^2 \nn
\\
& \geq - \nm{ x_b^{(i+1)} - x_b^{(i)} }_{M_b^{(i+1)}} \label{eq:seqBound:new}
\\
& \quad \cdot \nm{ \nabla_{x_b} f_b^{(i+1)} (\acute{x}_b^{(i+1)}) - \nabla_{x_b} f_b^{(i+1)} (x_b^{(i)}) }_{\left(\! M_b^{(i+1)} \!\right)^{-1}} \nn
\\
& \quad - \nm{  x_b^{(i+1)} - x_b^{(i)} }_{\widetilde{M}_b^{(i+1)}} \cdot \nm{ x_b^{(i)} - \acute{x}_b^{(i+1)} }_{\widetilde{M}_b^{(i+1)}} \nn
\\
& \quad + \frac{1}{2} \nm{ x_b^{(i+1)} - x_b^{(i)} }_{\widetilde{M}_b^{(i+1)} - M_b^{(i+1)}}^2 \nn
\\
& \geq - \nm{ x_b^{(i+1)} - x_b^{(i)} }_{M_b^{(i+1)}} \cdot \nm{ \acute{x}_b^{(i+1)} - x_b^{(i)} }_{M_b^{(i+1)}} \label{eq:seqBound:QM}
\\
& \quad - \nm{ x_b^{(i+1)} - x_b^{(i)} }_{\widetilde{M}_b^{(i+1)}} \cdot \nm{ x_b^{(i)} - \acute{x}_b^{(i+1)} }_{\widetilde{M}_b^{(i+1)}} \nn
\\
& \quad + \frac{1}{2} \nm{ x_b^{(i+1)} - x_b^{(i)} }_{\widetilde{M}_b^{(i+1)} - M_b^{(i+1)}}^2 \nn
\\
& \geq - \nm{ x_b^{(i+1)} - x_b^{(i)} }_{\widetilde{M}_b^{(i+1)} + M_b^{(i+1)}} \cdot \nm{ x_b^{(i)} - \acute{x}_b^{(i+1)} }_{M_b^{(i+1)}} 
\label{eq:seqBound:muM}
\\
& \quad - \nm{ x_b^{(i+1)} - x_b^{(i)} }_{\widetilde{M}_b^{(i+1)} + M_b^{(i+1)}} \cdot \nm{ x_b^{(i)} - \acute{x}_b^{(i+1)} }_{\widetilde{M}_b^{(i+1)}} \nn
\\
& \quad + \frac{1}{2} \nm{ x_b^{(i+1)} - x_b^{(i)} }_{\widetilde{M}_b^{(i+1)} - M_b^{(i+1)}}^2 \nn
\\
& \geq \frac{\lambda_b - 1}{4} \nm{ x_b^{(i+1)} - x_b^{(i)} }_{M_b^{(i+1)}}^2 \label{eq:seqBound:Young}
\\
& \quad - \frac{ (\lambda_b + 1)^2 }{\lambda_b - 1} \nm{ x_b^{(i)} - \acute{x}_b^{(i+1)} }_{M_b^{(i+1)}}^2 \nn
\\
& = \frac{\lambda_b - 1}{4} \nm{ x_b^{(i+1)} - x_b^{(i)} }_{M_b^{(i+1)}}^2 \label{eq:seqBound:xacute}
\\
& \quad - \frac{ (\lambda_b + 1)^2 }{\lambda_b - 1} \nm{ E_b^{(i+1)} ( x_b^{(i)} - x_b^{(i-1)} ) }_{ M_b^{(i+1)}}^2 \nn
\\
& \geq \frac{\lambda_b - 1}{4} \nm{ x_b^{(i+1)} - x_b^{(i)} }_{M_b^{(i+1)}}^2 \label{eq:seqBound:final}
\\
& \quad - \frac{ ( \lambda_b - 1 ) \delta^2}{4} \nm{ x_b^{(i)} - x_b^{(i-1)} }_{ M_b^{(i)}}^2 \nn
}
\endgroup
where the inequality \R{eq:seqBound:Cauchy} holds by Cauchy-Schwarz inequality, 
the inequality \R{eq:seqBound:new} holds by \R{eq:newIneq}, 
the inequality \R{eq:seqBound:QM} holds by \R{eq:QMbound} in Assumption 2, 
the inequality \R{eq:seqBound:muM} holds by \R{update:Mtilde}, 
the inequality \R{eq:seqBound:Young} holds by \R{update:Mtilde} and Young's inequality, i.e., $ab \leq \frac{a^2}{2 \varepsilon} + \frac{\varepsilon b^2}{2}$, where $a,b \geq 0$ and $\varepsilon > 0$, with $\varepsilon = 2 (\lambda_b + 1) (\lambda_b - 1)^{-1}$ (note that $\lambda_b  > 1$ via \R{update:Mtilde}), 
the equality \R{eq:seqBound:xacute} holds by \R{update:xacute}, 
and the inequality \R{eq:seqBound:final} holds by \R{cond:Wb} in Assumption 3. This completes the proof.

\section{Proof of Proposition~\ref{p:sqSum}} \label{sec:prf:p:sqSum}

Summing the following inequality of $F (x_b^{(i)}) - F (x_b^{(i+1)})$
\begingroup
\setlength{\thinmuskip}{1.5mu}
\setlength{\medmuskip}{2mu plus 1mu minus 2mu}
\setlength{\thickmuskip}{2.5mu plus 2.5mu}
\fontsize{9.5pt}{11.4pt}\selectfont
\allowdisplaybreaks
\eas{
&\,\, F (x_b^{(i)}) - F (x_b^{(i+1)}) 
\\
&= \sum_{b=1}^B F_b (x_b^{(i)}) - F_b (x_b^{(i+1)}) 
\\
&\geq \sum_{b=1}^B \frac{\lambda_b - 1}{4} \left( \nm{ x_b^{(i)} - x_b^{(i+1)} }_{M_b^{(i+1)}}^2 - \delta^2 \nm{ x_b^{(i-1)} - x_b^{(i)} }_{ M_b^{(i)}}^2 \right)
}
over $i = 0,\ldots, \mathrm{Iter}-1$, we obtain
\ea{
&\,\, F(x^{(0)}) - F(x^{(\mathrm{Iter}+1)}) \nn
\\
& \geq  \sum_{i=0}^{\mathrm{Iter}-1} \sum_{b=1}^B \frac{\lambda_b - 1}{4} \left( \nm{ x_b^{(i)} - x_b^{(i+1)} }_{M_b^{(i+1)}}^2 \right. \nn
\\
& \hspace{5.2em} \left. - \delta^2 \nm{ x_b^{(i-1)} - x_b^{(i)} }_{ M_b^{(i)}}^2 \right) \nn
\\
& \geq  \sum_{i=0}^{\mathrm{Iter}-1} \sum_{b=1}^B \frac{(\lambda_b - 1) (1 - \delta^2)}{4} \nm{ x_b^{(i)} - x_b^{(i+1)} }_{M_b^{(i+1)}}^2 \nn
\\
& \geq \min_{b\in[B]} \left\{ \frac{(\lambda_b - 1) m_{b} }{4} \right\} (1 - \delta^2) \sum_{i=0}^{\mathrm{Iter}-1} \nm{ x^{(i)} - x^{(i+1)} }_2^2 \label{eq:sqSum:mmin}
}
\endgroup
where the inequality \R{eq:sqSum:mmin} holds by Assumption 2.
Due to the lower boundedness of $F$ in Assumption 1 (i.e., $\inf_{x \in \dom(F)} F(x) > -\infty$), taking $\mathrm{Iter} \rightarrow \infty$ completes the proof.

\section{Proofs of Theorem~\ref{t:subseqConv}} \label{sec:prf:t:subseqConv}

The following proof extends that given in \cite[Thm.~1]{Xu&Yin:17JSC}.
Let $\bar{x}$ be a limit point of $\{ x^{(i+1)} : i \geq 0 \}$ and $\{ x^{(i_j+1)} \}$ be the subsequence converging to $\bar{x}$.
Using \R{eq:convg_to0}, $\{ x^{(i_j + \iota)} \}$ converges to $\bar{x}$ for any $\iota \geq 0$.
Note that, taking another subsequence if necessary, $M_b^{(i_j)}$ converges to some $\bar{M}_b$ as $j \rightarrow \infty$ for $b \in [B]$, since $M_b^{(i)}$ is bounded by Assumption 2.

We first observe that
\begingroup
\setlength{\thinmuskip}{1.5mu}
\setlength{\medmuskip}{2mu plus 1mu minus 2mu}
\setlength{\thickmuskip}{2.5mu plus 2.5mu}
\ea{
\label{eq:t:subseqConv:ij}
x_b^{(i_j+1)} = \argmin_{ x_b } &~ \ip{ \nabla_{x_b} f_b^{(i_j+1)} (\acute{x}_b^{(i_j+1)}) }{ x_b - \acute{x}_b^{(i_j+1)} } \nn
\\
& ~ + \frac{\lambda_b}{2} \nm{ x_b - \acute{x}_b^{(i_j+1)} }_{M_b^{(i_j+1)}}^2 + g_b (x_b),
}
\endgroup
for any $i_j$, since $\widetilde{M}_b^{(i+1)} = \lambda_b M_b^{(i+1)}$, $\forall i$. Since $f$ is continuously differentiable and $g_b$'s are lower semicontinuous, 
we have
\begingroup
\setlength{\thinmuskip}{1.5mu}
\setlength{\medmuskip}{2mu plus 1mu minus 2mu}
\setlength{\thickmuskip}{2.5mu plus 2.5mu}
\allowdisplaybreaks
\ea{
&\,\, g_b (\bar{x}_b) \nn
\\
& \leq \liminf_{j \rightarrow \infty} \bigg\{ \ip{ \nabla_{x_b} f_b^{(i_j+1)} (\acute{x}_b^{(i_j+1)}) }{ x_b^{(i_j+1)} - \acute{x}_b^{(i_j+1)} } \nn
\\
& \hspace{4.5em} + \frac{\lambda_b}{2} \nm{ x_b^{(i_j+1)} - \acute{x}_b^{(i_j+1)} }_{M_b^{(i_j+1)}}^2 + g_b (x_b^{(i_j+1)}) \bigg\} \nn
\\
& \leq \liminf_{j \rightarrow \infty} \bigg\{ \ip{ \nabla_{x_b} f_b^{(i_j+1)} (\acute{x}_b^{(i_j+1)}) }{ x_b - \acute{x}_b^{(i_j+1)} } \nn
\\
& \hspace{5.1em} + \frac{\lambda_b}{2} \nm{ x_b - \acute{x}_b^{(i_j+1)} }_{M_b^{(i_j+1)}}^2 + g_b (x_b) \bigg\} \nn
\\
& = \ip{ \nabla_{x_b} f_b (\bar{x}_b ) }{ x_b - \bar{x}_b } + \frac{\lambda_b}{2} \nm{ x_b - \bar{x}_b }_{\bar{M}_b}^2 + g_b (x_b), \nn
}
\endgroup
for all $x_b \in \dom(F)$, where the last equality holds by letting $j \rightarrow \infty$.
This result can be viewed by
\eas{
&\,\, \ip{ \nabla_{x_b} f_b (\bar{x}_b ) }{ \bar{x}_b - \bar{x}_b } + \frac{\lambda_b}{2} \nm{ \bar{x}_b - \bar{x}_b }_{\bar{M}_b}^2 + g_b (\bar{x}_b)
\\
& \leq \ip{ \nabla_{x_b} f_b (\bar{x}_b ) }{ x_b - \bar{x}_b } + \frac{\lambda_b}{2} \nm{ x_b - \bar{x}_b }_{\bar{M}_b}^2 + g_b (x_b), 
}
for all $x_b \in \dom(F)$.
Thus, we have
\bes{
\bar{x}_b  = \argmin_{x_b}  \ip{ \nabla_{x_b} f_b (\bar{x}_b ) }{ x_b - \bar{x}_b } +  \frac{\lambda_b}{2}  \nm{ x_b - \bar{x}_b }_{\bar{M}_b}^2 + g_b (x_b)
}
and $\bar{x}_b$ satisfies the first-order optimality condition:
\be{
\label{eq:t:subseqConv:1stOrd}
0 \in \nabla_{x_b} f(\bar{x}) + \partial g_b (\bar{x}_b).
}
Since \R{eq:t:subseqConv:1stOrd} holds for $b = 1,\ldots,B$, $\bar{x}$ is a critical point of \R{sys:multiConvx}. This completes the proof of the first result in Theorem~\ref{t:subseqConv}.

In addition, \R{eq:t:subseqConv:ij} implies
\eas{
&\,\, \ip{ \nabla_{x_b} f_b^{(i_j+1)} (\acute{x}_b^{(i_j+1)}) }{ x_b^{(i_j+1)} - \acute{x}_b^{(i_j+1)} } 
\\
&\, + \frac{\lambda_b}{2} \nm{ x_b^{(i_j+1)} - \acute{x}_b^{(i_j+1)} }_{M_b^{(i+1)}}^2 + g_b (x_b^{(i_j+1)}) 
\\
& \leq \ip{ \nabla_{x_b} f_b^{(i_j+1)} (\acute{x}_b^{(i_j+1)}) }{ \bar{x}_b - \acute{x}_b^{(i_j+1)} } 
\\
&\quad\, + \frac{\lambda_b}{2} \nm{ \bar{x}_b - \acute{x}_b^{(i_j+1)} }_{M_b^{(i+1)}}^2 + g_b (\bar{x}_b).
}
Applying limit superior to both sides of the above inequality over $j$ gives 
\be{
\label{eq:t:subseqConv:limsup}
\limsup_{j \rightarrow \infty}  g_b (x_b^{(i_j+1)}) \leq g_b (\bar{x}_b), \quad b = 1,\ldots,B.
}
Because $g_b$ is lower semi-continuous, 
\be{
\label{eq:t:subseqConv:liminf}
\liminf_{j \rightarrow \infty} g_b (x_b^{(i_j+1)}) \geq g_b (\bar{x}_b), \quad b = 1,\ldots,B.
}
Combining \R{eq:t:subseqConv:limsup} and \R{eq:t:subseqConv:liminf} gives
\bes{
\lim_{j \rightarrow \infty} g_b (x_b^{(i_j+1)})  = g_b (\bar{x}_b).
}
Considering the continuity of $f$ completes the proof of the second result in Theorem~\ref{t:subseqConv}.

For simplicity, our convergence analysis assumes a deterministically cyclic block update order. Similar to \cite{Xu&Yin:17JSC:supp}, one can extend our proofs in Sections~\ref{sec:prf:l:QM}--\ref{sec:prf:t:subseqConv} to the randomly shuffled update order (for each cycle).

\section{Summary of reBPEG-M} \label{sec:reBPGM-supp}

This section summarizes updates of reBPEG-M. See Algorithm~\ref{alg:reBPGM}.

\begin{algorithm}[H]
\caption{reBPEG-M: Restarting BPEG-M}
\label{alg:reBPGM}

\begin{algorithmic}
\REQUIRE $\{ x_b^{(0)} = x_b^{(-1)} : \forall b \}$, $\{ E_b^{(i)} \in [0, 1], \forall b,i \}$, $i=0$

\WHILE{a stopping criterion is not satisfied}

\FOR{$b = 1,\ldots,B$}

\begingroup
\setlength{\thinmuskip}{1.5mu}
\setlength{\medmuskip}{2mu plus 1mu minus 2mu}
\setlength{\thickmuskip}{2.5mu plus 2.5mu}
\fontsize{9.5pt}{11.4pt}\selectfont
\STATE Calculate  $M_b^{(i+1)}$, $\displaystyle \widetilde{M}_b^{(i+1)}$ by \R{update:Mtilde}, and $E_b^{(i+1)}$ by \R{update:Eb}
\STATE $\displaystyle \acute{x}_b^{(i+1)} =\, x_{b}^{(i)} + E_b^{(i+1)} \! \left( x_b^{(i)} - x_b^{(i-1)} \right)$
\STATE $\displaystyle x_b^{(i+1)} = \, \ldots$ \\ $\displaystyle \mathrm{Prox}_{g_b}^{\widetilde{M}_b^{(i+1)}} \!\! \bigg( \!\! \acute{x}_b^{(i+1)} \!-\! \Big( \! \widetilde{M}_b^{(i+1)} \! \Big)^{\!\!\!\!-1} \!\! \nabla f_b^{(i+1)} (\acute{x}_b^{(i+1)}) \!\! \bigg)$
\endgroup

\IF{restarting criterion \R{eq:restart:grad} is satisfied}

\begingroup
\setlength{\thinmuskip}{1.5mu}
\setlength{\medmuskip}{2mu plus 1mu minus 2mu}
\setlength{\thickmuskip}{2.5mu plus 2.5mu}
\fontsize{9.5pt}{11.4pt}\selectfont
\STATE $\displaystyle \acute{x}_b^{(i+1)} =\, x_{b}^{(i)}$
\STATE $\displaystyle x_b^{(i+1) = \, \ldots}$ \\ $\displaystyle \mathrm{Prox}_{g_b}^{\widetilde{M}_b^{(i+1)}} \!\! \bigg( \!\! \acute{x}_b^{(i+1)} \!-\! \Big( \! \widetilde{M}_b^{(i+1)} \! \Big)^{\!\!\!\!-1} \!\! \nabla f_b^{(i+1)} (\acute{x}_b^{(i+1)}) \!\! \bigg)$
\endgroup

\ENDIF

\STATE Update $\displaystyle e_b^{(i+1)}$ using \R{eq:mom_coeff}

\ENDFOR

\STATE $i = i+1$

\ENDWHILE

\end{algorithmic}
\end{algorithm}

\section{Proofs of Lemmas~\ref{l:MDdiag}--\ref{l:MDscaleI}} \label{sec:prf:l:MD}

We first introduce the following lemmas that are useful in designing majorization matrices for a wide class of (positive semidefinite) Hessian matrices:

\lem{[\!\!\!\protect{\cite[Lem.~S.3]{Chun&Fessler:18TIP}}]
\label{l:diag(|At|W|A|1)}
For a complex-valued matrix $A$ and a diagonal matrix $W$ with non-negative entries, $A^H W A  \preceq \diag( | A^H | W | A | 1 )$, where $| A |$ denotes the matrix consisting of the absolute values of the elements of $A$.
}

\lem{[\!\!\protect{\cite[Lem.~S.2]{Chun&Fessler:18TIP}}]
\label{l:diag(|A|1)}
For a complex-valued positive semidefinite Hermitian matrix $A$ (i.e., diagonal entries of a Hermitian matrix are nonnegative), $A \preceq \diag( | A | 1 )$.
}
The diagonal majorization matrix design in Lemma~\ref{l:MDdiag} is obtained by straightforwardly applying Lemma~\ref{l:diag(|At|W|A|1)}.
For the majorization matrix design in Lemma~\ref{l:MDscaleI}, we first observe that, for circular boundary condition, the Hessian $\sum_{l} \Psi_l^H \Psi_l$ in \R{sys:filter} is a (symmetric) Toeplitz matrix (for 2D, a block Toeplitz matrix with Toeplitz blocks).
Next, we approximate the Toeplitz matrix $\sum_{l} \Psi_l^H \Psi_l$ with a circulant matrix with a first row vector $\tilde{\psi}^H \in \bbC^R$ (similar to designing a preconditioner to a Toeplitz system):
\ea{
\label{eq:MDcirc}
\sum_{l=1}^L \Psi_l^H \Psi_l &\,\approx\, \mathrm{circ} \! \left( \tilde{\psi}^H \right), 
\\
\tilde{\psi} &:= \left[ \begin{array}{c} \left( \sum_{l=1}^L \ip{P_{B_1} \hat{x}_l}{P_{B_{1}} \hat{x}_l} \right)^{\!\!*} \\ \vdots \\ \left( \sum_{l=1}^L \ip{P_{B_1} \hat{x}_l}{P_{B_{R}} \hat{x}_l} \right)^{\!\!*} \end{array} \right], \nn
}
where $\mathrm{circ}(\cdot) \!:\! \bbC^n \!\rightarrow\! \bbC^{n \times n}$ constructs a circulant matrix from a row vector of size $n$.
Assuming that the circulant matrix $\mathrm{circ} ( \tilde{\psi}^H )$ in \R{eq:MDcirc} is positive definite (we observed that this holds for all the training datasets used in the paper) and using its circulant structure,
we design the scaled identity majorization matrix via Lemma~\ref{l:diag(|A|1)} as follows:
\bes{
M_D = \sum_{r=1}^R \left| \sum_{l=1}^L \ip{P_{B_1} \hat{x}_l}{P_{B_{r}} \hat{x}_l} \right| \cdot I_R.
}
This completes the proofs for Lemma~\ref{l:MDscaleI}.

\section{Proofs of Proposition~\ref{p:orth}} \label{sec:prf:p:orth}

The following proof is closely related to reduced rank Procrustes rotation \cite[Thm.~4]{Zou&Hastie&Tibshirani:06JCGS}; however, we shall pay 
careful attention to the feasibility of solution by considering the corresponding matrix dimensions.  
We rewrite the objective function of \R{p:eq:TF} by
\eas{
&\,\, \nm{ \widetilde{M}_D^{1/2} D  - \widetilde{M}_D^{1/2} \mathcal{V} }_{F}^2 
\\
&= \tr (D^H \widetilde{M}_D D) - 2\tr (D^H \widetilde{M}_D \mathcal{V}) + \tr(\mathcal{V}^T \widetilde{M}_D \mathcal{V})
\\
&= \frac{1}{R} \tr (\widetilde{M}_D) - 2\tr (D^H \widetilde{M}_D \mathcal{V}) + \tr(\mathcal{V}^T \widetilde{M}_D \mathcal{V}).
}
The second equality holds by the constraint $D D^H = \frac{1}{R} I$.
Then, we rewrite \R{p:eq:TF} as follows:
\ea{
\begin{split}
\label{p:prf:TFre}
\max_{D} ~ \tr (D^H \widetilde{M}_D \mathcal{V}), 
\quad \mathrm{subj.~to} ~ D D^H = \frac{1}{R} \cdot I.
\end{split}
}
Considering singular value decomposition (SVD) of $\widetilde{M}_D \mathcal{V} $, i.e., $\widetilde{M}_D \mathcal{V} = U \Lambda V^H$, observe that
\bes{
\tr (D^H \widetilde{M}_D \mathcal{V}) = \tr (D^H \widetilde{M}_D \mathcal{V}) = \tr ( \widetilde{D}^H U \Lambda )
}
where $\widetilde{D} = D V$. Because $V$ is unitary, we recast \R{p:prf:TFre}
\ea{
\begin{split}
\label{p:prf:TFre2}
\max_{\widetilde{D}} ~ \tr (\widetilde{D}^H U \Lambda), 
\quad \mathrm{subj.~to} ~ \widetilde{D} \widetilde{D}^H = \frac{1}{R} \cdot I.
\end{split}
}
Consider that $\Lambda \in \bbR^{R \times K}$ is (rectangular) diagonal, i.e., $\Lambda = [\widetilde{\Lambda}_{R \times R}, 0_{R \times (K-R)}]$ for $R \leq K$, in which $\widetilde{\Lambda}_{R \times R}$ is a ($R \!\times\! R$-sized) diagonal matrix with singular values.
Based on the structure of $\Lambda$, we rewrite $\tr (\widetilde{D}^H U \Lambda)$ in \R{p:prf:TFre2} as
\bes{
\tr (\widetilde{D}^H U \Lambda) = \sum_{r=1}^R (\widetilde{D}^H U)_{r,r} \widetilde{\Lambda}_{r,r},
}
Thus, \R{p:prf:TFre2} is maximized when the diagonals elements $( \widetilde{D}^H U )_{r,r}$'s are positive and maximized.
Under the constraint in \R{p:prf:TFre2}, the maximum is achieved by setting $\widetilde{D}^{\star} = \frac{1}{\sqrt{R}} U [I_R, 0_{R \times (K-R)}]$ for $R \leq K$. Combining this result with $\widetilde{D} = D V$ completes the proofs.

Note that the similar technique above in finding $\widetilde{D}^{\star}$ can be applied to the case of $R > K$; however, the constraint in \R{p:prf:TFre2} cannot be satisfied.
For $R > K$, observe that $\Lambda = [\widetilde{\Lambda}_{K \times K}, 0_{K \times (R-K)}]^T$, where $\widetilde{\Lambda}_{K \times K}$ is a diagonal matrix with singular values. 
With the similar reason above, $\widetilde{D}^{\star} = \frac{1}{\sqrt{R}} U [I_K, 0_{K \times (R-K)}]^T$ maximizes the cost function in \R{p:prf:TFre2}.
However, this solution does not satisfy the constraint $\widetilde{D} \widetilde{D}^H = \frac{1}{R} \cdot I$ in \R{p:prf:TFre2}.
On a side note, one cannot apply some tricks based on reduced SVD ($R > K$), because $U U^H = I$ does not hold.

\section{Accelerated Newton's Method To Solve \R{sys:prox:filter:div}} \label{sec:Newton}

The optimal solution to \R{sys:prox:filter2} can be obtained by the classical approach for solving a quadratically constrained quadratic program (see, for example, \cite[Ex.~4.22]{Boyd&Vandenberghe:book}):
\ea{
\label{eq:soln:prox_map:blk:FISTA:synthF_kernel}
d_k^{(i+1)} = &\, \left( G_k + \varphi_k I_R \right)^{-1} g_k, 
\\
G_k := &\, \widetilde{M}_D + \beta \Gamma_k, \nn
\\
g_k := &\, \widetilde{M}_D \nu_{k}^{(i)}
}
where the Lagrangian parameter is determined by $\varphi_k = \max \{ 0 , \varphi_k^{\star} \}$ and $\varphi_k^{\star}$ is the largest solution of the nonlinear equation $f (\varphi_k) = R^{-1}$, in which
\be{
\label{eq:secular1}
f (\varphi_k) := \nm{ \left( G_k + \varphi_k I_R \right)^{-1} g_k }_2^2,
}
for $k=1,\ldots,K$ (\R{eq:secular1} is the so-called \textit{secular} equation). 
More specifically, the algorithm goes as follows. 
First obtain $d_k^{(i+1)} =  G_k^{-1} g_k$ (note again that $G_k \succ 0$). If it satisfies the unit norm equality constraint in \R{sys:prox:filter:div}, it is optimal.
Otherwise, one can obtain the optimal solution $d_k^{(i+1)}$ through \R{eq:soln:prox_map:blk:FISTA:synthF_kernel} with the Lagrangian parameter $\varphi_k = \varphi_k^{\star}$, where $\varphi_k^{\star}$ is optimized by solving the secular equation $f (\varphi_k) = R^{-1}$ and $f (\varphi_k)$ is given as \R{eq:secular1}.
To solve $f (\varphi_k) = R^{-1}$, we first rewrite \R{eq:secular1} by
\be{
\label{eq:secular2}
f (\varphi_k) = \sum_{r=1}^R \frac{ | \tilde{g}_k |_{r}^2 }{ \left(  \varphi_k + ( \sigma_k )_{r} \right)^2 }.
}
where $\{ \tilde{g}_k = Q_k^{H} g_k : k = 1,\ldots,K \}$, $\{ G_k = Q_k \Sigma_k Q_k^{H} : k=1,\ldots,K \}$, $\{ ( \sigma_k )_{1} \geq \cdots \geq ( \sigma_k )_{R} > 0 \}$ is a set of eigenvalues of $G_k$ for $k=1,\ldots,K$ (note that $G_k + \varphi_k I_R \succ 0$ because $G_k \succ 0$).
To simplify the discussion, we assume that $\{ ( g_k )_{r} \neq 0 : k=1,\ldots,K, r=1,\ldots,R \}$ \cite{Elden:02BIT:supp}.
Noting that, for $\varphi_k > -( \sigma_k )_{R}$, $f (\varphi_k)$ monotonically decreases to zero as $\varphi_k \rightarrow \infty$), the nonlinear equation $f (\varphi_k) = R^{-1}$ has exactly one nonnegative solution $\varphi_k^{\star}$.  
The optimal solution $\varphi_k^\star$ can be determined by using the classical Newton's method. 
We apply the accelerated Newton's method in \cite{Reinsch:71NM:supp, Chun&Fessler:18TIP:supp} that solves $1/f (\varphi_k) = R$:
\be{
\label{eq:acc_newton}
\varphi_k^{(\iota+1)} = \varphi_k^{(\iota)} - 2 \frac{f (\varphi_k^{(\iota)})}{f' (\varphi_k^{(\iota)})} \left( \sqrt{f (\varphi_k^{(\iota)})}  - 1 \right)
}
where $f(\varphi_k)$ is given as \R{eq:secular2},
\bes{
f'(\varphi_k) =  -2 \sum_{r=1}^R \frac{ | \tilde{g}_k |_{r}^2 }{ \left( \varphi_k + ( \sigma_k )_{r} \right)^3 },
}
and $\varphi_k^{(0)} = -( \sigma_k )_{R} + 10^{-10}$. Note that \R{eq:acc_newton} approaches the optimal solution $\varphi_k^\star$ faster than the classical Newton's method.

\begin{figure}[t!]
\small\addtolength{\tabcolsep}{-4pt}
\renewcommand{\arraystretch}{1.1}
\centering

\begin{tabular}{cc}

\vspace{-0.1em}\includegraphics[scale=0.47, trim=3.3em 5.6em 0em 4.5em, clip]{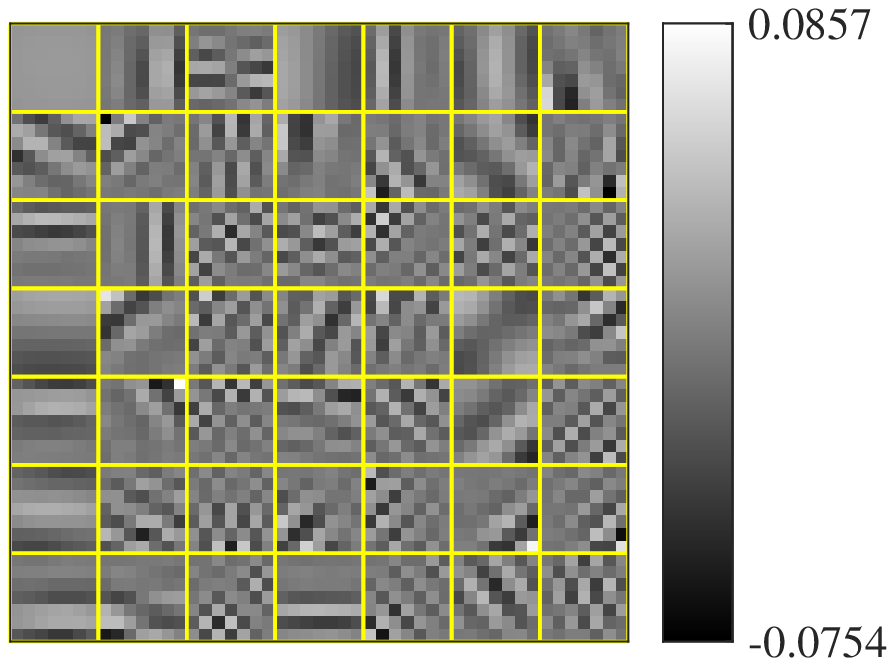}  &
\vspace{-0.1em}\includegraphics[scale=0.47, trim=3.3em 5.6em 0em 4.5em, clip]{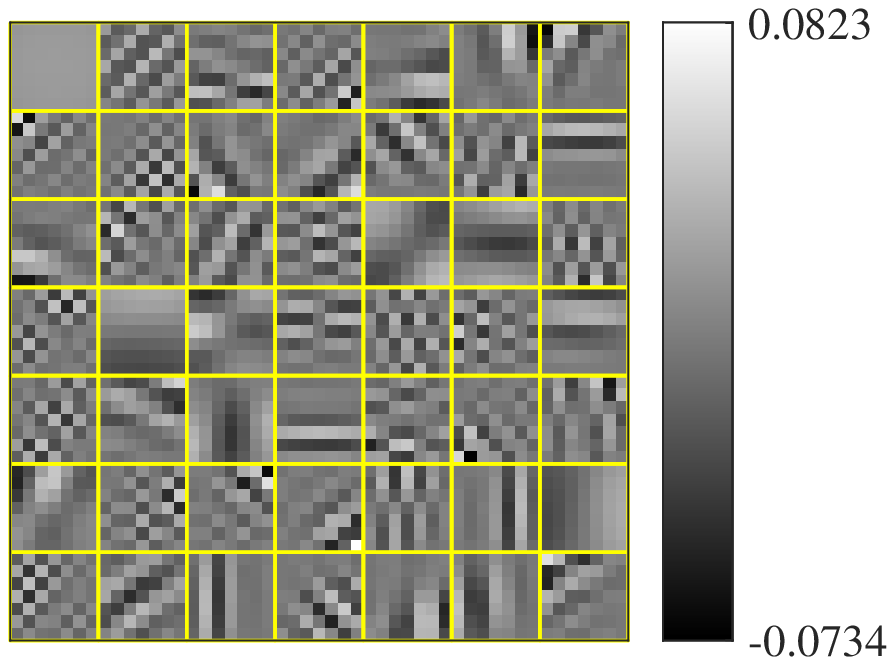}  \\

{\small (a1) Deterministic $\{ d_k^{(0)} \}$} & {\small (a2) Random $\{ d_k^{(0)} \}$} \\ 

\multicolumn{2}{c}{\small (a) The fruit dataset ($L = 10$, $N = 100 \!\times\! 100$)} \\

\vspace{-0.1em}\includegraphics[scale=0.47, trim=3.3em 5.6em 0em 4.5em, clip]{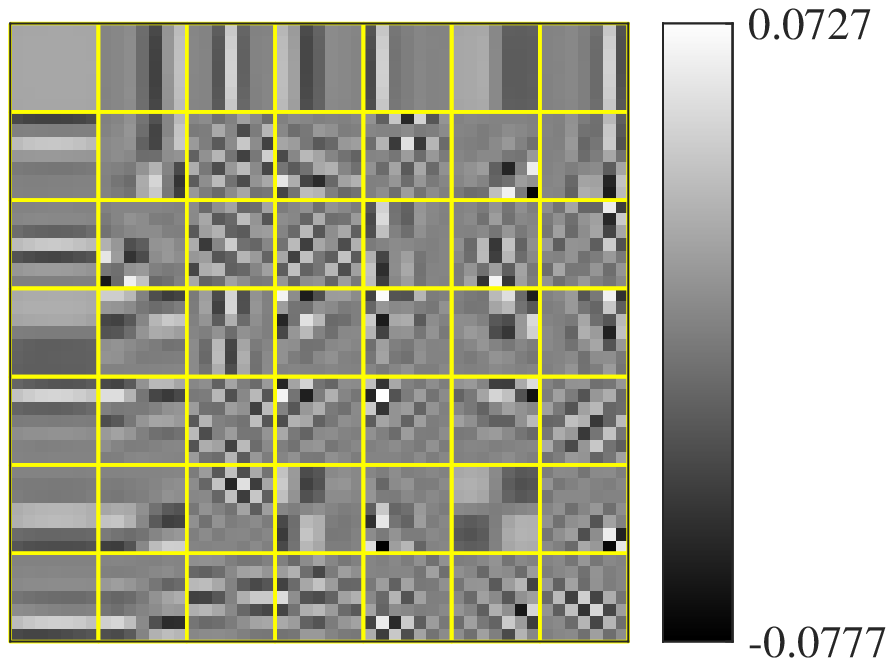}  &
\vspace{-0.1em}\includegraphics[scale=0.47, trim=3.3em 5.6em 0em 4.5em, clip]{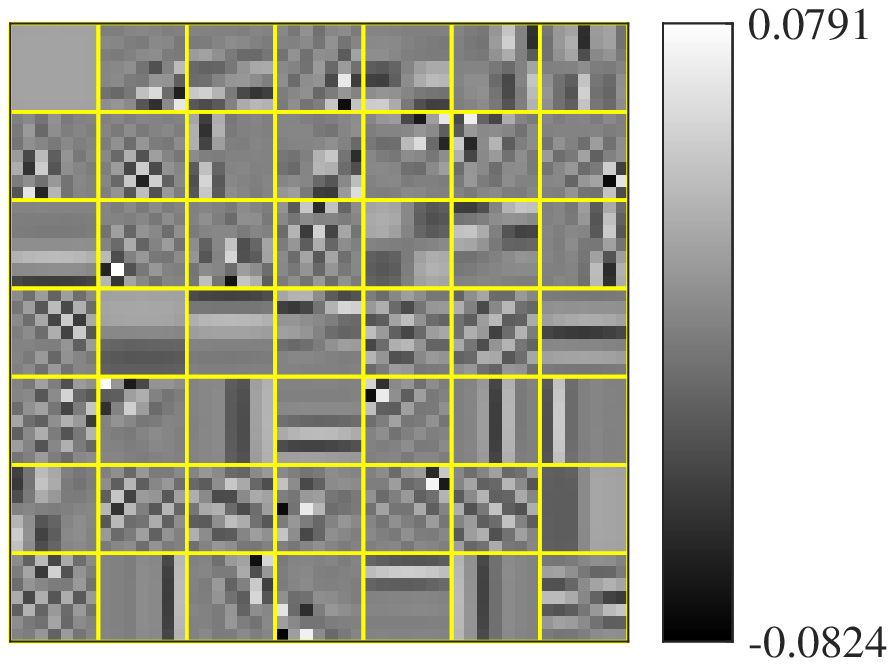}  \\

{\small (b1) Deterministic $\{ d_k^{(0)} \}$} & {\small (b2) Random $\{ d_k^{(0)} \}$} \\ 

\multicolumn{2}{c}{\small (b) The city dataset ($L = 10$, $N = 100 \!\times\! 100$)} 

\end{tabular}

\vspace{-0.25em}
\caption{Examples of learned filters via CAOL \R{sys:CAOL:orth} with different filter initialization, from different datasets (Proposition~\ref{p:MDexhess} was used for $M_D$; $R \!=\! K \!=\! 49$, $\alpha \!=\! 2.5 \!\times\! 10^{-4}$, and circular boundary condition).}
\label{fig:filters_BPGM}
\end{figure}

\begin{figure}[t!]
\small\addtolength{\tabcolsep}{-4pt}
\centering

\begin{tabular}{cc}

\vspace{-0.1em}\includegraphics[scale=0.47, trim=3.3em 5.6em 0em 4.5em, clip]{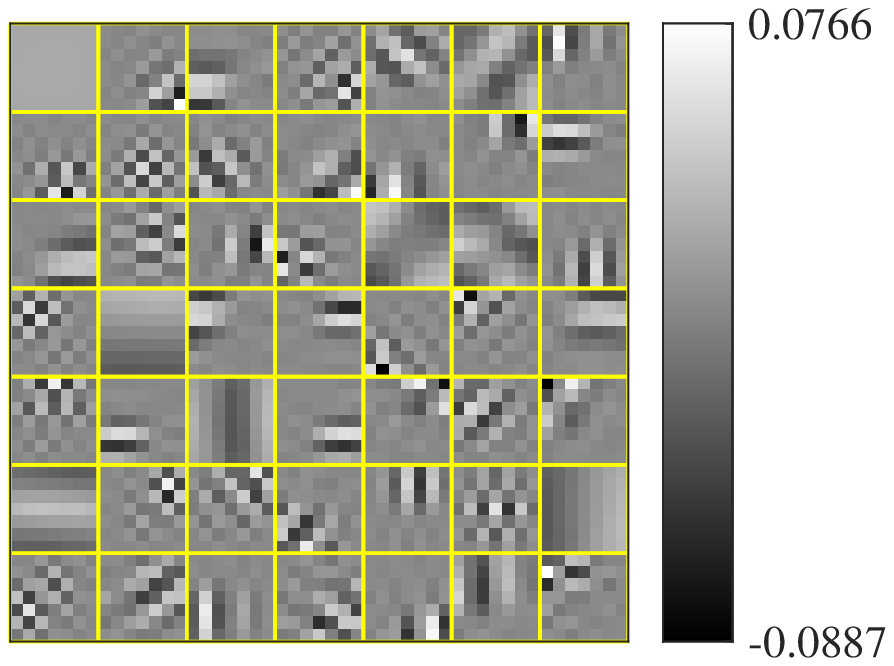} &
\vspace{-0.1em}\includegraphics[scale=0.47, trim=3.3em 5.6em 0em 4.5em, clip]{./Fig/filt_randD_Mexhess_fruit.eps} \\

{\small (a1) $\alpha = 2.5 \!\times\! 10^{-5}$} & {\small (a2) $\alpha = 2.5 \!\times\! 10^{-4}$} \\ 

\multicolumn{2}{c}{\small (a) The fruit dataset ($L = 10$, $N = 100 \!\times\! 100$)} \\

\vspace{-0.1em}\includegraphics[scale=0.47, trim=3.3em 5.6em 0em 4.5em, clip]{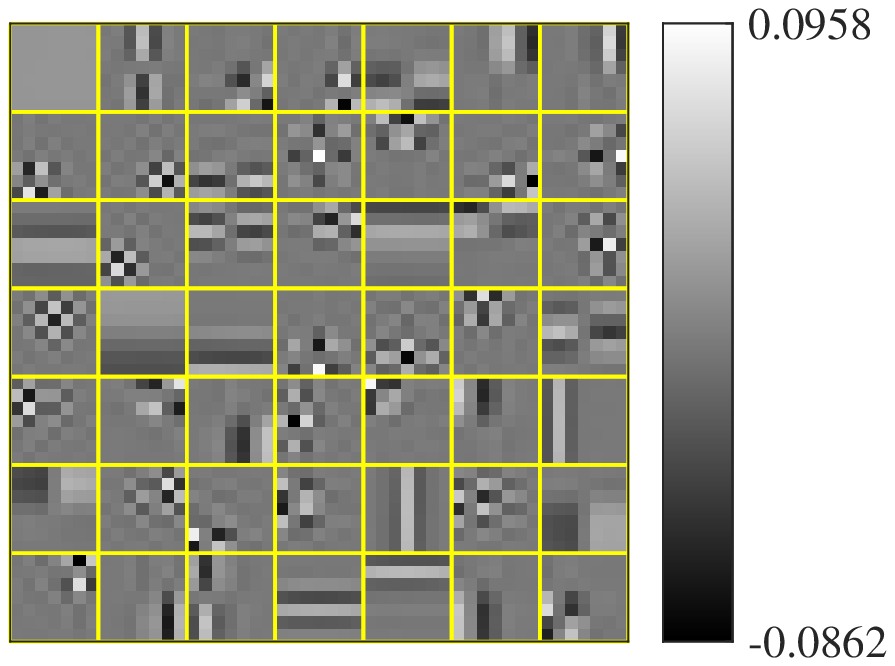} &
\vspace{-0.1em}\includegraphics[scale=0.47, trim=3.3em 5.6em 0em 4.5em, clip]{./Fig/filt_randD_Mexhess_city.eps} \\

{\small (b1) $\alpha = 2.5 \!\times\! 10^{-5}$} & {\small (b2) $\alpha = 2.5 \!\times\! 10^{-4}$} \\ 

\multicolumn{2}{c}{\small (b) The city dataset ($L = 10$, $N = 100 \!\times\! 100$)} \\

\vspace{-0.1em}\includegraphics[scale=0.47, trim=2em 4.8em 0em 4em, clip]{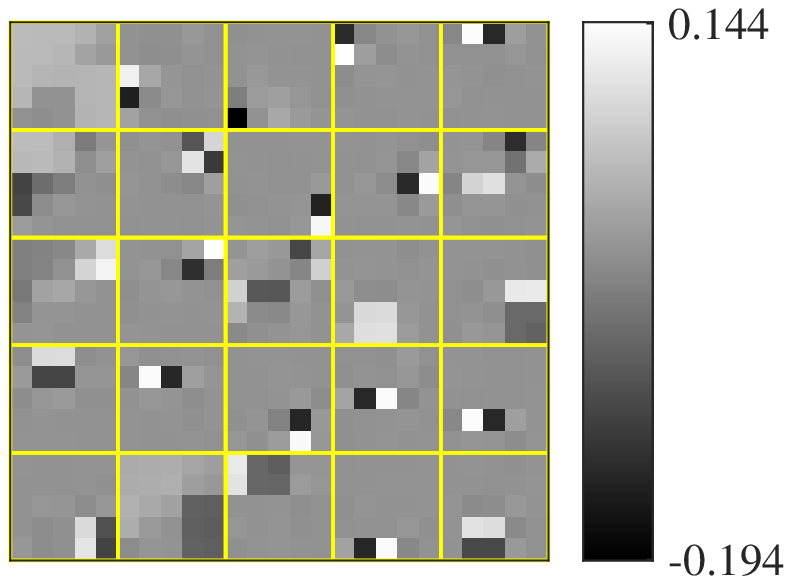} & 
\vspace{-0.1em}\includegraphics[scale=0.47, trim=2em 4.8em 0em 4em, clip]{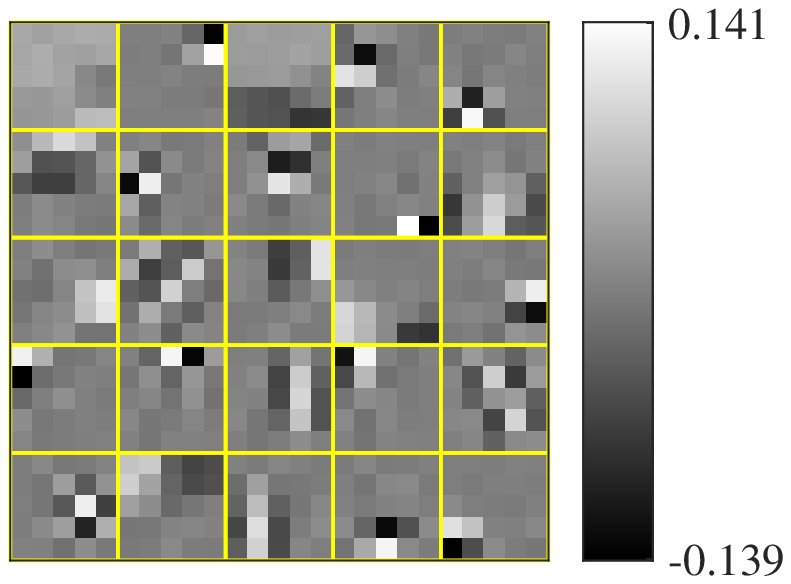} \\

{\small (c1) $\alpha = 10^{-4}$} & {\small (c2) $\alpha = 2 \!\times\! 10^{-4}$} \\ 

\multicolumn{2}{c}{\small (c) The CT-(\romnum{2}) dataset ($L = 10$, $N = 512 \!\times\! 512$)} 

\end{tabular}

\vspace{-0.25em}
\caption{Examples of learned filters via CAOL \R{sys:CAOL:orth} with different datasets and regularization parameters (Proposition~\ref{p:MDexhess} was used for $M_D$; $R \!=\! K \!=\! 49$ for the fruit and city datasets, and $R \!=\! K \!=\! 25$ for the CT-(\romnum{2}) dataset; circular boundary condition). We observed that the learned filters in (c2) give higher signal recovery accuracy than those in (c1) for CT MBIR \R{sys:CT&CAOL}. This implies that the diverse features captured in (c2) are useful to improve the performance of the proposed MBIR model \R{sys:CT&CAOL}.}
\label{fig:CAOL:filters:fruit&city&ct-ii}
\end{figure}

\section{Supplementary Results} \label{sec:result:supp}

\begin{figure}[t]
\centering

\begin{tabular}{c}
\vspace{-0.2em}\includegraphics[scale=0.58, trim=0 0.2em 2em 1.2em, clip]{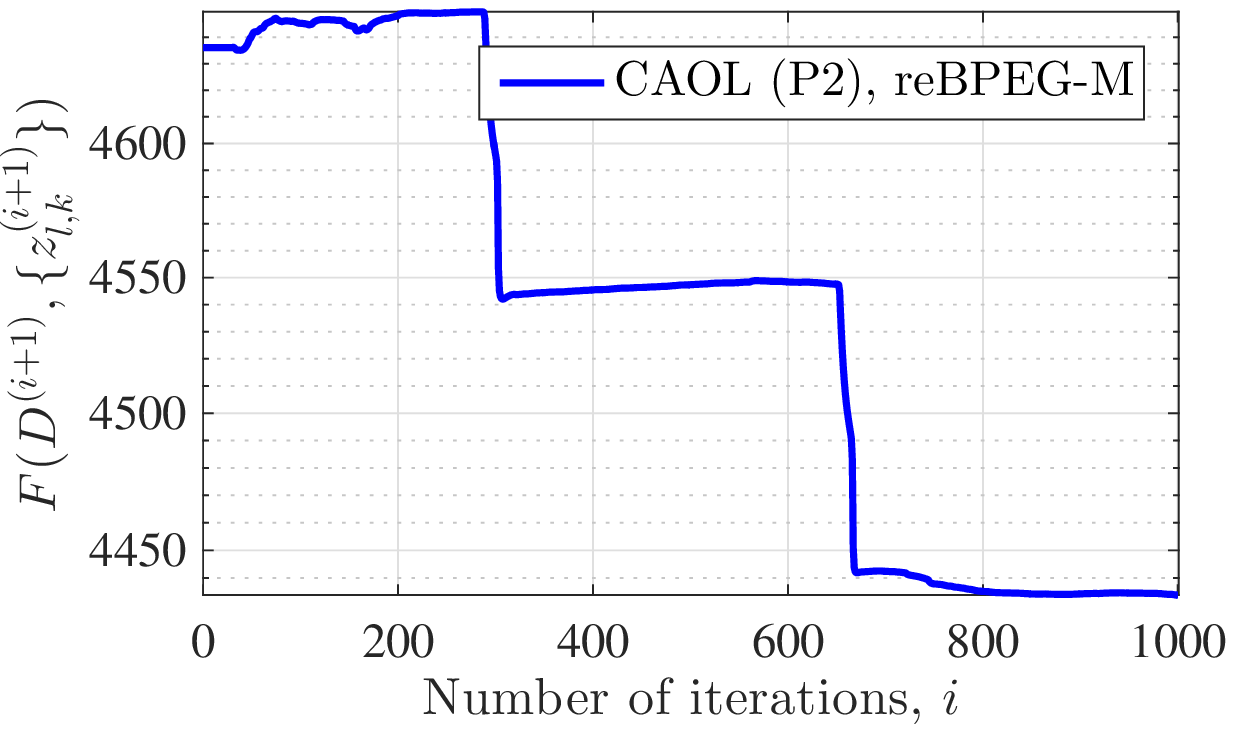} \\
\small{(a) $\alpha = 10^{-4}$, $\beta = 5 \!\times\! 10^{6}$} 
\\
\vspace{-0.2em}\includegraphics[scale=0.58, trim=0 0.2em 2em 1.2em, clip]{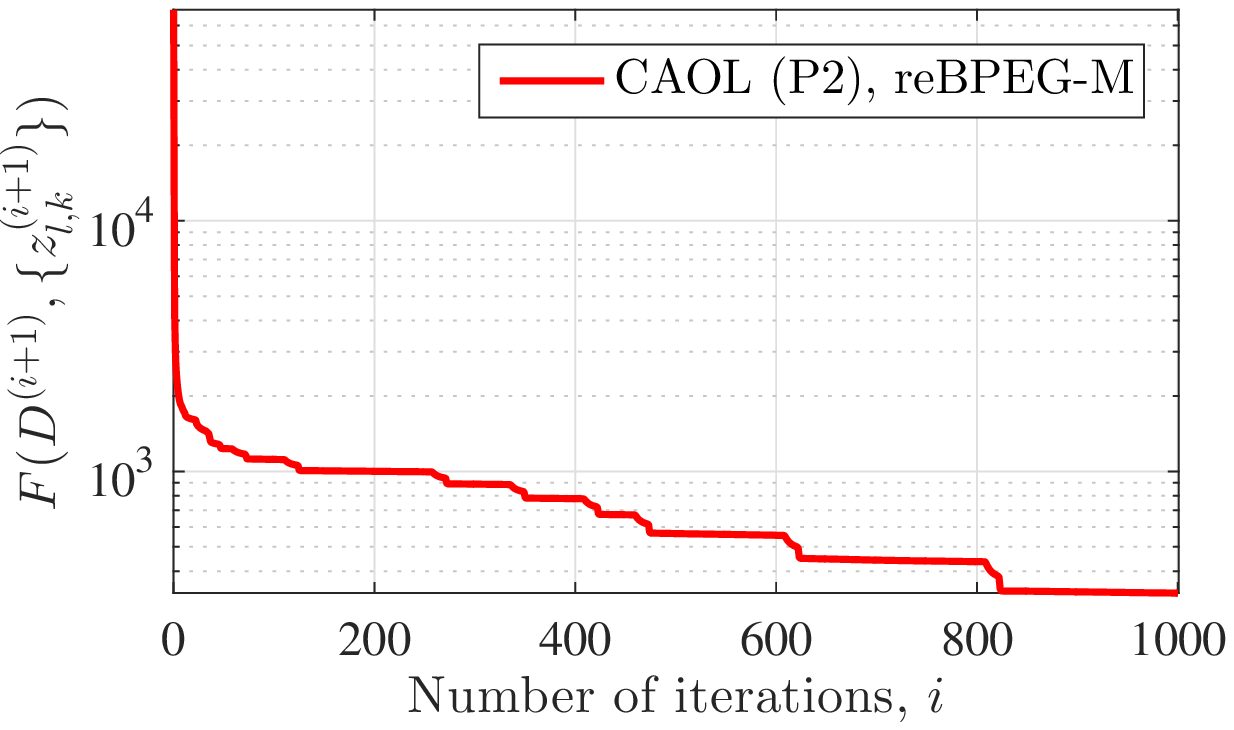} \\
\small{(b) $\alpha = 10^{-4}$, $\beta = 5 \!\times\! 10^{4}$}
\end{tabular}

\vspace{-0.25em}
\caption{Cost minimization in CAOL \R{sys:CAOL:div} with different diversity promoting regularization parameters ($R \!=\! 25$ and $K \!=\! 20$; Proposition~\ref{p:MDexhess} was used for $M_D$; the CT-(\romnum{1}) dataset). CAOL \R{sys:CAOL:div} can consider the case $R \!>\! K$ -- noting that CAOL \R{sys:CAOL:orth} only considers the case of $R \leq K$ for the efficient solution in Proposition~\ref{p:orth} -- and BPEG-M stably minimizes the corresponding cost function.}
\label{fig:CAOL:CT(ii):obj}
\end{figure}

\begin{figure}[t]
\centering
\small\addtolength{\tabcolsep}{-7.5pt}
\renewcommand{\arraystretch}{0.1}

\begin{tabular}{ccc}

	&
	\specialcell[c]{(a) Trained filters via \\ patch-domain AOL \cite{Ravishankar&Bressler:15TSP:supp} \\ ($R \!=\! K \!=\! 64$)}
	&
	\specialcell[c]{(b) Trained filters \\ via CAOL \R{sys:CAOL:orth} \\ ($R \!=\! K \!=\! 49$)}		
	\\
	
	\raisebox{-0.5\height}{\begin{turn}{+90}\small{(\romnum{1})~Trained filters} \end{turn}}~ &
        \raisebox{-0.5\height}{
        \includegraphics[trim={2em 5.2em 7em 4em},clip,scale=0.62]{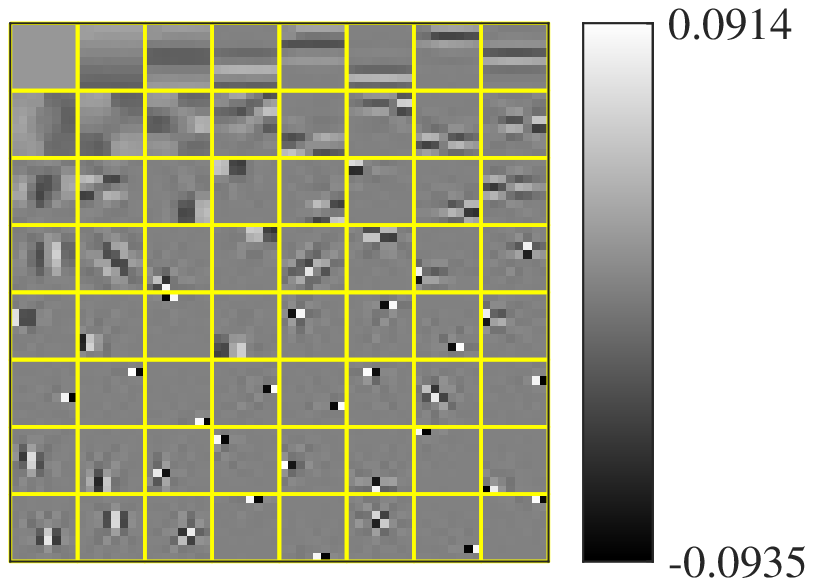}
        } 
        &
        \raisebox{-0.5\height}{
        \includegraphics[trim={2em 5.2em 7em 4em},clip,scale=0.5]{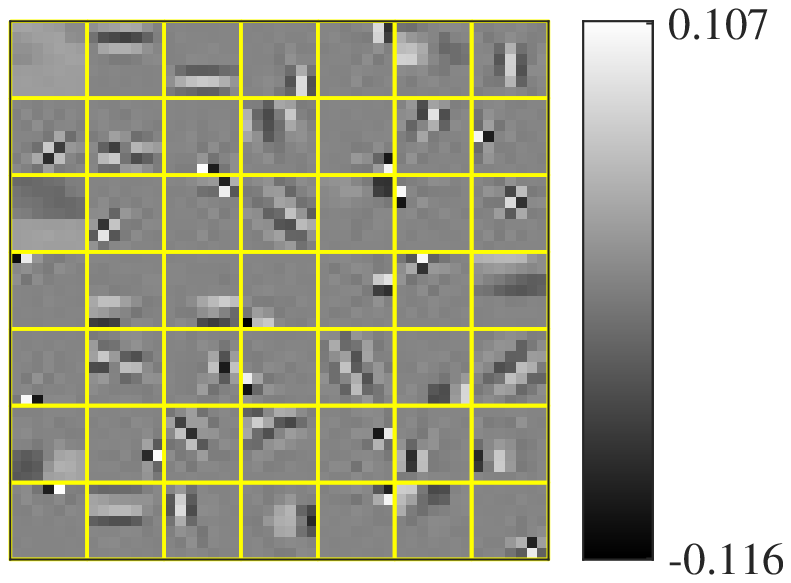}
        }
	\\
	
	\raisebox{-0.5\height}{\begin{turn}{+90} \small{(\romnum{2})~Reconstructed images} \end{turn}}~ &
	\raisebox{-0.5\height}{
        \begin{tikzpicture}
        		\begin{scope}[spy using outlines={rectangle,yellow,magnification=1.25,size=15mm,connect spies}]
       			\node {\includegraphics[viewport=35 105 275 345, clip, width=3.4cm,height=3.4cm]{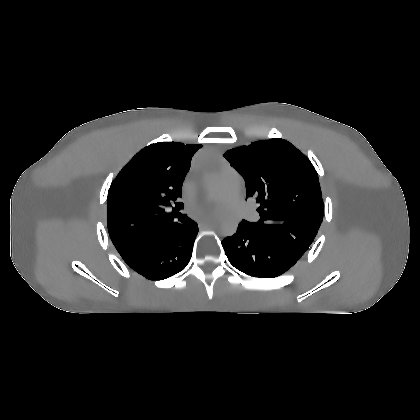}};
			\spy on (0.8,-0.95) in node [left] at (-0.3,-2.2);
                		\draw [->, thick, red] (0.1,0.85+0.3) -- (0.1,0.85);
                		\draw [->, thick, red] (-0.65+0.3,0.1) -- (-0.65,0.1);
                		\draw [->, thick, red] (-1.1,-1.3) -- (-1.1+0.3,-1.3);  
			\draw [red] (-1.35,0.15) circle [radius=0.28];
			\draw [red] (-1.35,-0.45) circle [radius=0.28];
			\node [black] at (0.7,-1.95) {\small $\mathrm{RMSE} = 35.1$};
		\end{scope}
        \end{tikzpicture} 
        }  
        &
        \raisebox{-0.5\height}{
        \begin{tikzpicture}
        		\begin{scope}[spy using outlines={rectangle,yellow,magnification=1.25,size=15mm,connect spies}]
       			\node {\includegraphics[viewport=35 105 275 345, clip, width=3.4cm,height=3.4cm]{Fig_CTrecon_v1/x123_caol7x7_old.png}};
			\spy on (0.8,-0.95) in node [left] at (-0.3,-2.2);
                		\draw [->, thick, red] (0.1,0.85+0.3) -- (0.1,0.85);
                		\draw [->, thick, red] (-0.65+0.3,0.1) -- (-0.65,0.1);
                		\draw [->, thick, red] (-1.1,-1.3) -- (-1.1+0.3,-1.3);  
			\draw [red] (-1.35,0.15) circle [radius=0.28];
			\draw [red] (-1.35,-0.45) circle [radius=0.28];
			\node [red] at (0.7,-1.95) {\small $\mathrm{RMSE} = 34.7$};
		\end{scope}
        \end{tikzpicture}   
        }
        \\

	\raisebox{-0.5\height}{\begin{turn}{+90}\small{(\romnum{3})~Error maps} \end{turn}}~ &
	\raisebox{-0.5\height}{
        \begin{tikzpicture}
        		\begin{scope}[spy using outlines={rectangle,yellow,magnification=1.25,size=15mm,connect spies}]
       			\node {\includegraphics[viewport=35 105 275 345, clip, width=3.4cm,height=3.4cm]{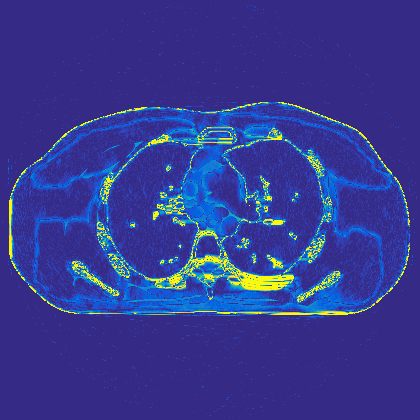}};
			\spy on (0.8,-0.95) in node [left] at (-0.3,-2.2);
                		\draw [->, thick, red] (0.1,0.85+0.3) -- (0.1,0.85);
                		\draw [->, thick, red] (-0.65+0.3,0.1) -- (-0.65,0.1);
                		\draw [->, thick, red] (-1.1,-1.3) -- (-1.1+0.3,-1.3);  
			\draw [red] (-1.35,0.15) circle [radius=0.28];
			\draw [red] (-1.35,-0.45) circle [radius=0.28];
			\node [black] at (0.7,-1.95) {\small $\mathrm{RMSE} = 35.1$};
		\end{scope}
        \end{tikzpicture}   
        }
        &
        \raisebox{-0.5\height}{
        \begin{tikzpicture}
        		\begin{scope}[spy using outlines={rectangle,yellow,magnification=1.25,size=15mm,connect spies}]
       			\node {\includegraphics[viewport=35 105 275 345, clip, width=3.4cm,height=3.4cm]{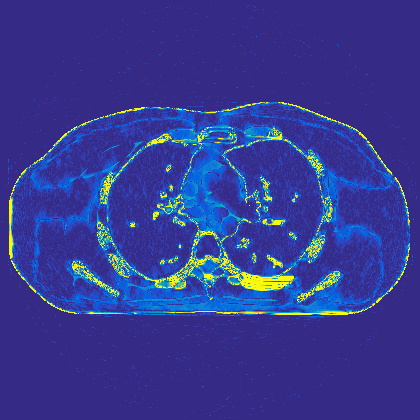}};
			\spy on (0.8,-0.95) in node [left] at (-0.3,-2.2);
                		\draw [->, thick, red] (0.1,0.85+0.3) -- (0.1,0.85);
                		\draw [->, thick, red] (-0.65+0.3,0.1) -- (-0.65,0.1);
                		\draw [->, thick, red] (-1.1,-1.3) -- (-1.1+0.3,-1.3);  
			\draw [red] (-1.35,0.15) circle [radius=0.28];
			\draw [red] (-1.35,-0.45) circle [radius=0.28];
			\node [red] at (0.7,-1.95) {\small $\mathrm{RMSE} = 34.7$};
		\end{scope}
        \end{tikzpicture}   
        }

\end{tabular}

\vspace{-0.25em}
\caption{Performance comparisons between MBIR models \R{sys:CT&CAOL} of which
use filters trained via the patch-domain AOL method \cite{Ravishankar&Bressler:15TSP:supp} and 
CAOL \R{sys:CAOL:orth} in sparse-view CT
($123$ views ($12.5$\% sampling); display window is $[0, 100]$ HU).
}
\label{fig:CTrecon:TL}
\end{figure}

\begin{figure*}[t!]
\centering
\small\addtolength{\tabcolsep}{-6.5pt}
\renewcommand{\arraystretch}{0.1}

    \begin{tabular}{cccc}

       \small{(a) Filtered back-projection} 
       & 
       \small{(b) EP} 
       & 
       \specialcell[c]{\small (c) Proposed MBIR \mbox{\R{sys:CT&CAOL}}, \\ \small with \R{eq:autoenc} of $R \!=\! K \!=\! 25$} 
       & 
       \specialcell[c]{\small (d) Proposed MBIR \mbox{\R{sys:CT&CAOL}}, \\ \small with \R{eq:autoenc} of $R \!=\! K \!=\! 49$} 
       \\
       
        \begin{tikzpicture}
        		\begin{scope}[spy using outlines={rectangle,yellow,magnification=1.25,size=15mm,connect spies}]
       			\node {\includegraphics[viewport=35 105 275 345, clip, width=3.4cm,height=3.4cm]{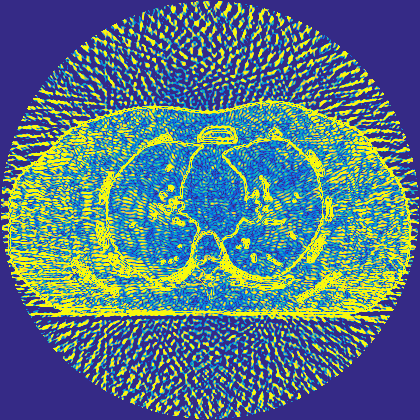}};
			\spy on (0.8,-0.95) in node [left] at (-0.3,-2.2);
			\node [black] at (0.7,-1.95) {\small $\mathrm{RMSE} = 82.8$};
		\end{scope}
        \end{tikzpicture}   
        &
        \begin{tikzpicture}
        		\begin{scope}[spy using outlines={rectangle,yellow,magnification=1.25,size=15mm,connect spies}]
       			\node {\includegraphics[viewport=35 105 275 345, clip, width=3.4cm,height=3.4cm]{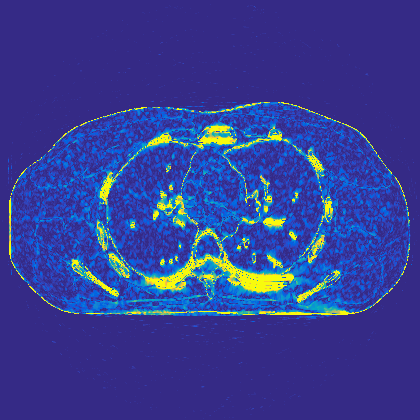}};
			\spy on (0.8,-0.95) in node [left] at (-0.3,-2.2);
                		\draw [->, thick, red] (0.1,0.85+0.3) -- (0.1,0.85);
                		\draw [->, thick, red] (-0.65+0.3,0.1) -- (-0.65,0.1);
                		\draw [->, thick, red] (-1.1,-1.3) -- (-1.1+0.3,-1.3);  
			\node [black] at (0.7,-1.95) {\small $\mathrm{RMSE} = 40.8$};
		\end{scope}
        \end{tikzpicture}   
        &
        \begin{tikzpicture}
        		\begin{scope}[spy using outlines={rectangle,yellow,magnification=1.25,size=15mm,connect spies}]
       			\node {\includegraphics[viewport=35 105 275 345, clip, width=3.4cm,height=3.4cm]{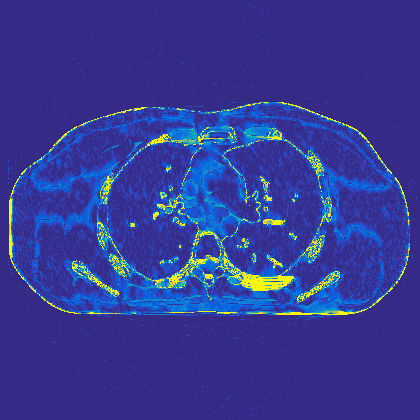}};
			\spy on (0.8,-0.95) in node [left] at (-0.3,-2.2);
                		\draw [->, thick, red] (0.1,0.85+0.3) -- (0.1,0.85);
                		\draw [->, thick, red] (-0.65+0.3,0.1) -- (-0.65,0.1);
                		\draw [->, thick, red] (-1.1,-1.3) -- (-1.1+0.3,-1.3);  
			\draw [red] (-1.35,0.15) circle [radius=0.28];
			\draw [red] (-1.35,-0.45) circle [radius=0.28];
			\node [blue] at (0.7,-1.95) {\small $\mathrm{RMSE} = 35.2$};
		\end{scope}
        \end{tikzpicture}   
        &
        \begin{tikzpicture}
        		\begin{scope}[spy using outlines={rectangle,yellow,magnification=1.25,size=15mm,connect spies}]
       			\node {\includegraphics[viewport=35 105 275 345, clip, width=3.4cm,height=3.4cm]{Fig_CTrecon_v1/err_caol7x7.png}};
			\spy on (0.8,-0.95) in node [left] at (-0.3,-2.2);
                		\draw [->, thick, red] (0.1,0.85+0.3) -- (0.1,0.85);
                		\draw [->, thick, red] (-0.65+0.3,0.1) -- (-0.65,0.1);
                		\draw [->, thick, red] (-1.1,-1.3) -- (-1.1+0.3,-1.3);  
			\draw [red] (-1.35,0.15) circle [radius=0.28];
			\draw [red] (-1.35,-0.45) circle [radius=0.28];
			\node [red] at (0.7,-1.95) {\small $\mathrm{RMSE} = 34.7$};
		\end{scope}
        \end{tikzpicture}   
       
    \end{tabular}

 \vspace{-0.25em}
\caption{
Error map comparisons of reconstructed images from different reconstruction methods for sparse-view CT ($123$ views ($12.5$\% sampling); for the MBIR model \R{sys:CT&CAOL}, convolutional regularizers were trained by CAOL \R{sys:CAOL:orth} -- see \cite[Fig.~2]{Chun&Fessler:18Asilomar}; display window is $[0, 100]$ HU) \cite{Chun&Fessler:18Asilomar:supp}.  
The MBIR model \R{sys:CT&CAOL} using convolutional sparsifying regularizers trained via CAOL \R{sys:CAOL:orth} shows higher image reconstruction accuracy compared to the EP reconstruction; see red arrows and magnified areas. 
For the MBIR model \R{sys:CT&CAOL}, the autoencoder (see Remark~\ref{r:autoenc}) using the filter dimension $R \!=\! K \!=\! 49$ improves reconstruction accuracy of that using $R \!=\! K \!=\! 25$; compare the results in (c) and (d).
In particular, the larger dimensional filters improve the edge sharpness of reconstructed images; see circled areas.
}
\label{fig:CTrecon:err}
\end{figure*}

This section provides additional results to support several arguments in the main manuscript.
Examples of additional results include Figs.~\ref{fig:filters_BPGM}, 
\ref{fig:CAOL:filters:fruit&city&ct-ii}, 
\ref{fig:CAOL:CT(ii):obj},
\ref{fig:CTrecon:TL},
and \ref{fig:CTrecon:err}.

We compare sparse-view CT reconstruction performances between MBIR models \R{sys:CT&CAOL} 
that use filters trained via the patch-domain AOL \cite{Ravishankar&Bressler:15TSP:supp} and CAOL \R{sys:CAOL:orth}:

\begin{itemize}
\item The filters $\{ d_k^\star \in \bbR^{49} \!:\! k \!\in\! [49] \}$ trained via CAOL \R{sys:CAOL:orth} 
and filters $\{ w_k^\star \!\in\! \bbR^{64} \!:\! k \!\in\! [64] \}$ trained via the patch-domain AOL method \cite{Ravishankar&Bressler:15TSP:supp} 
provided similar reconstruction quality in RMSE values, when applied to the MBIR model \R{sys:CT&CAOL}.\footnote{
The filters $\{ w_k^\star \!\in\! \bbR^{64} \!:\! k \!\in\! [64] \}$ 
trained via the patch-domain AOL method \cite{Ravishankar&Bressler:15TSP:supp} 
achieved state-of-the-art performance for CT MBIR optimization; see, e.g., \cite{Zheng&etal:19TCI:supp}.
In running the BPEG-M algorithm for MBIR \R{sys:CT&CAOL} using $\{ w_k^\star \!:\! k \!\in\! [64] \}$,
we normalized them to satisfy $\max_{k\in[64]} \| w_k^\star \|_2^2 = 1/64$ 
(indeed, they became $\| w_k^\star \|_2^2 \!\approx\! 1/64$, $\forall k$), and 
selected the thresholding parameter $\alpha'$ as $0.4 \cdot ( 2 \!\times\! 10^{-10} )$ 
by considering the energy of the filter,
where we chose $\alpha'$ as $2 \!\times\! 10^{-10}$ and $0.5 \cdot (2 \!\times\! 10^{-10})$
for the filters $\{ d_k^\star \in \bbR^{25} \!:\! k \!\in\! [25] \}$ and $\{ d_k^\star \in \bbR^{49} \!:\! k \!\in\! [49] \}$ 
trained via CAOL \R{sys:CAOL:orth},
respectively \cite[Sec.~\Romnum{5}-A]{Chun&Fessler:18Asilomar}.
}
See RMSE values in Figs.~\ref{fig:CTrecon:TL}(\romnum{2})--(\romnum{3}).
However, even with larger parameter dimensions, $\{ w_k^\star \!:\! k \!\in\! [64] \}$
gave more blurry edges in some soft tissue and bone areas, 
compared to $\{ d_k^\star \!:\! k \!\in\! [49] \}$. 
See red-circled areas and yellow-magnified areas in Fig.~\ref{fig:CTrecon:TL}(\romnum{2}).
In particular, 
Fig.~\ref{fig:CTrecon:TL}(\romnum{1}) shows that
$\{ d_k^\star \!:\! k \!\in\! [49] \}$
are more diverse and less redundant compared to $\{ w_k^\star \!:\! k \!\in\! [64] \}$,
and this implies that learning diverse (i.e., incoherent) filters
is important in improving signal recovery quality in MBIR using learned convolutional regularizers. 

\item At each BPEG-M iteration, 
MBIR \R{sys:CT&CAOL} using $\{ w_k^\star \!\in\! \bbR^{64} \!:\! k \!\in\! [64] \}$ 
-- trained via patch-domain AOL \cite{Ravishankar&Bressler:15TSP:supp}  --
uses more computations compared to MBIR \R{sys:CT&CAOL} using $\{ d_k^\star \!\in\! \bbR^{49} \!:\! k \!\in\! [49] \}$
-- trained via CAOL \R{sys:CAOL:orth}.
In particular, the former uses a $O(64^2 \cdot N')$-involved convolution operator \emph{three} times per BPEG-M iteration;
the latter uses a $O(49^2 \cdot N')$-involved convolution operator \emph{two} times per BPEG-M iteration.
(Both methods use identical computations involved with $f(x;y)$, i.e., (back-)projections by $A$ (and $A^T$).)
Different from $\{ d_k^\star \!\in\! \bbR^{49} \!:\! k \!\in\! [49] \}$, 
$\{ w_k^\star \!\in\! \bbR^{64} \!:\! k \!\in\! [64] \}$ does not satisfy the TF condition \R{eq:CAOL:TFcond}
and thus, each image update problem  requires a (diagonal) majorizer 
for the entire term $f(x;y) + \mu \sum_{k = 1}^{64} \| w_k^\star \circledast x - z_k \|_2^2$ 
to have easily computable proximal mapping.
Consequently, calculating the gradient of the above term with respect to $x$ at the extrapolated point $\acute{x}^{(i+1)}$ 
uses a $O(64^2 \cdot N')$-involved convolution operator two times;
each sparse code update $\{ z_k^{(i+1)} \!:\! k \!\in\! [64] \}$ uses an additional $O(64^2 \cdot N')$-involved convolution operator.

\end{itemize}

\section{Discussion Related to Modeling Mean Subtraction in \R{sys:CT&CAOL}} \label{sec:mean-sub}

In \R{sys:CT&CAOL}, the exact mean value for the unknown signal $x$ is unknown, and thus we do not model the mean subtraction operator. 
We observed that including the mean subtraction operator to \R{sys:CT&CAOL} with the \emph{exact} mean value does not improve the reconstruction accuracy. 
Since we have a DC filter among the TF filters learned via CAOL~\R{sys:CAOL:orth} (see examples in Fig.~\ref{fig:CAOL:filters:fruit&city&ct-ii}(c) and \cite[Fig.~2]{Chun&Fessler:18Asilomar}), the mean subtraction operator is not required to shift the sparse codes $\{ z_k^{(i+1)} : \forall k, i \}$ to have a zero mean.

\bibliographystyleSupp{IEEEtran}
\bibliographySupp{referencesSupp_Bobby}

\end{document}